\documentclass{aa}
\usepackage{graphicx,epsfig,natbib}
\usepackage{graphicx,epstopdf}
\usepackage{float}

\def\h{$^h$~}
\def\m{$^m$~}
\def\s{$^s$~}
\def\deg{$^{\circ}$~}
\def\arcmin{$^{\prime}$~}
\def\arcsec{$^{\prime\prime}$\,}

\def\gtsim{\raise 2pt \hbox {$>$} \kern-0.6em \lower 4pt \hbox {$\sim$}}

\begin{document}

\title{Diffuse Radio Sources in a Statistically Complete Sample of High-Redshift Galaxy Clusters}

\author{G. Giovannini$^{1,2}$\thanks{E-mail:
ggiovann@ira.inaf.it}, M. Cau $^{1,2}$, A. Bonafede$^{1,2}$,  H. Ebeling$^{3}$,
L. Feretti$^{2}$, M. Girardi$^{4}$, M. Gitti$^{1,2}$, F. Govoni$^{5}$, 
 A. Ignesti$^{1,2}$,  M. Murgia$^{5}$, G.B. Taylor$^{6}$, V. Vacca$^{5}$}

 \institute{
Dipartimento di Fisica e Astronomia, University of Bologna, Via Gobetti 93/2, 4019 Bologna, Italy
\and
INAF - Istituto di Radioastronomia, Via P. Gobetti 101, 40129 Bologna, Italy
\and
Institute for Astronomy, University of Hawaii, 2680 Woodlawn Drive, Honolulu, HI 96822, USA
\and
Dipartimento di Fisica, Università di Trieste
\and
INAF - Osservatorio Astronomico di Cagliari, Via della Scienza 5, I-09047 Selargius (CA), Italy
\and
Department of Physics and Astronomy, University of New Mexico, Albuquerque, NM 87131, USA
}

\abstract{
{\it Aims.}
Non-thermal properties of galaxy clusters have been studied with detailed and deep radio images in comparison with X-ray data. While much progress has been made, most of the studied clusters are at a relatively low redshift (z $<$ 0.3).  We here investigate the evolutionary properties of the non-thermal cluster emission using two statistically complete samples at z $>$ 0.3.

{\it Methods.}
We obtained short JVLA observations at L-band of the statistically complete sample of very X-ray luminous clusters from the Massive Cluster Survey (MACS) presented by Ebeling et al.\ (2010), namely 34 clusters in the redshift range of 0.3--0.5 and with nominal X-ray fluxes in excess of 2 $\times$ 10$^{-12}$ erg s$^{-1}$ cm$^{-2}$ (0.1--2.4 keV) in the ROSAT Bright Source Catalogue. We add to this list the complete  sample of the 12 most distant MACS clusters ($z>0.5$) presented in Ebeling et al.\ (2007). 

{\it Results.}
Most clusters show evidence of emission in the radio regime. We present the radio properties of all clusters in our sample and show images of newly detected diffuse sources. A radio halo is detected in 19 clusters, and five clusters contain a relic source. Most of the brightest cluster galaxies (BCG) in relaxed clusters show radio emission with powers typical of FRII radio galaxies, and some are surrounded by a radio mini-halo.

{\it Conclusions.}
The high frequency of radio emission from the BCG in relaxed clusters suggests that BCG feedback mechanisms are in place already at z $\sim$ 0.6. The properties of radio halos and the small number of detected relics suggest redshift evolution in the properties of diffuse sources. The radio power (and size) of radio halos could be related to the number of past merger events in the history of the system. In this scenario, the presence of a giant and high-power radio halo is indicative of an evolved system with a large number of past major mergers.
}


\authorrunning{Giovannini et al.}
\titlerunning{Non-thermal diffuse sources in high-redshift clusters}

\maketitle

\section{Introduction}

Clusters of galaxies are characterized by X-ray emission from the hot intra-cluster medium (ICM, $T \sim 2{-}10$ keV). Thermal emission is a common property of all clusters of galaxies and has (albeit at lower energies) been detected even in poor groups as well as in filaments connecting rich clusters \citep[see, e.g.,][]{Boeringer2010}.

In some clusters, the ICM is also characterized by non-thermal diffuse emission in the form of giant radio sources with a spatial extent similar to that of the hot ICM, which are called radio halos or relics according to their morphology and location in the cluster \citep[see, e.g.,][]{Feretti2012,Weeren2019}. These sources are not directly associated with the activity of individual galaxies and are related to physical properties of the whole cluster. In a few cases, observations have revealed that diffuse non-thermal  emission can extend beyond clusters, as  demonstrated by the discovery of radio filaments connecting massive galaxy clusters \citep{Vacca2018b,Govoni2019}.

Radio halos and relics originate from mergers of massive clusters. A small fraction of the gravitational energy released in these events is converted into magnetic-field amplification and into the acceleration of relativistic particles in the ICM, giving rise to diffuse radio sources \citep[see, e.g.,][]{Feretti2012,Brunetti2014}.

Relaxed clusters are characterized by the radio activity of the Brightest Cluster Galaxy (BCG) and by its interaction with the cluster's cool core. Moreover, in some relaxed clusters, radio observations show the presence of diffuse synchrotron radio sources that extend far from the dominant radio galaxy, named mini-halos. These sources are typically moderately extended ($\sim$ 300--500 kpc) and, like halos and relics, exhibit a low surface brightness and a steep spectrum. Their emission originates from relativistic particles and magnetic fields, which are believed to be deeply mixed with the thermal intracluster gas in the dense central regions of cool-core clusters \citep[e.g.,][]{Feretti2012,Giacintucci2019}. Recently, LOFAR observations conducted by \cite{Savini2019} found cluster-scale radio emission with a steep spectrum in a few relaxed clusters of galaxies, suggesting that the sloshing of the cluster's cool core could trigger particle reacceleration on a scale larger than that of classical mini-halos.

The properties of halos, relics, and mini-halos prove that relativistic electrons and magnetic fields are present within the cluster volume. Hence,  studies of these features provide a unique opportunity to probe the strength and structure of the magnetic field on Mpc scales \citep[e.g.,][]{Vacca2018}. As importantly, the location and properties of these diffuse non-thermal sources can be related to cluster characteristics derived from optical and X-ray observations, and are tightly connected to the cluster's evolutionary history. In particular halos and relics are always located in clusters showing merging activity,  even if not all merging clusters feature diffuse radio emission. Conversely, mini-halos are found at the center of cool-core clusters with active BCGs, and their properties are connected to local cool-core properties \citep[e.g.,][]{Bravi2016}.

Although the increasing interest in these sources has led to significant improvements in our understanding of non-thermal cluster properties, we do not understand yet why some merging clusters host a radio halo while others do not show any extended emission. Moreover, the correlation of the non-thermal  cluster properties with cluster evolution remains unclear.
 
To better investigate these issues we need radio observations of a homogeneous sample of clusters across a wide range of redshifts. Most of the clusters studied to date are relatively nearby ($z<0.3$). Among the exceptions are the El Gordo cluster at $z = 0.87$, where a halo and three relics have been found \citep {Lindner2014}, and the halo in PLCK147.3$-$16.6 at $z = 0.65$ \citep{Weeren2014}, confirming that non-thermal emission is likely present also at high redshift. This is an important point: since galaxy clusters are the most massive objects in the Universe, their numbers evolve rapidly with time, and their masses double on average from $z\sim 0.5$ \citep{Boylan2009,Haines2018} to the present epoch. This growth is driven by mergers, in agreement with clear X-ray evidence of substructure in about 40\% of local clusters \citep[e.g.,][]{Jones1999,Mann2012}. 

To investigate these dependencies, we have used the statistically complete sample of very X-ray luminous clusters from the Massive Cluster Survey \citep[MACS,][]{Ebeling2001} presented by \cite{Ebeling2010}. This sample comprises all 34 MACS clusters at $z=0.3-0.5$ with nominal X-ray fluxes (in the 0.1--2.4 keV band) in excess of 2 $\times$ 10$^{-12}$ erg s$^{-1}$ cm$^{-2}$ in the ROSAT Bright Source Catalogue. We add to this list the complete sample of the 12 most distant MACS clusters ($z>0.5$) presented in \cite{Ebeling2007}; the most distant MACS cluster is found at $z\sim 0.7$. Comprising both `dynamically active' (i.e., merging) clusters and `relaxed' systems, this sample stands to provide key insights on the non-thermal properties of high-redshift clusters, thus allowing us to study evolutionary trends.

We have observed all of these clusters with the JVLA in the L band (project 17A-025), excluding the few well studied systems presented by \cite{Feretti2012} and in more recent papers. We also used archival JVLA data.

Our paper is structured as follows. New observations (and the applied data-reduction techniques) are described in Sect.~2. In Sect.~3 we present our results with a short description of each cluster, before providing a discussion of the overall results in Sect.~4. We draw conclusions in Sect.~5.

Intrinsic source properties and parameters quoted in this paper are computed for a $\Lambda$CDM cosmology with $H_0$ = 71 km s$^{-1}$ Mpc$^{-1}$,
$\Omega_m = 0.27$, and $\Omega_{\Lambda} = 0.73$.

\begin{table*}
\caption{List of Targets with new observations or archival data}
\begin{footnotesize}
\begin{center}
\begin{tabular}{lccccccc}
\hline
\noalign{\smallskip}
Name & array & on time & obs. date & array & on time & obs. date & archival\\
     &      & minutes & dd-mm-yy  &       & minutes & dd-mm-yy  & number \\
\noalign{\smallskip}
\hline
\noalign{\smallskip}
MACS J0011.7$-$1523 &  C & 25 & 17-07-17 &   &    &          & \\
MACS J0018.5$+$1626 &  C & 24 & 01-07-17 &   &    &          & \\
MACS J0025.4$-$1222 &  C & 24 & 17-07-17 &   &    &          & \\
MACS J0035.4$-$2015 &  C & 24 & 17-07-17 &   &    &          & \\
MACS J0152.5$-$2852 &  C & 24 & 09-07-17 &   &    &          & \\
MACS J0159.8$-$0849 &  C & 24 & 09-07-17 & D & 10 & 08-04-17 & \\
MACS J0242.5$-$2132 &  C & 26 & 06-07-17 & D & 13 & 04-03-17 & \\
MACS J0257.1$-$2325 &  C & 25 & 06-07-17 & D & 10 & 04-03-17 & \\
MACS J0257.6$-$2209 &  C & 25 & 06-07-17 & D & 10 & 04-03-17 & \\
MACS J0308.9$+$2645 &  C & 24 & 01-07-17 & D & 10 & 20-02-17 & \\
MACS J0358.8$-$2955 &  C & 25 & 06-07-17 & D & 10 & 04-03-17 & \\
MACS J0404.6$+$1109 &  B &180 & 18-07-12 & C & 70 & 02-02-12 & 11B.018 \\
MACS J0417.5$-$1154 &  B &150 & 19-08-12 & C & 90 & 01-02-12 & 11B-018 \\
                  &  D &150 & 05-10-11 &   &    &          & 11B-018 \\
MACS J0429.6$-$0253 &  C & 25 & 08-07-17 &   &    &          & \\
MACS J0454.1$-$0300 &  C & 25 & 08-07-17 &   &    &          & \\
MACS J0520.7$-$1328 &  C & 23 & 01-07-17 &   &    &          & \\
MACS J0547.0$-$3904 &    &    &          & D & -- & ----     & NVSS \\
MACS J0647.7$+$7015 &  C & 23 & 24-07-17 & D & 12 & 14-02-17 & \\
MACS J0744.8$+$3927 &  C & 25 & 22-07-17 & D & 10 & 14-02-17 & \\
MACS J0911.2$+$1746 &  C & 25 & 22-07-17 & D & 12 & 14-02-17 & \\
MACS J0947.2$+$7623 &  C & 22 & 25-06-17 & D & 10 & 14-02-17 & \\
MACS J0949.8$+$1708 &  C & 24 & 22-07-17 & D & 10 & 14-02-17 & \\
MACS J1115.8$+$0129 &    &    &          & D & 10 & 06-05-17 & \\
MACS J1149.5$+$2223 &  B & 60 & 17-11-13 & C &150 & 24-06-13 & 13A.056 \\
                  &    &    &          & D & 60 & 14-02-13 & 13A.056 \\
MACS J1206.2$-$0847 & B/C& 90 & 14-05-12 & D & 09 & 06-05-17 & 12A.164 (B/C)\\
MACS J1319.9$+$7003 &  C & 24 & 28-08-17 &   &    &          & \\
MACS J1423.8$+$2404 &  C & 25 & 17-07-17 &   &    &          & \\
MACS J1427.6$-$2521 &    &    &          & D & 09 & 06-05-17 & \\
MACS J1532.8$+$3021 &  C & 25 & 17-07-17 &   &    &          & \\  
MACS J1720.2$+$3536 &  C & 24 & 17-07-17 &   &    &          & \\
MACS J1731.6$+$2252 &  C & 60 & 08-03-12 &   &    &          & 11B-018 \\
MACS J1931.8$-$2634 &  C & 23 & 26-05-17 & D & 15 & 12-02-17 & \\
MACS J2049.9$-$3217 &  C & 22 & 26-05-17 & D & 11 & 12-02-17 & \\
MACS J2129.4$-$0741 &  C & 24 & 24-05-17 & D & 10 & 12-02-17 & \\
MACS J2140.2$-$2339 &    &    &          & D & 11 & 12-02-17 & \\
MACS J2211.7$-$0349 &  C & 23 & 24-05-17 & D & 10 & 12-02-17 & \\
MACS J2214.9$-$1359 &    &    &          & D & 11 & 12-02-17 & \\
MACS J2228.5$+$2036 & C & 26 & 21-05-17 & D & 12 & 13-02-17 & \\
MACS J2229.7$-$2755 &    &    &          & D & 10 & 12-02-17 & \\
MACS J2245.0$+$2637 &  C & 24 & 21-05-17 & D & 10 & 13-02-17 & \\
MACS J2311.5$+$0338 &  C & 24 & 01-07-17 & D & 11 & 20-02-17 & \\
                  &   B & 20 & 01-01-14 & C & 60 & 08-06-13 & 13A.268 \\
\noalign{\smallskip}
\hline
\noalign{\smallskip}
\label{tab1}
\end{tabular}
\end{center}
\end{footnotesize}
\end{table*}

\begin{table*}
\caption{Results for the complete sample}
\begin{footnotesize}
\begin{center}
\begin{tabular}{lcccclccrll}
\hline
\noalign{\smallskip}
Name & z & Kpc/\arcsec & L$_{X-500}$ & Code & Source & S$_{1.5}$ & LogP$_{1.5}$ & LLS & Ref & Notes \\
     &   &    & 10$^{44}$ erg~s$^{-1}$    &   & type   & mJy      & W~Hz$^{-1}$         & Mpc &     &       \\
\noalign{\smallskip}
\hline
\noalign{\smallskip}
MACS J0011.7$-$1523 & 0.379 &5.17 & 8.9 &   1 &   P & 19.8$\pm$0.1 &25.00     & --  &     &     \\
MACS J0014.3$-$3022 & 0.308 &4.50 &13.6 &   4 &   H & 57.1$\pm$1.1 &  25.24     & 1.89& G01 &A2744 \\
                 &       &     &     &     &   R1 & 18.2$\pm$0.4 & 24.71     & 1.62& G01 & \\
                 &       &     &     &     &   R2 & 2.18$\pm$0.17&23.98     & 1.15& P17 &  \\
                 &       &     &     &     &   R3 & 1.46$\pm$0.14&23.67     & 1.1 & P17 &  \\
                 &       &     &     &     &   R4 & 0.88$\pm$0.05&23.48     & 0.05& P17 &  \\
MACS J0018.5$+$1626 & 0.5456&6.37 &19.6& 3 & H      & 9.95$\pm$0.02&25.07 &1.57& *   & CL0016+1609\\   
MACS J0025.4$-$1222 & 0.5843&6.59 & 8.8& 3 &     R1 & ----         &24.11 &0.6   & R17&  \\

                  & & &  &  &                  R2 & ----         &24.25 & 0.6   & R17&  \\

MACS J0035.4$-$2015 & 0.352 &4.93 &11.9 &3 &  --    & --           & --  &     &     &  \\ 
MACS J0152.5$-$2852 & 0.413 &5.45 & 8.6 &   2 &   H & 2.34$\pm$0.15&24.15 & 0.49& *    & \\
MACS J0159.8$-$0849 & 0.406 &5.39 &16.0 &   1 &  mH & 2.2 $\pm$0.2 & 24.10        & 0.09&     & \\
MACS J0242.5$-$2132 & 0.314 &4.56 &14.2 &   1 &   P &1140$\pm$60& 26.56        & --  &     &     \\
MACS J0257.1$-$2325 & 0.5049&6.11 &13.7 &   2 &   P &4.78$\pm$0.18& 24.67 & -- &    &  \\
MACS J0257.6$-$2209 & 0.322 &4.64 & 7.0 &   2 & H   &22.0$\pm$0.3 & 24.83        & 0.42&  *   & A402 \\
MACS J0308.9$+$2645 & 0.356 &4.96 &14.7 &   2 &   H &9.7$\pm$3.4 & 24.62        & 0.77& *    &  \\
MACS J0358.8$-$2955 & 0.425 &5.54 &18.9 &   4 &   H &2.7$\pm$0.2 & 24.24   & 0.22&  *   & A3192 \\
MACS J0404.6$+$1109 & 0.352 &4.93 & 4.3 &   4 &   P &14.7$\pm$0.3 & 24.90     & --  &     & \\  
MACS J0417.5$-$1154 & 0.443 &5.68 &29.1 &   3 &   H &33.7$\pm$0.7 & 25.38       & 1.20 & *    & \\
MACS J0429.6$-$0253 & 0.399 &5.34 &10.9 &   1 &   P &132.0$\pm$3.0 & 25.89      & --  &     &   \\ 
MACS J0454.1$-$0300 & 0.5377&6.32 &16.8 &   2 &   H &0.79$\pm$0.05& 23.20 &0.3 & *   & \\
MACS J0520.7$-$1328 & 0.336 &4.78 & 7.9 &   2 &   R &2.04$\pm$0.05& 23.88  & 1.1 & *    & \\ 
MACS J0547.0$-$3904 & 0.319 &4.61 & 6.4 &   2 &   P &85.0 $\pm$2.0& 25.45        & --  &NVSS &  \\
MACS J0647.7$+$7015 & 0.5907&6.62 &15.9&    2 &   H &0.46$\pm$0.05 & 23.82 &0.2 & *   & \\ 
MACS J0717.5$+$3745 & 0.5458&6.37 &24.6&    4 &   H &118.0$\pm$3.0 & 26.20 &1.5 & B09& \\
MACS J0744.8$+$3927 & 0.6976&7.13 &22.9&    2 &   P &0.87$\pm$0.03 & 24.27 & -- &    & \\
MACS J0911.2$+$1746 & 0.5049&6.11 & 7.8& 4 & -- & --    & -- &    & \\
MACS J0947.2$+$7623 & 0.354 &4.94 &20.0 &   1 &  mH &12.5$\pm$0.2 & 24.72    & 0.15&     & RBS0797 \\
MACS J0949.8$+$1708 & 0.384 &5.21 &10.6 &  2 &    H & ---         & 24.10 &$\sim$1 &B15 & Z2661\\
MACS J1115.8$+$0129 & 0.355 &4.95 &14.5 &  1 &   mH &9.21$\pm$0.08&24.59   & 0.05&  *   & \\ 
MACS J1131.8$-$1955 & 0.306 &4.48 &13.7 &   4 &   H &21.0$\pm$0.4 & 24.78    & 1.3 & R99 & A1300 \\
                  &       &     &     &     &   R &20.0$\pm$0.4 & 24.76  & 0.70& R99 & \\
MACS J1149.5$+$2223 & 0.5444&6.36 &17.6 &   4 &   H &0.9$\pm$0.1 & 24.01 &0.4 &  *  &  \\
                  &       &     &     &     &   F &3.76$\pm$0.03 & 24.65 & 0.6&    &  \\
MACS J1206.2$-$0847 & 0.439 &5.65 &21.1 &   2 &   P &108.8$\pm$0.2 & 25.88  & -   &     & \\
MACS J1319.9$+$7003 & 0.327 &4.69 & 4.2 &   2 & --  & --           &        & --  &     & A1722 \\
MACS J1347.5$-$1144 & 0.451 &5.74 &42.2 &   1 & mH  &25.2$\pm$0.5  & 25.27 &0.57 & G07 & RXJ1347.5-1145 \\
MACS J1423.8$+$2404 & 0.5431&6.35 &16.5 &   1 & P  &5.47$\pm$0.02 & 27.21 & -- &    & \\ 
MACS J1427.6$-$2521 & 0.318 &4.60 & 4.1 &   1 & P   &4.1$\pm$0.1  & 24.13  & -   &     &  \\
MACS J1532.8$+$3021 & 0.363 &5.03 &19.8 &   1 & mH  &4.4$\pm$0.3 & 24.29  &0.2  &     &  \\
MACS J1720.2$+$3536 & 0.387 &5.23 &10.2 &   1 & P  &17.2$\pm$0.3 & 24.95  & -   &   &  Z8201 \\
MACS J1731.6$+$2252 & 0.389 &5.25 & 9.3 &   4 & H  &3.23$\pm$0.03 & 24.23    &0.24 & *    & \\
MACS J1931.8$-$2634 & 0.352 &4.93 &19.7 &   1 & mH  &50.0$\pm$4.0& 25.32   &0.1  &     & \\
MACS J2049.9$-$3217 & 0.323 &4.65 & 6.1 &   3 & P &0.45$\pm$0.05& 23.18  & -- & * & \\ 
MACS J2129.4$-$0741 & 0.5889&6.67 &15.7 &   3 & (H)  &0.33$\pm$0.30& 23.67 &0.2 &  * & \\
MACS J2140.2$-$2339 & 0.313 &4.55 &11.1 &   1 & P  &3.8$\pm$0.2& 24.08  & --   & Y18&  \\
MACS J2211.7$-$0349 & 0.397 &5.37 &24.0 &   2 & (H)  &0.58$\pm$0.05 & 23.50 &0.22 &  *   & \\ 
MACS J2214.9$-$1359 & 0.5027&6.10 &14.1&    2 & -- & --           & -- &  &  & \\
MACS J2228.5$+$2036 & 0.411 &5.43 &13.3 &   4 & H &15.0$\pm$0.1 & 25.00 &1.09 & * & RXJ2228.6+2037\\ 
MACS J2229.7$-$2755 & 0.324 &4.66 &10.0 &   1 & P &4.3$\pm$0.2 & 24.17  & P   &     & \\ 
MACS J2243.3$-$0935 & 0.447 &5.71 &15.2 &   3 & H & --- & 24.48   & 0.9 & C16 &  \\
                  &       &     &     &     & R & --- & 24.14  &  0.68 & C16 & \\
MACS J2245.0$+$2637 & 0.301 &4.43 & 7.6 &   1 & P & 4.5$\pm$0.1 & 24.11   & P   &     & \\
MACS J2311.5$+$0338 & 0.305 &4.47 &12.9 &   3 & (H) & 0.32$\pm$0.02 & 23.12  & 0.15& * & A2552\\
\noalign{\smallskip}
\hline
\label{tab2}
\end{tabular}
\end{center}
\end{footnotesize}

{
Col. 1: name; Col. 2: redshift \citep{Ebeling2007,Ebeling2010}; 
Col. 3: angular-to-linear scale conversion; 
Col. 4: L$_{X-500}$ in units of 10$^{44}$ erg~s$^{-1}$ \citep{Ebeling2007,Ebeling2010}; 
Col. 5: Morphological code \citep{Ebeling2007}: 
1 pronounced cool core; 2 good optical/X-ray 
alignment; 3 small-scale substructure; 4 multiple peaks;
Col. 6: radio-emission classification: P = unresolved, H
= halo, (H) = halo candidate, R = relic, F = filament, 
mH = mini-halo; 
Col. 7 = flux density $\pm$ 1$\sigma$ uncertainty;
Col. 8: logarithmic radio power at 1.5 GHz in W Hz$^{-1}$,
no K-correction applied; 
Col. 9:  largest linear size in Mpc; Col. 10: reference to
radio data when from the literature: B09=\cite{Bonafede2009}, 
B15=\cite{Bonafede2015}, C16=\cite{Cantwell2016},
G01=\cite{Govoni2001}, G07=\cite{Gitti2007}, P17=\cite{Pearce2017}, 
R99=\cite{Reid1999}, R17=\cite{Riseley2017}, Y18 = \cite{Yu2018};
* = figure shown; 
Col.11: alternative name.}
\end{table*}

\section{Radio observations and data reduction}

All observations were carried out at 1.5 GHz. Table 1 lists all targets, the JVLA configuration, the on-source time (in minutes) and the observing date. Information on archival data used in this work is also provided. 

Observations were performed in the period February to May 2017 (D-configuration) and May to July 2017 (C-configuration). To aid the inclusion of our observations in the instrument observing plan, we prepared schedules in blocks of 1.5 to 2 hours, each consisting of four to five targets close together in the sky. A primary JVLA flux-density scale and bandwidth calibrator was observed during each  run. Phase calibrators were observed every 20 to 30 minutes. Because of the relatively short time on source, we did not attempt to calibrate and to derive polarization information. All observations were obtained with a 1 GHz bandwidth (from 1.008 to 2.032 GHz), divided into 16 spectral windows (IFs) of 64 channels each.

All data were reduced using the Common Astronomy Software Applications  package (CASA); some complex and intriguing cases were additionally reduced with the Astronomical Image Processing System (AIPS) for a check and comparison. Results were found to be in good agreement.

In CASA, we started with the calibration of the data using the standard pipeline. We obtained images and checked if more flagging was necessary. In general both C- and D-configuration data suffered from strong interference mainly on short baselines. Images obtained with natural weighting were used to run a few iterations of phase-only self-calibration and, if necessary, a final amplitude and phase self-calibration.

In AIPS, we started with un-calibrated data and applied the standard calibration procedure. Final images were obtained after self-calibration in phase and gain. Archival data were calibrated and reduced with AIPS.

\section{Results}

The sample studied here consists of 46 high-redshift clusters, derived from the two complete samples described in the Introduction. From a comparison between optical and X-ray data, \cite{Ebeling2007} performed  a morphological classification of these clusters and assigned a morphological code ranging from 1 to 4, to indicate fully virialized to heavily disturbed clusters. A code of 1 signals: pronounced cool core, very good alignment of X-ray peak and (single) cD galaxy; 2: good optical/X-ray alignment, concentric contours; 3: small-scale substructure, non-concentric contours; and 4: poor optical/X-ray alignment, multiple peaks, no cD galaxy.

The clusters in our sample span a wide range of morphologies with no obvious bias in favour of either relaxed or merging systems: there are 15 clusters of code 1, 14 clusters of code 2, eight clusters of code 3, and nine clusters of code 4. The results of the radio observations are reported in Table 2 for all clusters. Note that the radio powers listed in Table~2 were obtained without applying K corrections since, for the great majority of sources, we do not know the spectral index. 
As, however, the spectral index of radio halos and relics is typically close to 1, the K corrections for these sources are expected to be small. Only four clusters do not show any evidence of radio emission at the cluster center. A central radio halo is detected in 19 clusters. A relic source is present in five clusters, three of which also contain a radio halo. In 21 clusters, radio emission from the BCG is detected, sometimes surrounded by a diffuse mini-halo. 

In the following, we provide a short description and comments for each individual cluster. Figures are presented for clusters with diffuse sources that are either new detections or improved images of known sources. They are indicated by an asterisk in Col.\ 9 of Table 2. 

Because of the geometry and expansion of the Universe, the surface brightness of resolved objects becomes fainter at high redshift, scaling as $(1 + z)^{-4}$. As a result, detections at high redshift are biased in favour of sources objects featuring high surface brightness. The cosmological-dimming factor at $z = 0.1$ is 0.7; for the redshift range studied here, this factor drops to 0.35 ($z = 0.3$) and 0.12 ($z = 0.7$). However, we note that we measure the source surface brightness in mJy per beam. The half-power beam width (HPBW) in most images of nearby radio halos is 40\arcsec--50\arcsec (VLA D-array at 20 cm), corresponding to 70--90 kpc at $z = 0.1$. By contrast, most of our images have a HPBW of $\sim$20\arcsec (C-array at 20 cm), corresponding to 100--120 kpc at the redshifts of our targets. We therefore partially compensate for surface-brightness dimming by integrating over a larger area. Moreover, in most cases (see Table 1) we have data obtained with both the C- and D-configurations and can check that no resolved source is missed by combining the two data-sets or comparing the flux densities in images at the two different angular resolutions. The large bandwidth and higher sensitivity of JVLA data with respect to classic VLA data will also help compensate for the brightness-dimming effect.

For a comparison between extended radio features and the X-ray emission, we extracted X-ray data from the Chandra Data Archive. Images were reprocessed using the CIAO 4.9 software package to obtain a filtered image in the 0.1--2.4 keV energy range. Images were subsequently smoothed using a Gaussian function to improve the contrast with respect to the superposed radio images. These images were produced only for a morphological comparison with the radio distribution. The X-ray luminosities reported in Table 2 are from \cite{Ebeling2007} and \cite{Ebeling2010}. The slightly different cosmology used with respect to this paper produces small changes which have no influence on our results.

The optical images shown were obtained in the $i$ filter during the PanSTARRS $3\pi$ survey \citep{Chambers2016} and reach a magnitude limit of $m_{\rm AB}=23.1$. 

\subsection{Comments on individual clusters}

{\bf MACS J0011.7$-$1523} (no figure presented; morphology class 1) -- In our images of this relaxed cluster we detected an unresolved source, with a flux density of $(19.8\pm 0.1)$ mJy at RA: 00\h 11\m 42.85\s, DEC: --15\deg 23\arcmin 21.8\arcsec, coincident with the  cluster center.

{\bf MACS J0014.3$-$3022} (no figure presented; A2744; morphology class 4) -- This heavily disturbed cluster hosts a well known diffuse radio halo source and peripheral relic structures. We have not observed it and refer to \cite{Feretti2012} and \cite{Pearce2017}.

\begin{figure}[ht]
\includegraphics[scale=0.60, angle = 0]{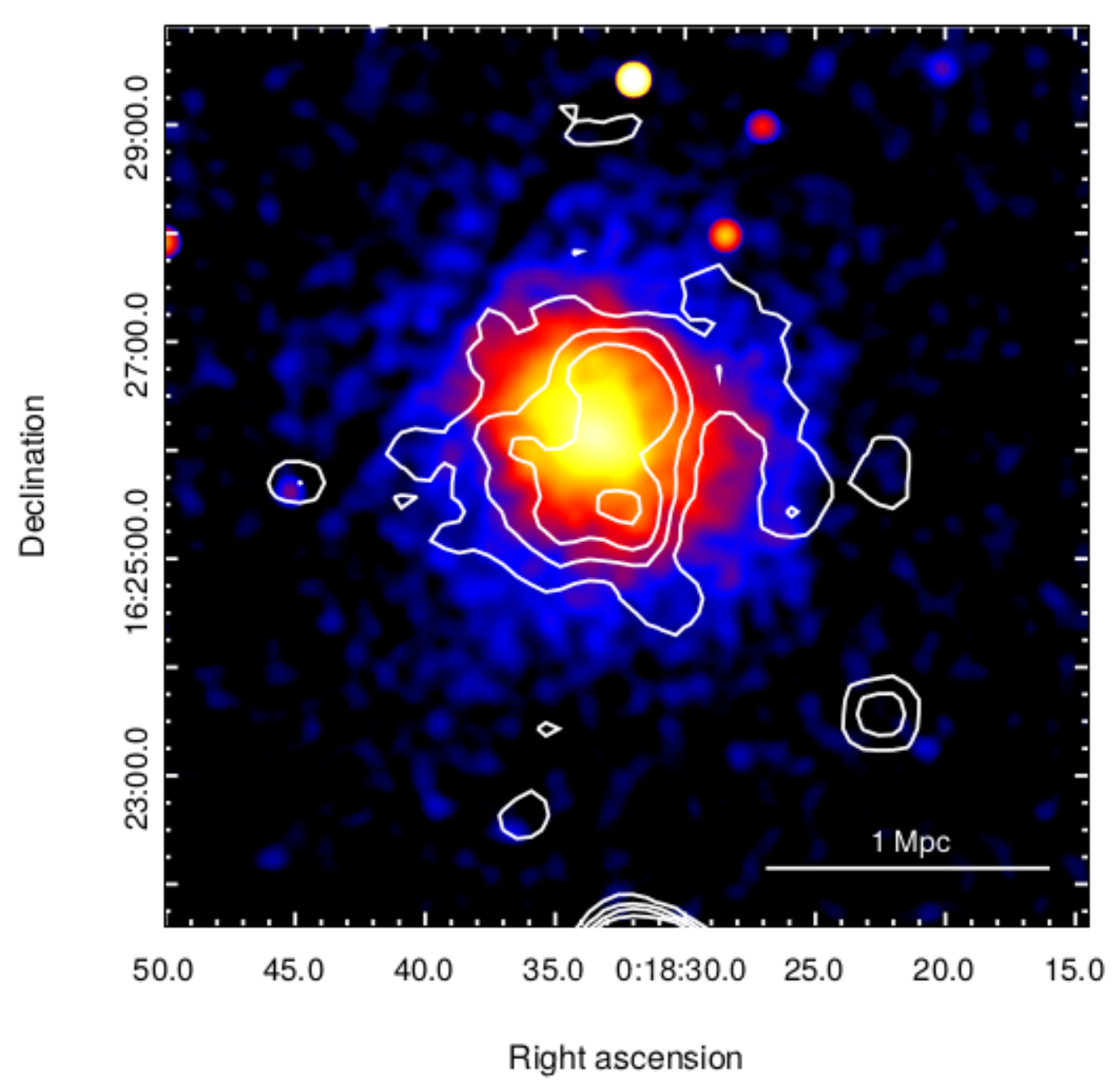}
\caption {Radio emission from the giant radio halo at the center of MACS J0018.5$+$1626 (CL0016$+$26), shown in contours, superposed on the X-ray image obtained by Chandra (colour). The HPBW is 40\arcsec, the noise level is 0.05 mJy per beam, and the shown contour levels are 0.3, 0.5, 0.7, 1 mJy per beam.}
\label{j0018}
\end{figure}

{\bf MACS J0018.5$+$1626} (CL0016$+$26; morphology class 3) -- This cluster was previously studied in the radio domain and is known to host a radio halo \citep{Moffet1989,Giovannini2000}.  The higher angular resolution of the data presented here show a a giant and regular radio halo (Fig.~\ref{j0018}). We do not detect any discrete source within the diffuse radio emission. The emission seen in the West could be connected with the filamentary structure surrounding this cluster studied by \cite{Geach2010}, who found a large number of galaxies in the filament with enhanced star formation possibly due to turbulence effects.

\begin{figure}[ht]
\centering
\minipage{0.4\textwidth}
\includegraphics[width=\linewidth]{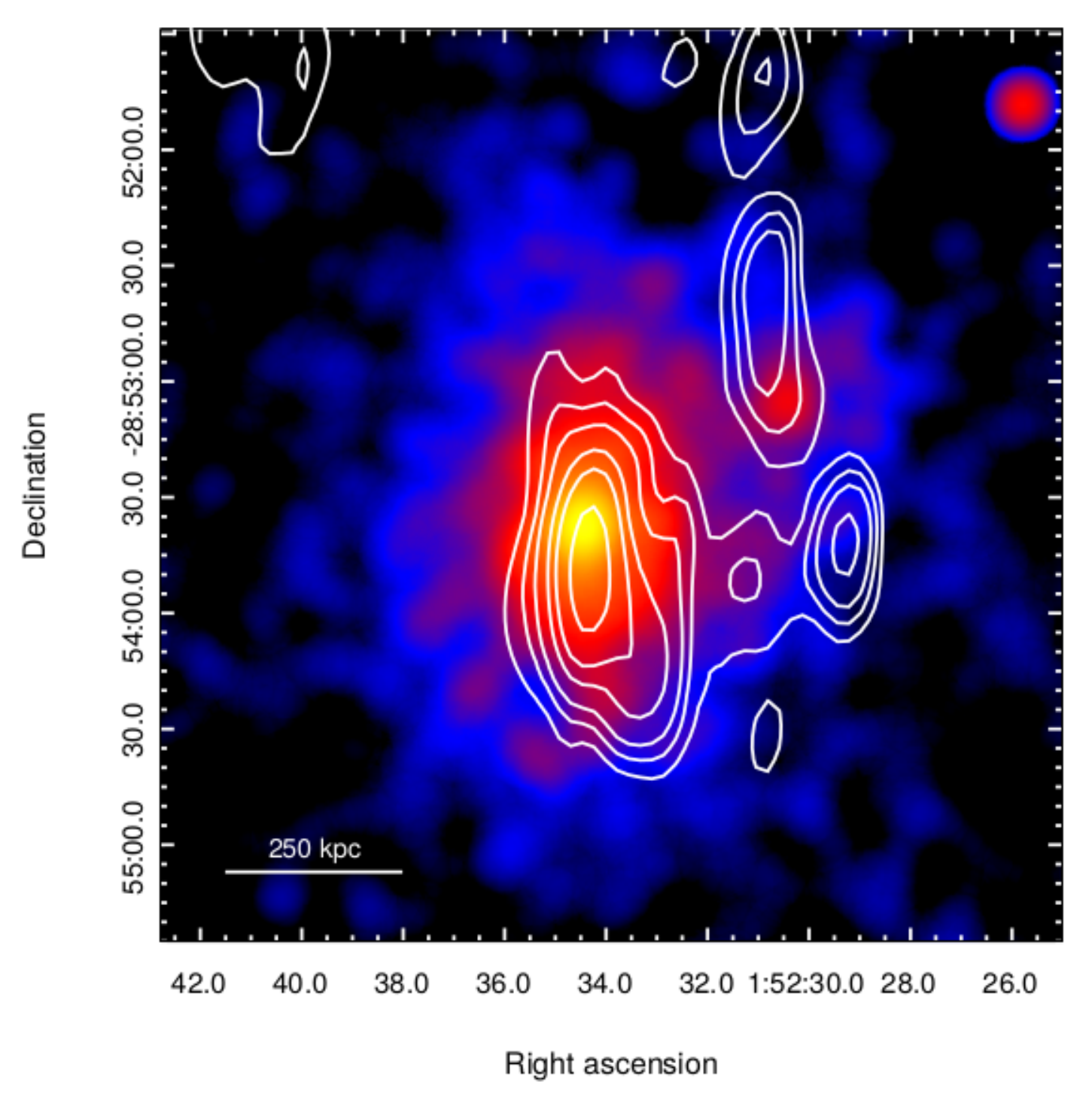}
\endminipage\hfill
\minipage{0.4\textwidth}
\includegraphics[width=\linewidth]{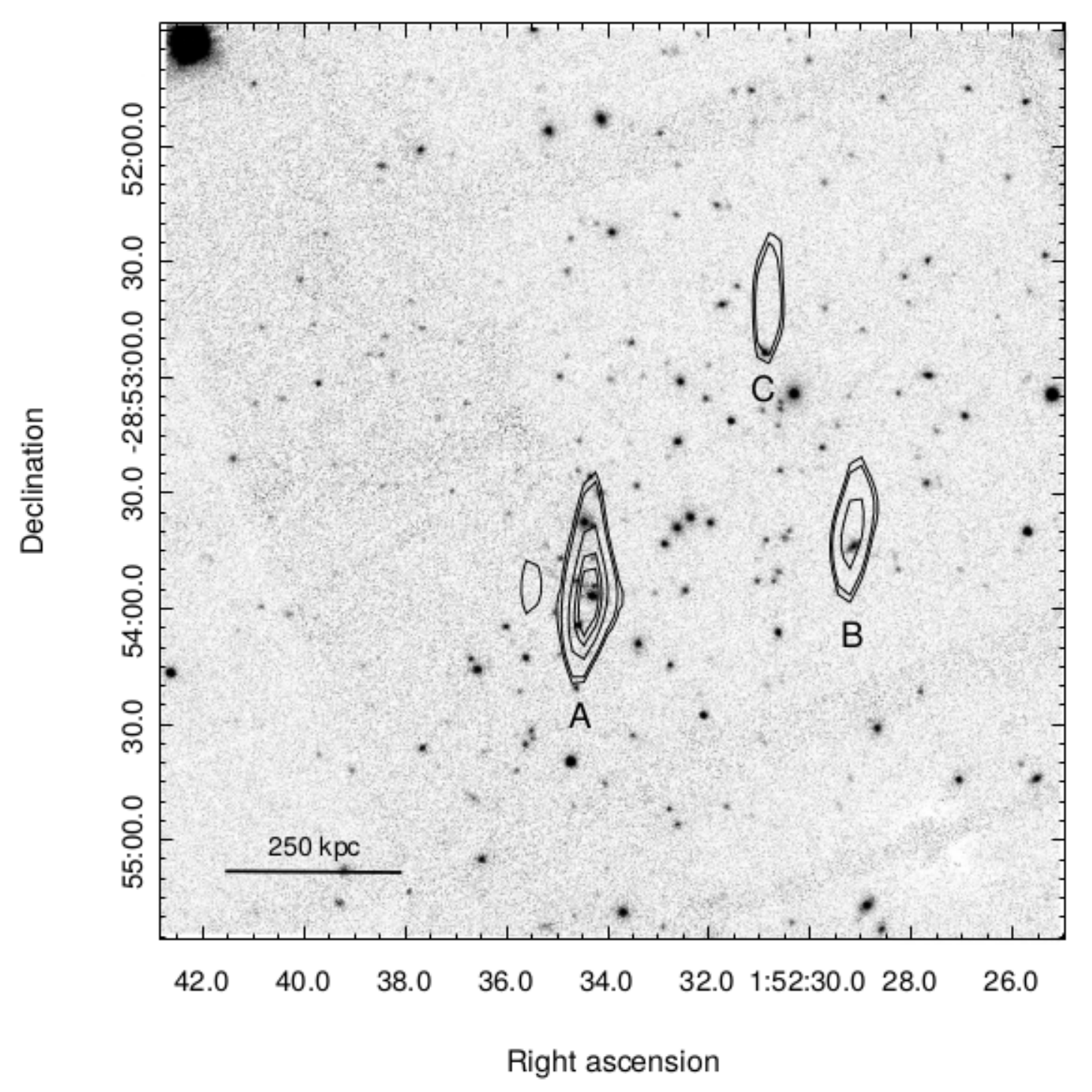}
\endminipage\hfill
\caption{\footnotesize \emph{Top panel}: total-intensity radio contours (C+D array) of MACS J0152.5$-$2852 at 1.5 GHz overlaid on the Chandra X-ray image. The radio image has an FWHM of 
36.2\arcsec$\times$ 12.3\arcsec (PA$=-6^\circ$). Contour levels are at (3, 6, 9, 12, 24, 48)$\times\sigma$ with rms noise $\sigma=$0.034 mJy per beam. \emph{Bottom panel}: C-array contours using only the longest baselines. The beam is 26.3\arcsec$\times$ 6.9\arcsec (PA$=-5.5^\circ$). Contour levels are at (3.6, 4, 8, 9, 10)$\times\sigma$ with rms noise $\sigma=$0.03 mJy per beam. The radio image is superposed on the optical image. Point sources are labeled as A, B, C.}
\label{j0152}
\end{figure} 

{\bf MACS J0025.4$-$1222} (no figure presented; morphology class 3) -- This cluster was studied in detail by \cite{Riseley2017} through GMRT observations at 323 MHz. After subtraction of discrete sources they find double-relic diffuse emission, symmetric with respect to the cluster center.  Assuming a spectral index\footnote{S$(\nu) \propto \nu^{-\alpha}$} of $\alpha=1.3$ they quote radio powers of 1.29 $\times$ 10$^{24}$ W Hz$^{-1}$ and 1.76 $\times$ 10$^{24}$ W Hz$^{-1}$ for the NW and SE relics, respectively. The size of each relic is approximately 600 kpc.

In our images, we detect all discrete sources found by \cite{Riseley2017}, but no diffuse emission is present at a level of 0.05 mJy/beam (one sigma) with a HPBW $\sim$= 25\arcsec suggesting a steep radio spectrum ($\alpha$ $>$ 1.2) for the two relic sources in agreement with the spectral index assumed by \cite{Riseley2017}.

{\bf MACS J0035.4$-$2015} (no figure presented; morphology class 3) -- In our images no radio emission (discrete or diffuse) was detected in the cluster region at a noise level of 0.18 mJy per beam and a circular HPBW of 45\arcsec. The noise is higher than expected because of a nearby, unrelated strong radio source and a high percentage of bad data, but the low resolution of our image provides strong evidence against the presence
of a diffuse source.

{\bf MACS J0152.5$-$2852} (morphology class 2) -- Although this massive cluster exhibits good optical/X-ray alignment and concentric contours, the X-ray image (see Fig. \ref{j0152}, top) is slightly irregular, suggesting a not fully relaxed status.

The radio images show diffuse emission of low surface brightness and a few discrete sources, one of them at the cluster center (A in Fig.~\ref{j0152}, bottom). We produced an image at high resolution using only the longest baselines and subtracted the point-like source A, $(0.19 \pm 0.03)$ mJy, and the nearby source B, $(0.12 \pm 0.03)$ mJy. We find residual diffuse emission with a flux density of $(2.34\pm 0.15)$ mJy and a size of $\sim$ 1.5\arcmin, corresponding to a radio power of 1.41 $\times$ 10$^{24}$ W Hz$^{-1}$ and 490 kpc in size. The bridge connecting the halo to source B as well as the southern extension of source C could be part of the radio halo, although their structure needs to be confirmed.

\begin{figure}[ht]
\centering
\minipage{0.4\textwidth}
\includegraphics[width=\linewidth]{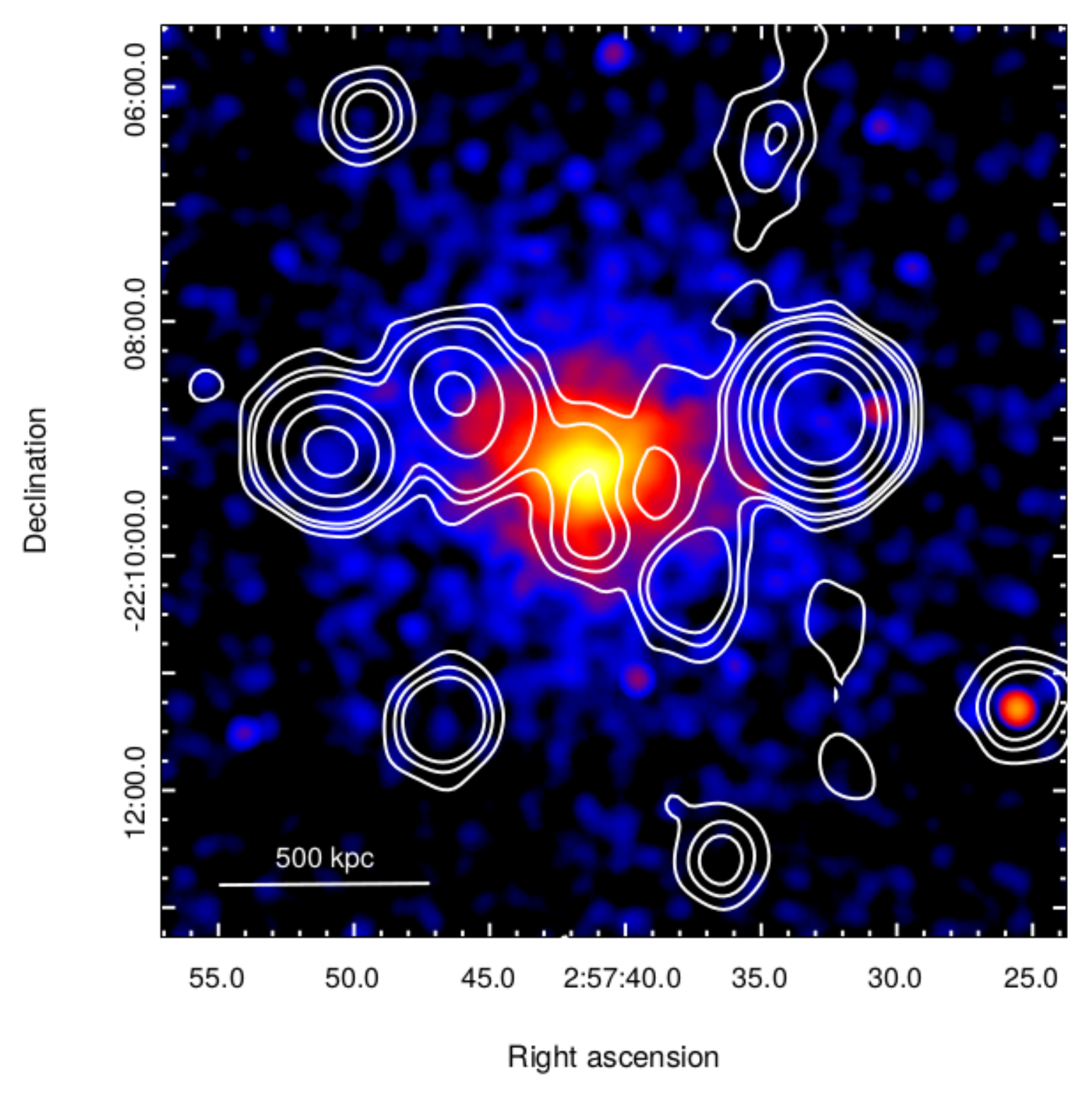}
\endminipage\hfill
\minipage{0.4\textwidth}
\includegraphics[width=\linewidth]{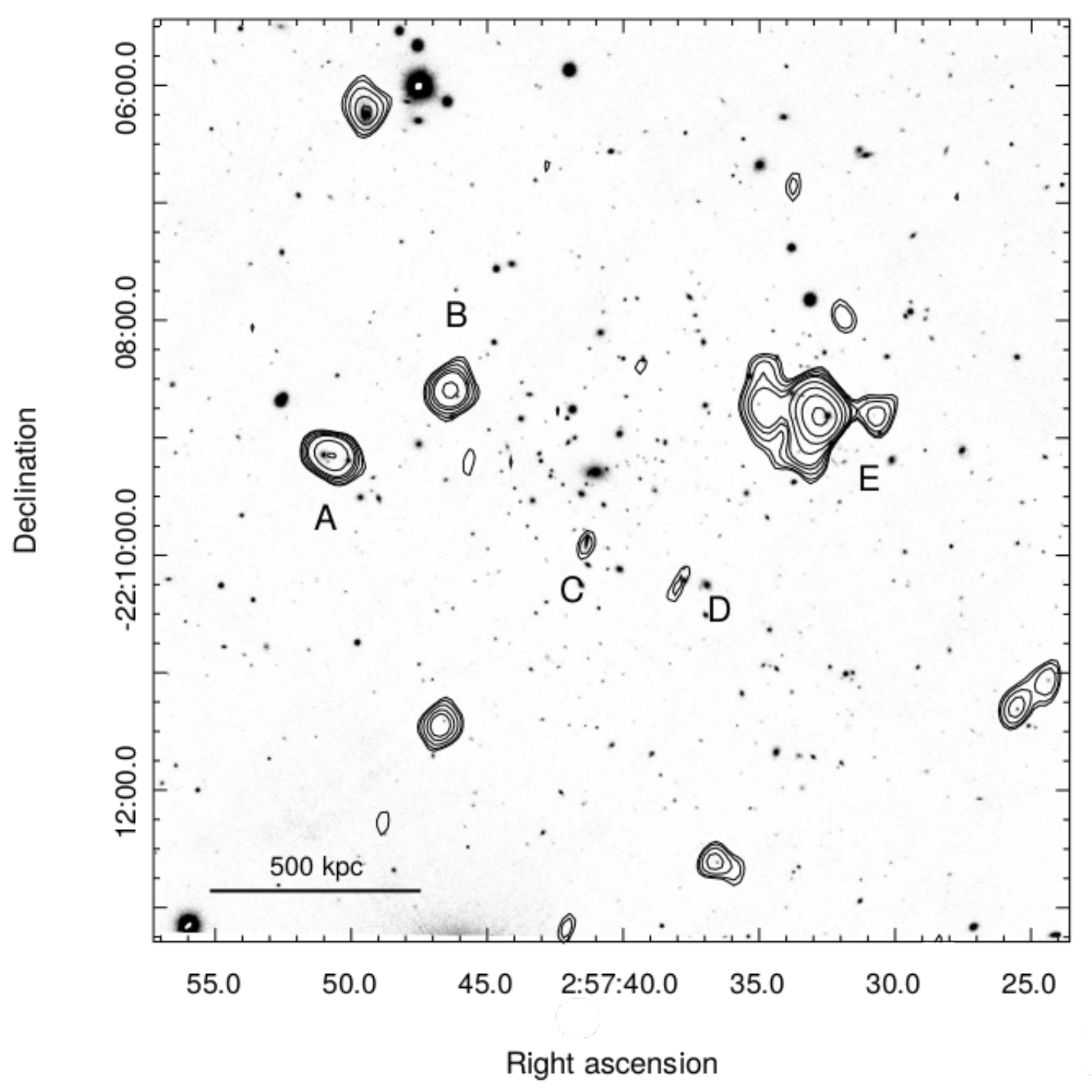}
\endminipage\hfill
\caption{\footnotesize \emph{Top panel}: MACS J0257.6$-$2209 (A402) -- radio contours at 1.5 GHz (total intensity) as detected with the JVLA in C+D configuration overlaid on the Chandra X-ray image (color). The radio HPBW is 37\arcsec (circular). Contour levels are 3, 6, 9, 12, ... $\times \sigma$ with $\sigma = 0.048$ mJy per beam. \emph{Bottom panel}: C array image of compact sources in the field of MACS J0257.6$-$2209, obtained with the longest baselines only, overlaid on the optical image. The HPBW is 15\arcsec (circular). Contour levels are 3, 4, 6, 9, 24, 38, 48, 78, 96 $\times \sigma$ with $\sigma = 0.04$ mJy per beam. Discrete sources are labelled as A, B, C, D, and E.}
\label{j0257}
\end{figure} 

{\bf MACS J0159.8$-$0849} (no figure presented; morphology class 1) -- This cluster shows a regular X-ray morphology, centred on the BCG, and is accordingly classified as a cool-core cluster. \cite{Giacintucci2014} reported a candidate mini-halo. After subtraction of the point-like source at the center (flux density = 35 mJy), we find a residual flux density of $(2.2\pm 0.2)$ mJy, consistent with the value of 2.4 mJy estimated by \cite{Giacintucci2014}.
 
{\bf MACS J0242.5$-$2132} (no figure presented; morphology class 1) -- In our images of this fully relaxed cluster we found an unresolved bright source at RA: 02\h 42\m 35.91\s, DEC: --21\deg 32\arcmin 26.6\arcsec coincident with the cluster center. The measured flux density of $(1.14 \pm 0.06)$ Jy is in good agreement with the value of $(1.26 \pm 0.07)$ Jy reported by \cite{Hlavacek2013}.

{\bf MACS J0257.1$-$2325} (no figure presented; morphology class 2) -- In our images we detected an unresolved component with a flux density of $(4.78 \pm 0.18)$ mJy coincident with the cluster center. Since the flux density measured in the C-array data is consistent with that measured at lower resolution (D-array data), we assume that most of the radio emission is related to the activity of the central galaxy.

\begin{figure}[ht]
\centering
\minipage{0.4\textwidth}
\includegraphics[width=\linewidth]{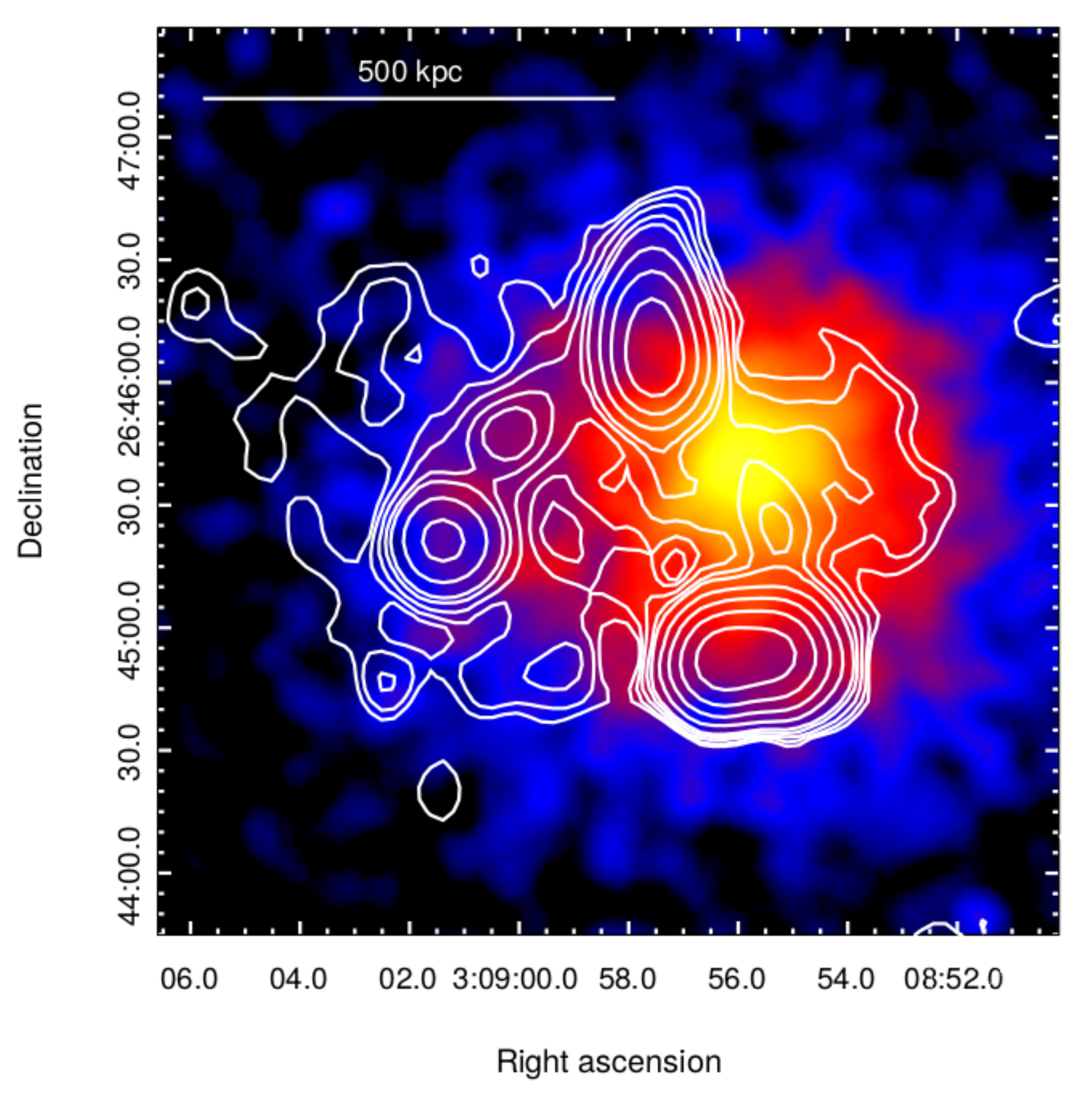}
\endminipage\hfill
\minipage{0.4\textwidth}
\includegraphics[width=\linewidth]{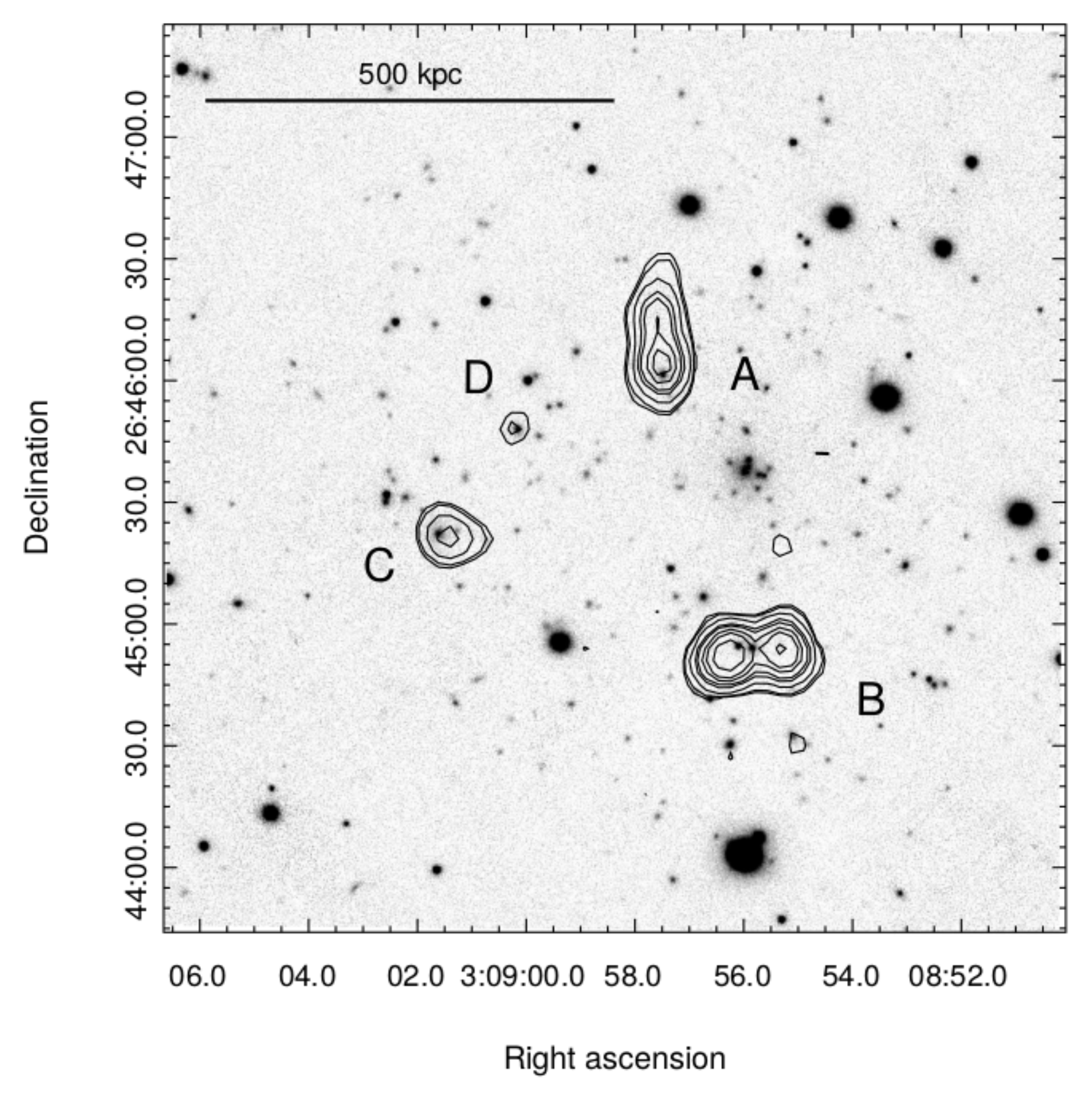}
\endminipage\hfill
\caption{\footnotesize \emph{Top panel}:  Radio contours at 1.5GHz of the cluster MACS J0308.9$+$2645 overlaid on the Chandra X-ray image. The radio image has a FWHM of 16.2\arcsec$\times$14.6\arcsec (PA $-6^\circ$). The contour levels are (3, 6, 9, 12, 24, 48, ...)$\times\sigma$ with rms noise $\sigma=$0.034 mJy per beam. \emph{Bottom panel}: JVLA C-array contours of MACS J0308.9$+$2645 at 1.52 GHz using only the longest baselines. Contours are placed at (3, 4, 9, 20, 25, 40, 55, 90) $\times$ 0.062 mJy per beam and the beam size is 8.7\arcsec$\times$7.7\arcsec (PA $-0.4^\circ$). 
The radio image is superposed on the optical image. Point sources are labeled as A, B, C, and D.}
\label{j0308}
\end{figure} 

{\bf MACS J0257.6-2209} (A402; morphology class 2) -- The radio images show a complex radio morphology (Fig.~\ref{j0257}) with faint diffuse radio emission at the cluster center and many discrete sources possibly related to cluster galaxies. The discrete sources, labelled  A, B, C, D, and E, are well visible in the high-resolution image obtained with the longest baselines only (bottom panel). Their fluxes are 2.50 mJy (A), 1.64 mJy (B) 0.14 mJy (C), 0.16 mJy (D), and 17.31 mJy (E), with an uncertainty of $\pm$0.04 mJy. By subtracting these sources from the total flux density of $(43.5\pm 2.2)$ mJy of the whole radio structure, we find the flux density of the radio halo to be $(22\pm 3)$ mJy, corresponding to a radio power of $6.72\times 10^{24}$ W Hz$^{-1}$. We obtain the same result when subtracting the point sources from the (u,v) data.

\begin{figure}[ht]
\centering
\minipage{0.4\textwidth}
\includegraphics[width=\linewidth]{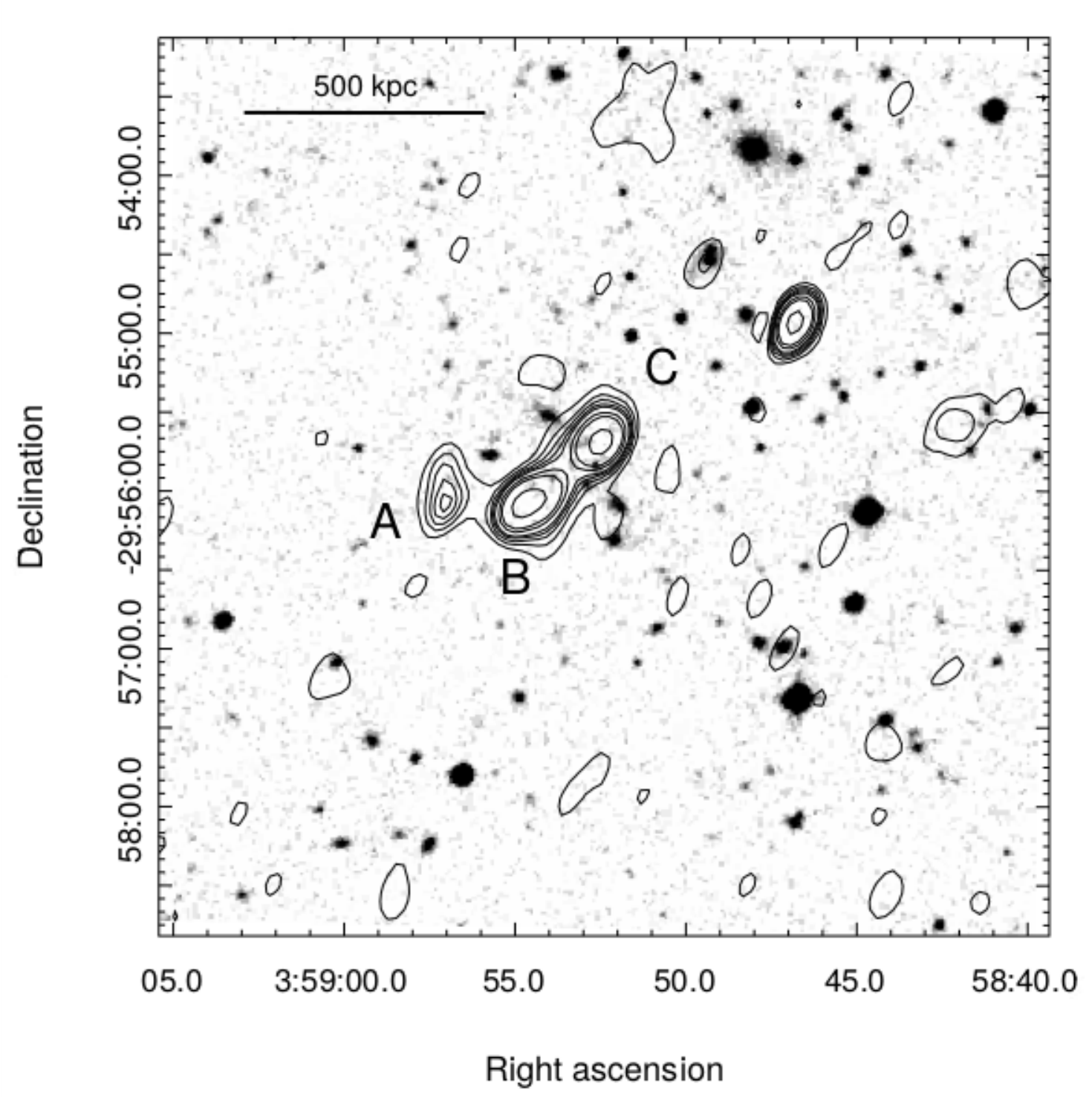}
\endminipage\hfill
\minipage{0.4\textwidth}
\includegraphics[width=\linewidth]{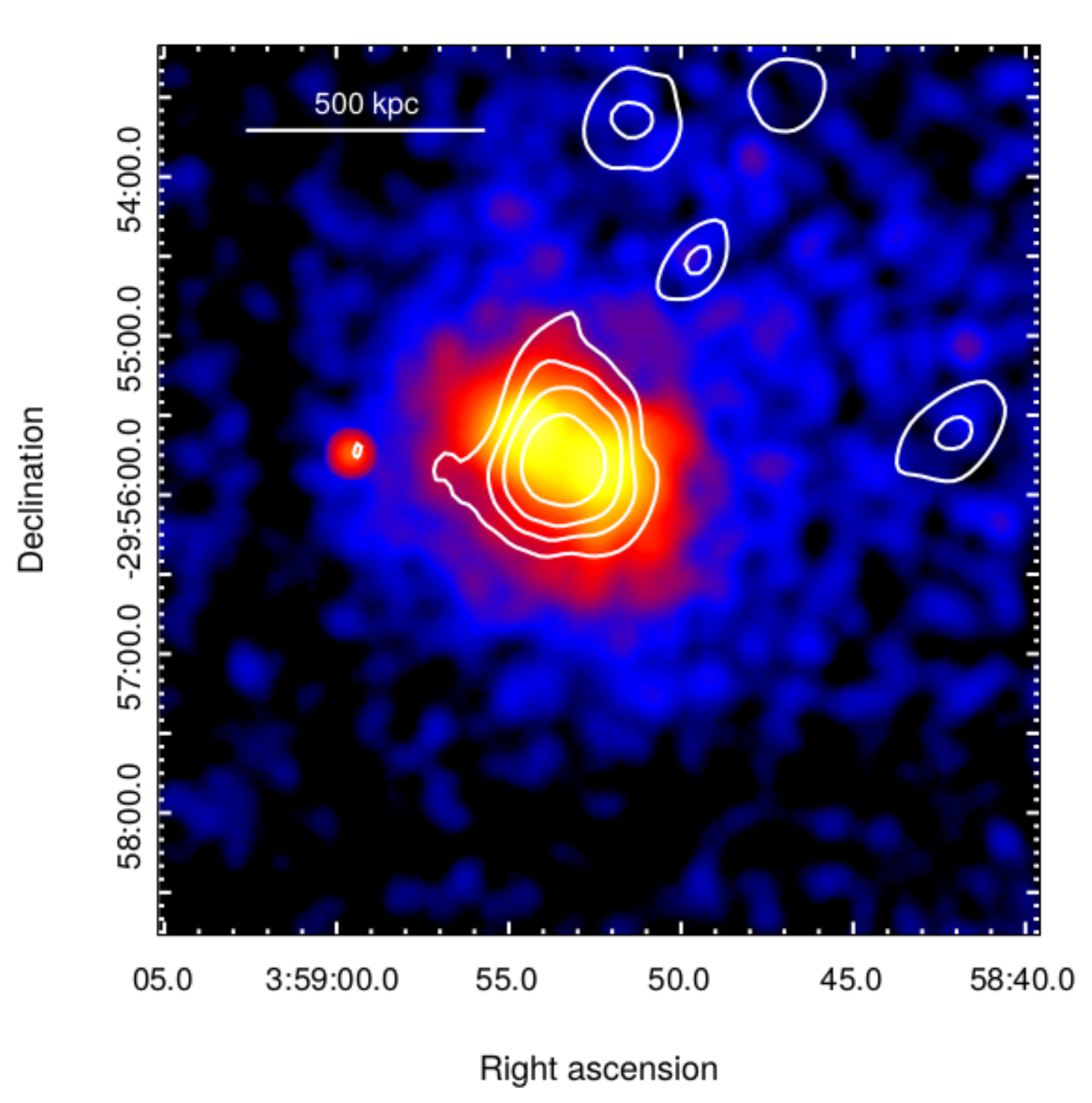}
\endminipage\hfill
\caption{\footnotesize \emph{Top panel}: total-intensity contours for the high-resolution image of MACS J0358.8$-$2955 (A3192) at 1.53 GHz overlaid on the optical image. Point sources are labeled as A, B, C, and D. The image has an FWHM of 10\arcsec$\times$15\arcsec at PA 25$^\circ$. The shown contour levels are (-0.3, 0.3, 0.2 ... 3, 6, 9, 12, 24, 48, ...)$\times\sigma$ with rms noise $\sigma=$0.1 mJy per beam. \emph{Bottom panel}: JVLA low-resolution radio contours after subtraction of discrete sources, superposed on the Chandra X-ray image. The radio HPBW is 35\arcsec with a noise level of 0.11 mJy per beam. Contours are placed at 0.3, 0.5, 0.7, and 1 mJy per beam.} 
\label{j0358}
\end{figure} 

{\bf MACS J0308.9$+$2645} (morphology class\footnote{We note that \cite{Ebeling2010} list an erroneous value of 4 for the morphological code of this system; the correct value of 2 is provided by \cite{Mann2012}.} 2) -- Using Chandra and GMRT data,  \cite{Parekh2017} found this cluster to have a relaxed and regular X-ray morphology. However, its temperature distribution is rather complex showing two cool cores that were suggested to possibly be part of two merging clusters. Diffuse radio emission was detected only at 610 MHz, located between the two cool cores.

Our radio images show a complex radio morphology with four unrelated discrete sources embedded in the diffuse emission (Fig.~\ref{j0308}). The radio emission overlaps the cluster center and is more extended than the X-ray emission to the East. We find the integrated flux density of the system to be $(39\pm 2)$ mJy. The halo flux density after subtracting sources A (8.8 mJy), B (18.4 mJy), C (2.0 mJy), and D (0.4 mJy) is $(9.7\pm 3.4)$ mJy. The radio halo in our images has a size of $\sim$2.6' ($\sim 770$ kpc) and a radio power of $4.12\times 10^{24}$ W Hz$^{-1}$. The spectral index of the radio halo between 610 MHz \citep{Parekh2017} and 1.5 GHz is 1.3.

{\bf MACS J0358.8$-$2955} (A3192; morphology class 4) -- This massive high-temperature system was observed by \cite{Parekh2017} with the GMRT at 610 MHz, but no diffuse emission was reported.

In our images we detected  a complex structure in the cluster center, consisting of three discrete sources (A, B, and C in Fig. \ref{j0358}, top) and an additional discrete source (D) towards the NW. Diffuse low-brightness emission is marginally visible around the cluster center. The Chandra X-ray image (Fig.~\ref{j0358}, bottom) shows a double structure perpendicular to the direction of the  double radio structure (B and C). We produced an image without short baselines with HPBW = 10\arcsec $\times$ 15\arcsec and subtracted clean components of the four discrete sources from the (u,v) data. In the new image with a HPBW of 35\arcsec\ the discrete sources are no longer present (Fig.~\ref{j0358}, bottom), and the diffuse emission is easily visible and resolved. This is a small radio halo  with a total flux density of $(2.7\pm 0.2)$ mJy. The noise level in this image is 0.11 mJy per beam.

{\bf MACS J0404.6$+$1109} (no figure presented; morphology class 4) -- \cite{Mann2012} noted that the X-ray peak is located between two BCGs, which show very small separation (about 100 kpc), suggesting either a line-of-sight merger or a very recent collision.  We did not observe this cluster, but analyzed archival data in the B and C configuration (L-band).  Only a bright point-like source (14.7 mJy) is visible at the position RA: 04\h 04\m 32.733\s, DEC: +11\deg 08\arcmin 03.85\arcsec coincident with the Northern BCG.  The lack of diffuse emission, despite the disturbed X-ray morphology, and the bright radio emission typical of BCG in relaxed clusters suggest a very recent/ongoing collision. The noise level in our images is 0.01 mJy per beam with a HPBW of 4.5\arcsec (or 0.04 mJy per beam with a HPBW $\sim10$\arcsec).

\begin{figure}[ht]
\centering
\minipage{0.4\textwidth}
\includegraphics[width=\linewidth]{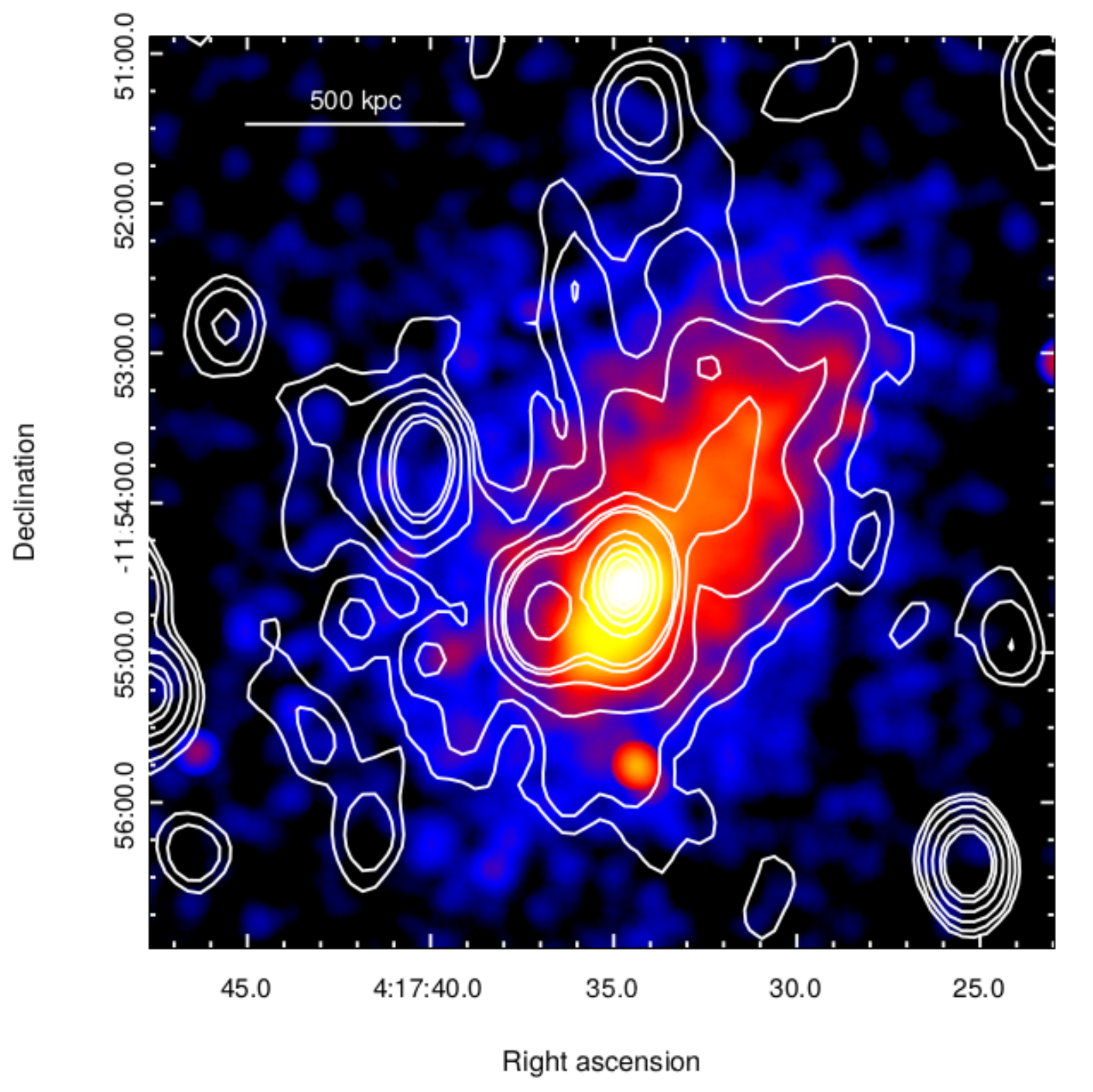}
\endminipage\hfill
\minipage{0.4\textwidth}
\includegraphics[width=\linewidth]{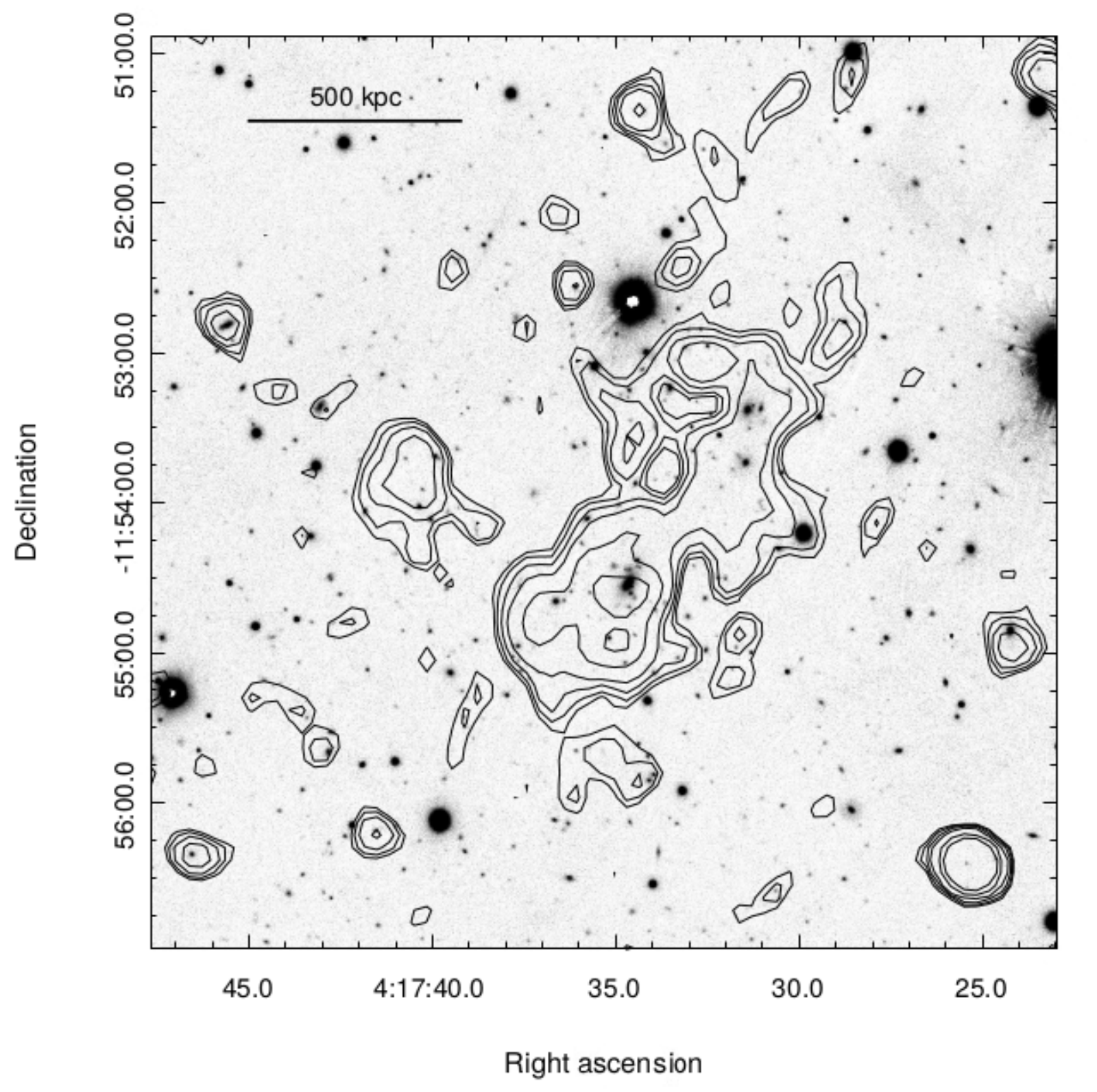}
\endminipage\hfill
\caption{\footnotesize \emph{Top panel}: Radio contours at 1.5 GHz for the cluster MACS J0417.5$-$1154 overlaid on the Chandra X-ray image.  Discrete sources have not been subtracted. The HPBW is 26.7\arcsec $\times$ 18.8\arcsec (PA = $-1.2^\circ$). Contours are placed at 0.05, 0.1, 0.2, 0.4, 0.7, 0.9, 3, 5, 7, 9 mJy per beam; the noise level is 0.02 mJy per beam \emph{Bottom panel}: Contours of the diffuse radio source classified as a radio halo at the cluster center superposed on the optical image. Discrete sources in the halo region have been subtracted.  The HPBW is 15\arcsec. The noise level is 0.03 mJy per beam; the shown contours mark $-$0.07, 0.05, 0.07, 0.1, 0.15, 0.3, 0.5 mJy per beam.}
\label{j0417}
\end{figure} 

{\bf MACS J0417.5$-$1154} (morphology class 3) -- The structure of this system was recently discussed in detail by \cite{Pandge2018}. The mass distribution identifies the cluster as a dissociative merger, as one of the merging clusters lost its gas content after the pericentric passage. The peak of the SZ decrement is displaced from the X-ray emission center because of the merger dynamics.

Diffuse radio emission in this cluster was reported by \cite{ Dwarakanath2011}, based on GMRT observations at 230 and 610 MHz.  \cite{Parekh2017} performed radio observations of this halo at 235 MHz and 610 MHz (GMRT) and at 1.5 GHz (JVLA). 

We did not observe this cluster, but analyzed archival data in the L-band and the B, C, and D configurations. We note that these are the same data used by \cite{Parekh2017} at 1.5 GHz; however, these authors only used the C-array data in their analysis, while we used C- and D-array data to image the halo source and B-array data to subtract unrelated sources. In Fig.~\ref{j0417} (top) we present our radio image of the cluster center superposed on the Chandra X-ray image (shown in colour). The X-ray image is elongated and irregular, in agreement with the presence of a major merger. The radio image, obtained combining C- and D-configuration data, shows discrete sources at the cluster center embedded in diffuse radio emission in very good agreement with the diffuse X-ray emission.

Using B-array data, we produced a list of clean components of discrete sources in the halo region: the main cluster's dominant galaxy, a head-tail galaxy near the cluster center, and two head-tail galaxies in the secondary group (see below) in agreement with the merger geometry. These sources were then subtracted from the (u,v) data for the C- and D-configuration to obtain an image of only the diffuse emission. The detection of a halo source agrees with the findings of \cite{Parekh2017}, but it is more extended and covers the same region as the number-density distribution of red cluster galaxies and the mass distribution shown by \cite{Pandge2018} (their Fig.~2). The total flux density (after subtraction of discrete sources) is 33.7 mJy with a maximum linear size of 1.2 Mpc.

In Fig.~\ref{j0417} (bottom) we present an image of the halo source at slightly higher resolution (obtained by giving larger weight to the C-array data), superposed on the optical image.  The main halo is well visible and coincident with the emission shown in the Chandra image. Moreover, less extended diffuse radio emission is present to the East (near RA: 04\h 17\m 40\s, DEC: --11\deg 54\arcmin 00\arcsec) that could be a small relic source. However, in conflict with this interpretation, the optical image shows a group of galaxies at the center of the radio emission, with magnitudes similar to that of the galaxies belonging to the main cluster. We therefore suggest that the small region of eastern diffuse emission is associated with a group that hosts a radio halo and is merging with the main cluster, although we note the lack of X-ray emission from this region.

\cite{Parekh2017} find the diffuse emission to have a steep and complex spectrum. We note that the radio images from the GMRT show only the brightest halo region, which explains the large discrepancy with our flux density at 1.5 GHz. Moreover, the 610 MHz image is affected by large negative regions (see their Figure 1(b)). For this reason we compared our image only with their 235 MHz data and estimated the flux densities at these two frequencies in the same area, finding a spectral index of the brightest region of the radio halo between 235 MHz and 1.5 GHz of 1.01$\pm$0.05.

\begin{figure}[ht]
\centering
\minipage{0.4\textwidth}
\includegraphics[width=\linewidth]{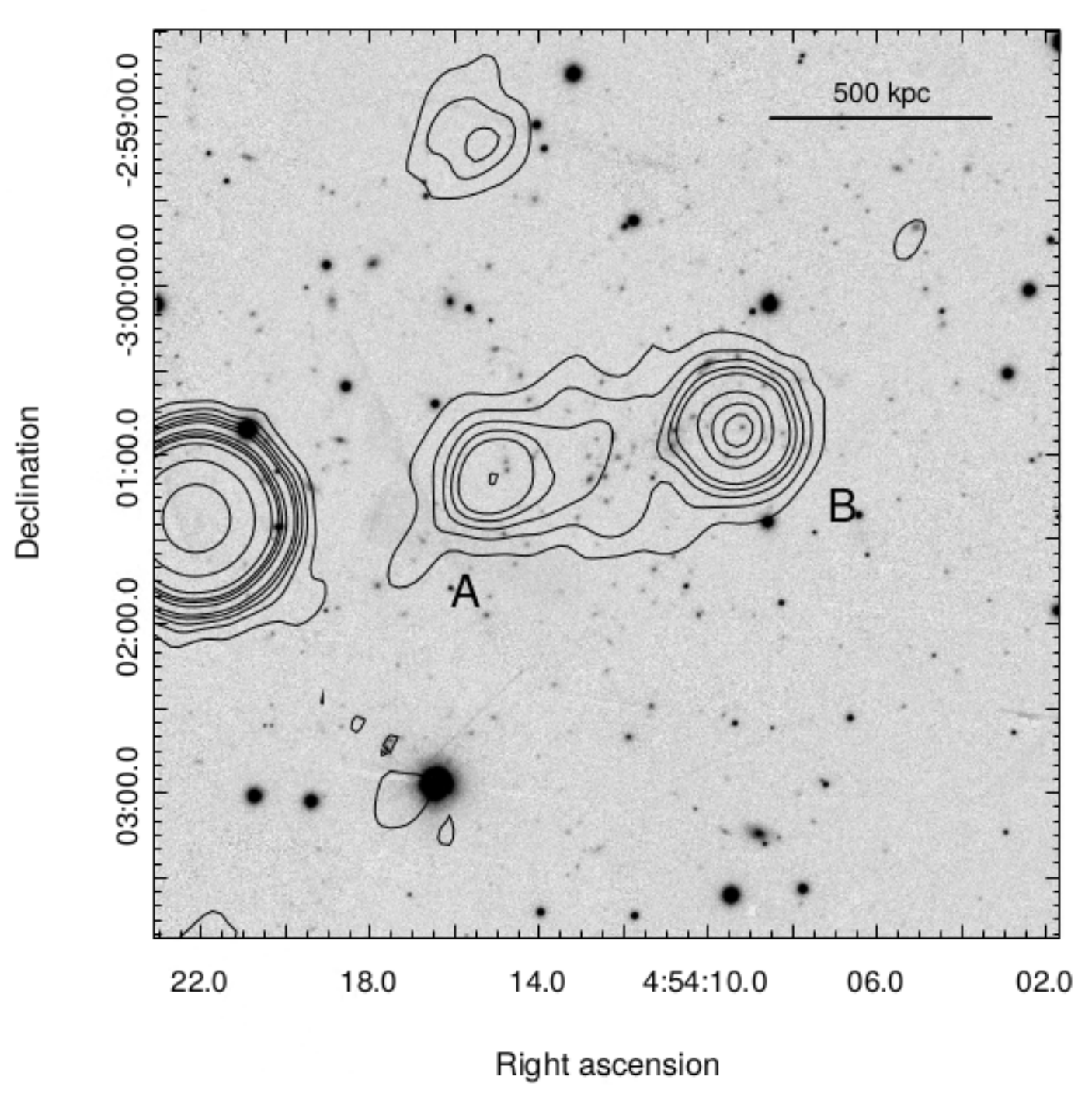}
\endminipage\hfill
\minipage{0.4\textwidth}
\includegraphics[width=\linewidth]{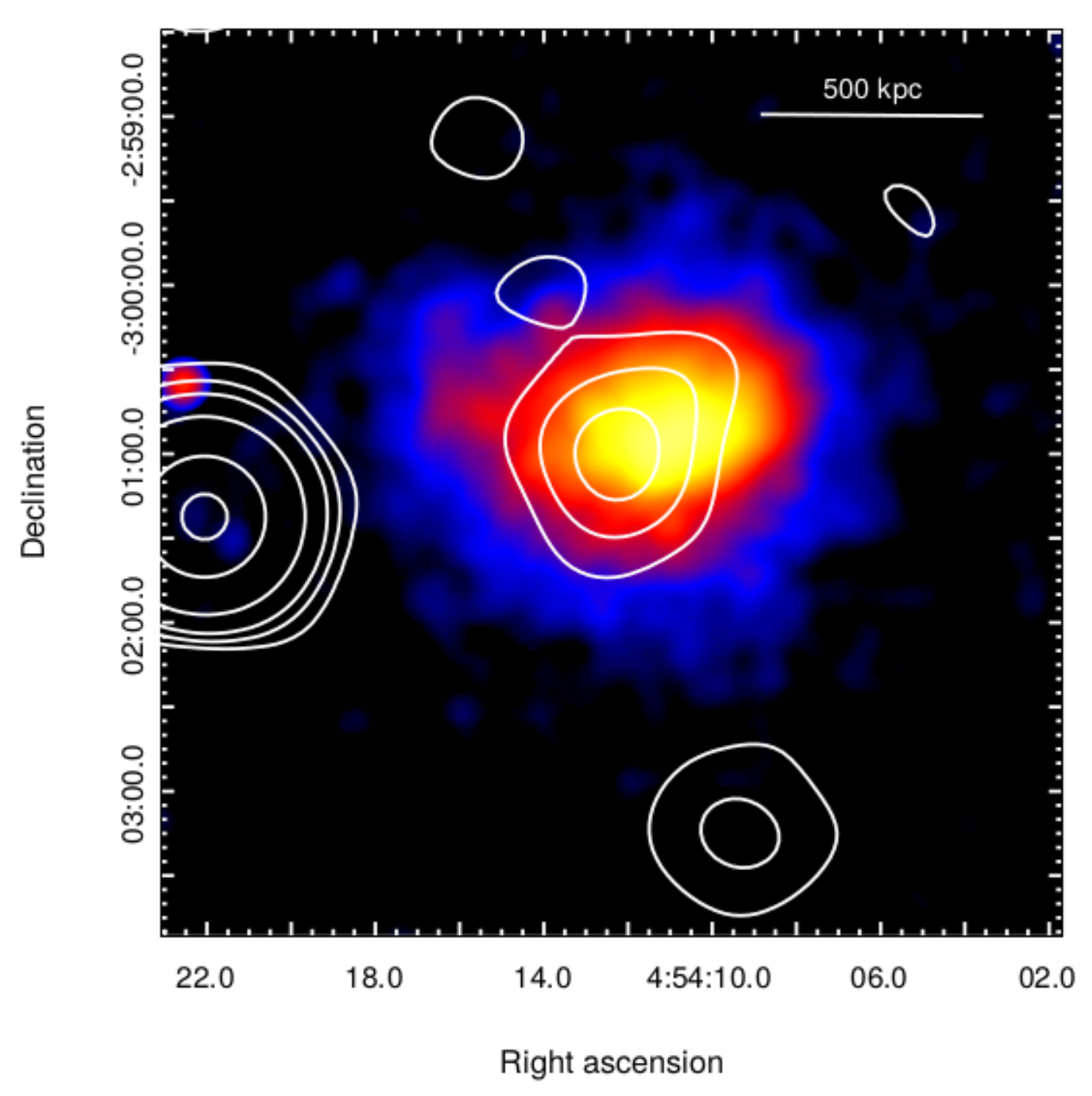}
\endminipage\hfill
\caption{\footnotesize \emph{Top panel}: Radio image of MACS J0454.1$-$0300 at 1.5 GHz superposed on the optical image. Discrete sources are labeled as A, B. The image has a FWHM of 33\arcsec (circular). The contour levels are (3, 6, 9, 12, 24, 48)$\times\sigma$ with $\sigma=0.043$ mJy per beam. \emph{Bottom panel}: Radio contours of the halo at the center of MACS J0454.1$-$0300 overlaid on the X-ray image obtained by Chandra (colour).  The HPBW of the radio data is 40\arcsec, and the noise level is 0.04 mJy per beam. The shown contours mark 0.1, 0.4, 1.2, 4.8, 9.6 mJy per beam.} 
\label{j0454}
\end{figure}

\begin{figure}[ht]
\centering
\includegraphics[scale=0.5, angle = 0]{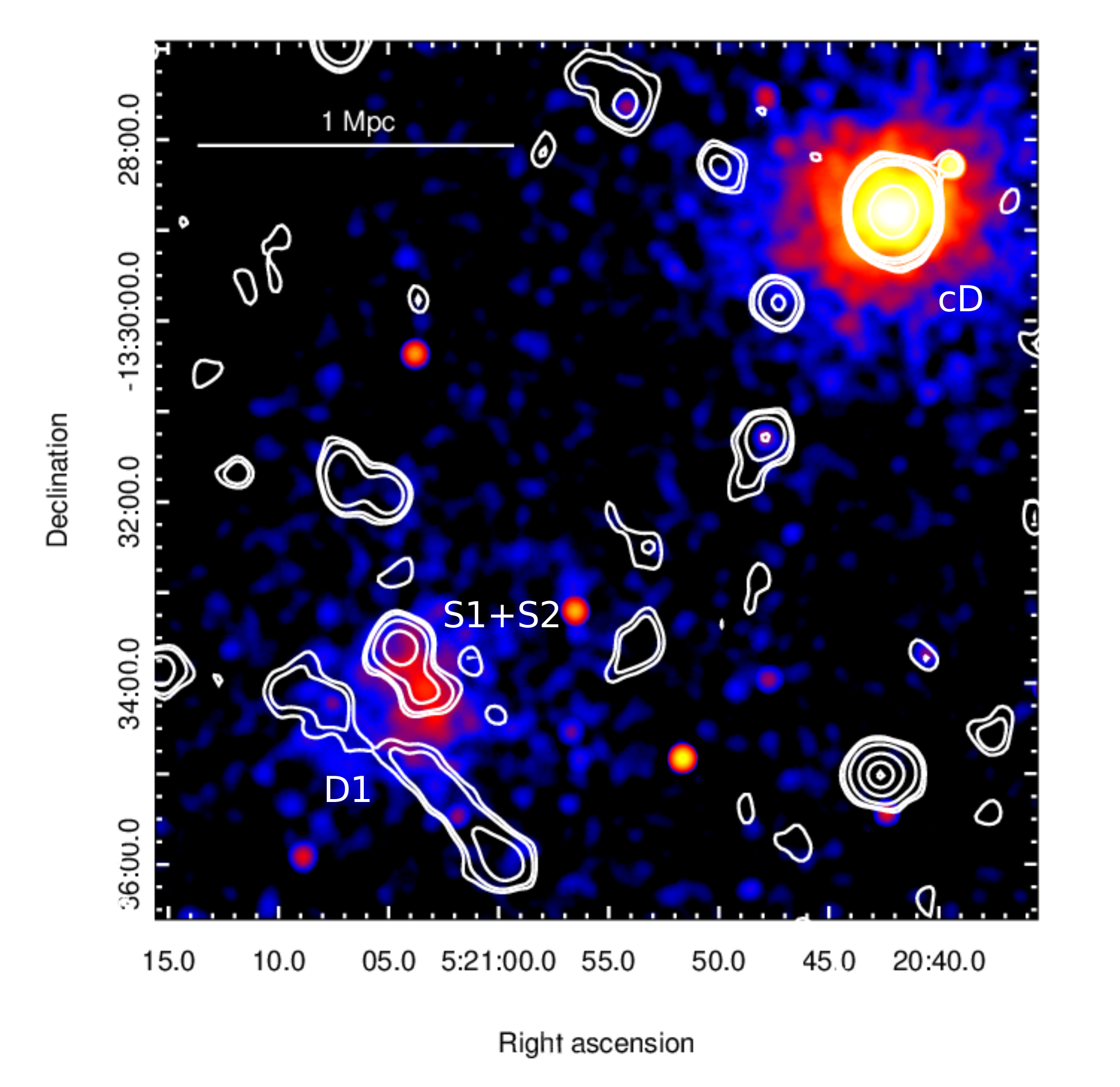}
\caption{Radio contours in the field of MACS J0520.7$-$1328, including 1WGA J0521.0$-$1333. The cD galaxy at the center of MACS J0520.7$-$1328 (top-right in the figure) is a strong radio and X-ray source; the diffuse X-ray emission of 1WGA J0521.0$-$1333 is centred near the S1+S2 position. D1 is the relic source discussed in the text. The HPBW is 24\arcsec. Contours are drawn at (3, 6, 9, 12, 24, 48)$\times \sigma$ with a noise level of $\sigma$ = 0.023 mJy per beam.}
\label{j0520}
\end{figure}

{\bf MACS J0429.6$-$0253} (no figure presented; morphology class 1) -- This relaxed cluster shows an unresolved source associated with the BCG. The flux density is $(132\pm 3)$ mJy in agreement with \cite{Hlavacek2013} (139 mJy at 1.4 GHz), and the location of the peak is RA: 04\h 29\m 36.019\s, DEC: $-$02\deg 53\arcmin 06.77\arcsec.

{\bf MACS J0454.1$-$0300} (morphology class 2) -- The Chandra image of this cluster shows clearly elongated emission suggesting a non-relaxed structure. In our radio image we find two strong  sources (A, B) with diffuse emission in between, coincident with the cluster centre (Fig.~\ref{j0454}, top). We subtracted the point-like sources from the (u,v) data to obtain the image presented in Fig.~\ref{j0454} (bottom), at a resolution of 40\arcsec. We classify the diffuse emission as a small halo source with a total flux density of $(0.79\pm 0.05)$ mJy (the noise level is 0.04 mJy per beam). The halo size ($\sim$300 kpc) was estimated by fitting a Gaussian convolved with the beam size. We note that the radio emission is slightly displaced from the X-ray peak, which could be real physical effect, as found in several clusters \citep[e.g.,][]{Govoni2012}, or due to noise in this faint radio source.

{\bf MACS J0520.7$-$1328} (morphology class 2) -- This almost relaxed cluster is dominated by radio emission from the BCG that was found to be slightly extended by \cite{Macario2014} in GMRT radio observations at 323 MHz. They also found faint diffuse and elongated emission about 8\arcmin SE of the cluster center (named D1), with an extension of about 2.8\arcmin along its major axis. Two discrete sources (named S1 and S2) were detected nearby. Based on the overlay between Chandra and GMRT data, they report the discovery of an unclassified extended X-ray source (1WGAJ0521.0$-$1333), identified as an irregular and disturbed galaxy cluster at $z = 0.34$, in the same region as the D1 and S1+S2 sources. An interaction between these two clusters is very likely, given that they are at the same redshift.

In our images (see Fig.~\ref{j0520}), we detected all the features described above and found the sources S1 and S2 to be nearly coincident with the X-ray emission from the poor cluster 1WGAJ0521.0$-$1333, while D1 appears as an elongated radio structure at the periphery of the X-ray emission. Therefore,  we tentatively identify D1 as a relic source related to the interaction between the main cluster (MACS J0520.7$-$1328) and the 1WGAJ0521.0$-$1333 satellite. This morphology is reminiscent of the prototype relic 1253+275 at the Coma cluster periphery, related to the merging between the group of NGC 4839 and the main cluster \citep{Giovannini1991}. Note that this scenario is not in conflict with the almost relaxed state of MACS J0520.7$-$1328. A peripheral merger with a galaxy group can produce a peripheral shock resulting in a relic source with no change in the physical properties at the cluster center.

\cite{Macario2014} also present a low-resolution radio image at the same frequency (323 MHz), which shows a filamentary structure between D1 and the BCG, named D2 by these authors, as well as two more patches of diffuse emission (D3 and D4). We did not detect any radio emission in the D2, D3, and D4 regions in low-resolution images created by us.

{\bf MACS J0547.0$-$3904} (no figure presented; morphology class 2) -- This relaxed cluster, characterized by the presence of cavities, was not observed by us, and we did not find published radio information. NVSS images show an unresolved source coincident with the cluster centre, with a flux density of $(85\pm 2)$ mJy.

\begin{figure}[ht]
\centering
\includegraphics[scale=0.7, angle = 0]{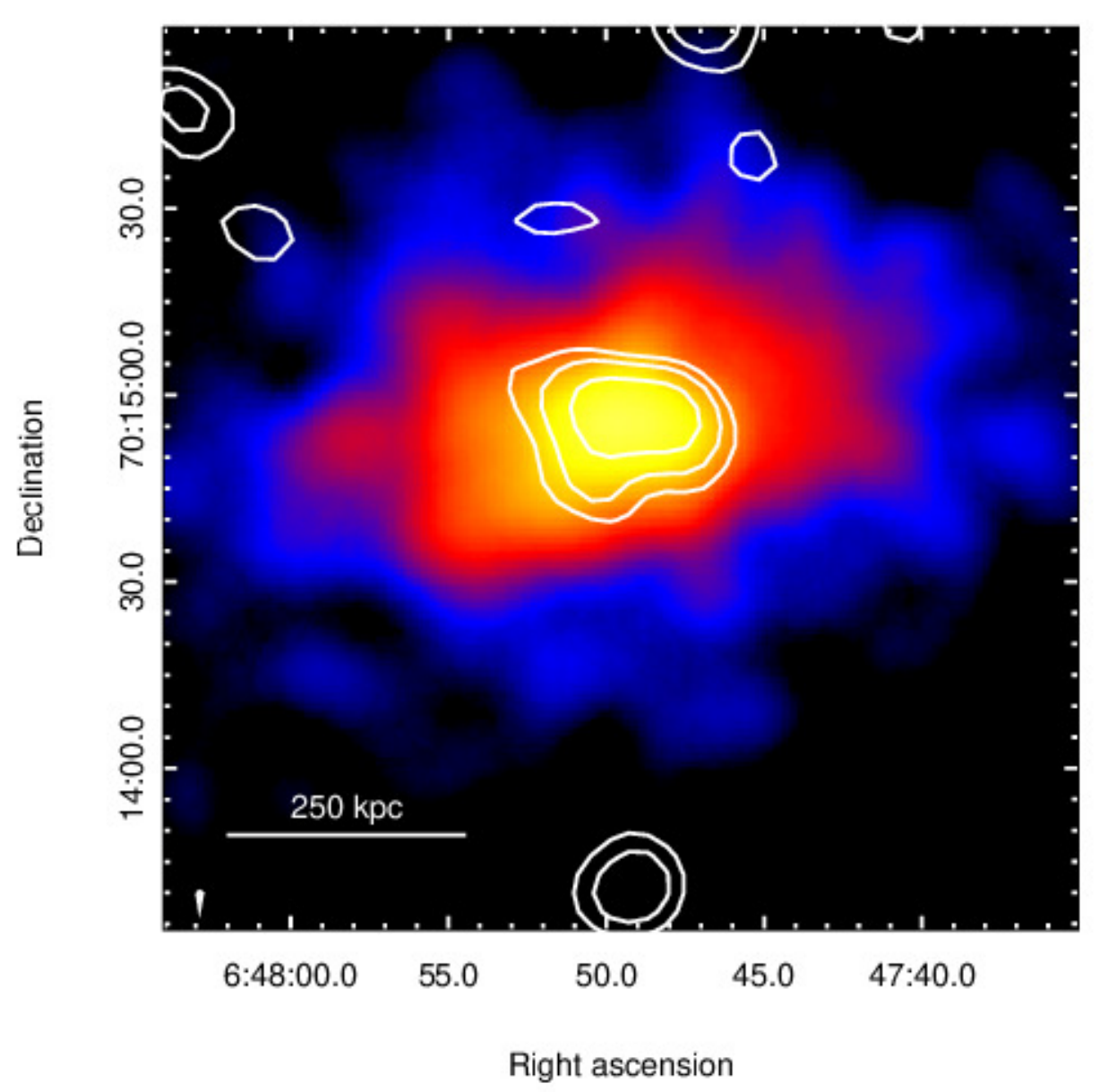}
\caption{\footnotesize Radio image (contours) of MACS J0647.7$+$7015 at 1.5 GHz overlaid on the Chandra X-ray image. The radio HPBW is 15\arcsec and the noise level 0.04 mJy per beam. The contour levels are: 0.07, 0.10, 0.15, 0.25 mJy per beam. An unrelated discrete source has been subtracted (see text).}
\label{j0647}
\end{figure}

{\bf MACS J0647.7$+$7015} (morphology class 2) -- Optical and X-ray images of this cluster suggest a disturbed structure with no evidence of a cool core. In our image (Fig.~\ref{j0647}), we detect a resolved source at the cluster center, elongated in the same E-W direction as the X-ray emission. An unrelated discrete source at 06\h 47\m 47.42\s, +70\deg 14\arcmin 20.2\arcsec (flux density 0.35 mJy) was subtracted. We classify the diffuse emission as a faint, small radio halo with a flux density of $(0.46\pm 0.05)$ mJy.

{\bf MACS J0717.5$+$3745} (no figure presented; morphology class 4)  -- This extreme merger is the most massive cluster in the MACS sample \citep{Ebeling2007} and well studied across the  electromagnetic spectrum \citep[e.g.,][]{Ma2009,Feretti2012,Limousin2016}. It hosts a radio halo that was recently discussed in detail by \cite{Weeren2017}  and \cite{Bonafede2018}.

{\bf MACS J0744.8$+$3927} (no figure presented; morphology class 2) -- LOFAR observations of this cluster, at $z = 0.6976$ the most distant one in our sample, led \cite{Wilber2018} to suggest the existence of a shock that is, however, too weak to accelerate electrons from the intracluster medium. In our images we find point-like emission (at an angular resolution of 15\arcsec) with a flux density of 0.87 mJy at the position of the X-ray image peak (07\h 44\m 52.85\s, +39\deg  27\arcmin 26.6\arcsec).  

{\bf MACS J0911.2$+$1746} (no figure presented;  morphology class 3) -- No diffuse or discrete radio source was detected in the field of this disturbed cluster. In our best image the noise level is 0.055 mJy per beam with a HPBW of 27.7\arcsec $\times$ 24.6\arcsec (PA = $-64^\circ$).

{\bf MACS J0947.2$+$7623} (no figure presented;  RBS797; morphology class 1) -- Multifrequency observations of this cool-core cluster by \cite{Gitti2006} and \cite{Doria2012} showed an irregular mini-halo and two pairs of radio jets misaligned by $\sim$ 90$^{\circ}$ emanating from the same radio core. These features suggest the presence of a double super-massive black hole, a scenario that is supported by VLBI observations \citep{Gitti2013}. Our images are in agreement with the published data.

{\bf MACS J0949.8$+$1708} (no figure presented;  morphology class 2) -- Radio observations of this cluster at 610 MHz by \cite{Venturi2008} found positive residuals after the subtraction of discrete sources and identified the system as a candidate radio halo. \cite{Bonafede2015} detected diffuse radio emission located at the cluster centre in new GMRT observations at 323 MHz, with an LLS of about 1 Mpc and a flux density of $(21.0 \pm 2.2)$ mJy, confirming the positive residuals detected by \cite{Venturi2008} as part of more extended radio emission.  \cite{Bonafede2015} classified the emission as a radio halo, elongated in the SW-NE direction, and commented that it does not follow the emission of the cluster thermal gas.

\begin{figure}[ht]
\centering
\includegraphics[scale=0.6, angle = 0]{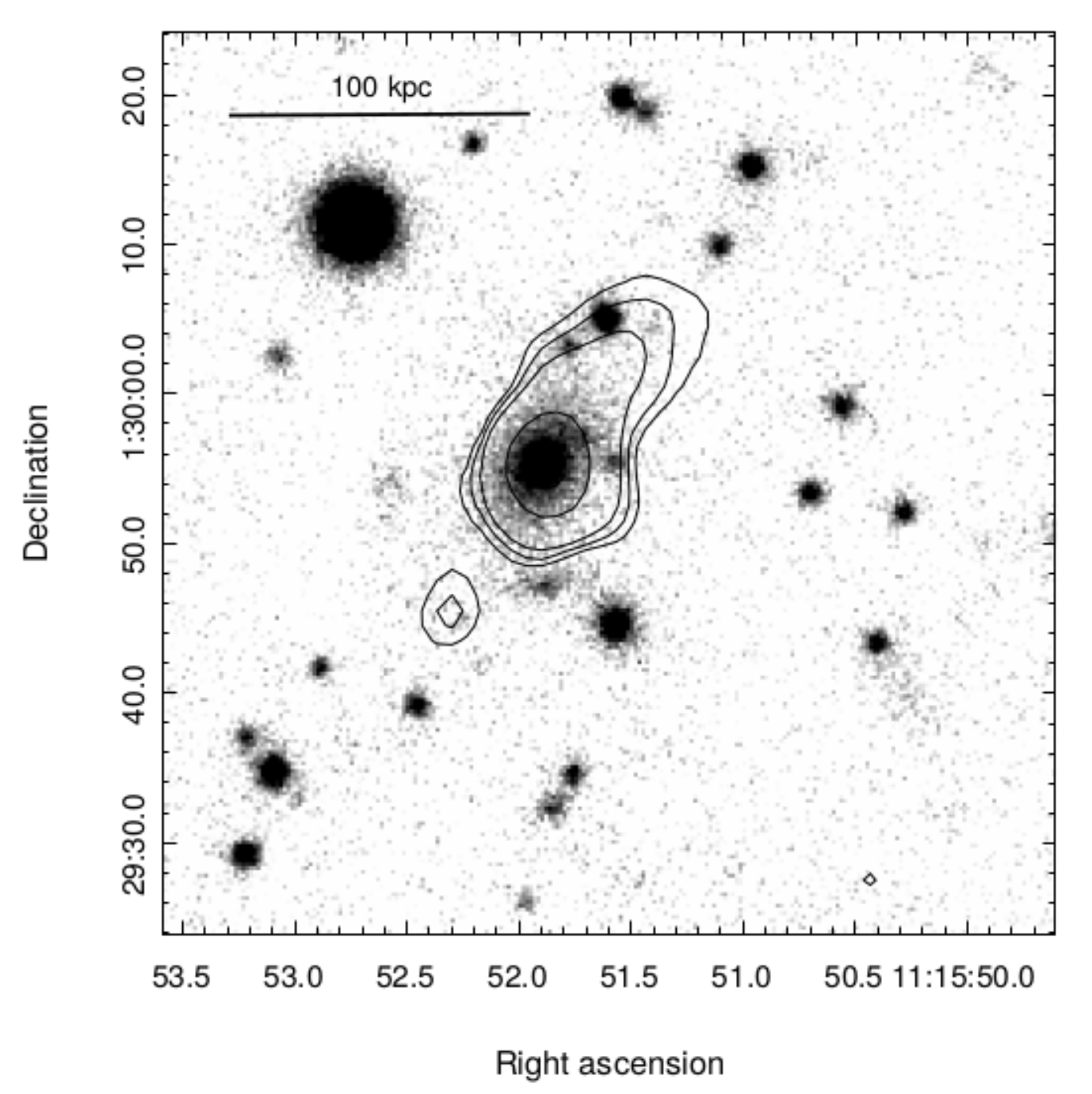}
\caption {FIRST radio contours of the central region of MACS J1115.8$+$0129. The peak of the extended structure is coincident with the brightest galaxy in the underlying optical image. The HPBW is 6.4\arcsec $\times$ 5.4\arcsec at a PA of 0$^\circ$. Contour levels are 0.5, 0.7, 1, 3, 5 mJy per beam.}
\label{j1115}
\end{figure}

Our images at 1.5 GHz show a faint (0.2 mJy) point-like source at the cluster center that is possibly associated with the cluster BCG. Discrete sources detected by \cite{Bonafede2015} are all readily visible in our images, but no diffuse emission is present at a noise level of 0.07 mJy per beam with a HPBW of 25\arcsec suggesting  a steep spectral index for the two relics of $\alpha> 1.3$. The data listed in Table~2 are from \cite{Bonafede2015} and have been scaled to 1.5 GHz using a spectral index of 1.4.

{\bf MACS J1115.8$+$0129} (morphology class 1) -- In our JVLA data (D-configuration) we found an unresolved source coincident with the cluster centre at RA = 11\h 15\m 51.80\s, DEC = +01\deg 29\arcmin 55.3\arcsec  with a flux density of $(17.6\pm 0.1)$ mJy. Unfortunately, C-configuration data, as well as images at higher resolution, are not available. The FIRST image of the same region shows an extended diffuse source with a total flux density of $(9.21\pm 0.02)$ mJy as well as a faint radio source of 1.20 mJy towards the SE. The peak of the extended structure is 5.9 mJy at 11\h 15\m 51.90\s, +01\deg 29\arcmin 55.5\arcsec, coincident with the position of the brightest cluster galaxy from the PanSTARRS image (Fig.~\ref{j1115}). We note a large difference between the flux density in our image and the FIRST total flux density in the same region (7 mJy are missing in the FIRST image with respect to the low-resolution image). We therefore suggest that the detected radio sources represent the brightest regions of more extended diffuse emission that we classify as a mini-halo.

{\bf MACS J1131.8$-$1955} (no figure presented; A1300; morphology class 4) -- This disturbed cluster hosts a well known diffuse radio halo, a peripheral relic, and another possible relic candidate. We have not observed it and refer to \cite{Reid1999,Feretti2012}, and \cite{Venturi2013}.


\begin{figure}[ht]
\centering
\includegraphics[scale=0.5, angle = 0]{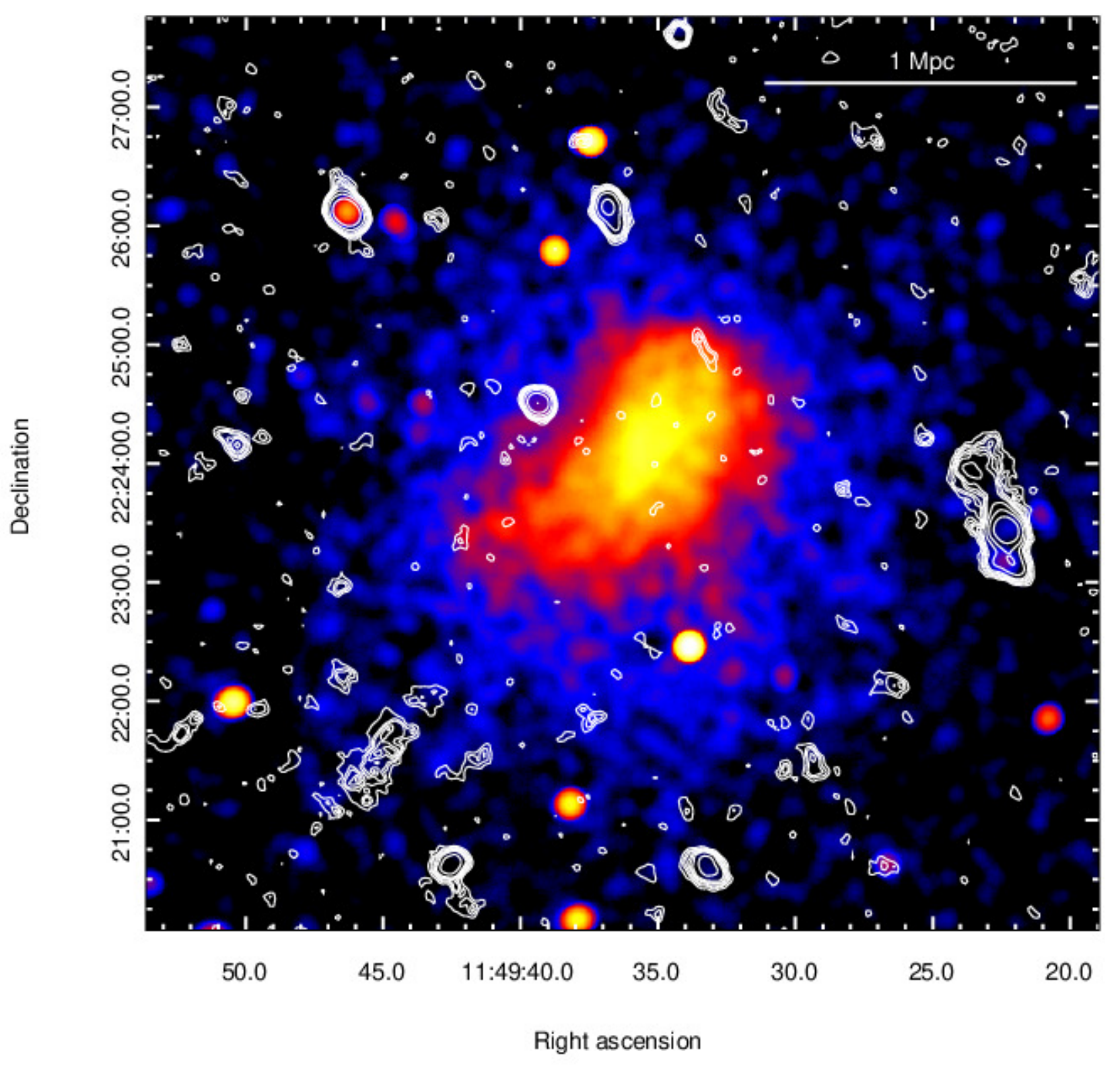}
\caption {Radio contours in the central region of MACS J1149.5$+$2223 superposed on the X-ray image obtained by Chandra. The radio HPBW is 10.8\arcsec$\times$ 9.6\arcsec (PA=49$^\circ$), and the noise level is 0.034 mJy per beam. The shown contours indicate 0.07, 0.1, 0.15, 0.2, 0.25, 0.3, 0.5, 1, and 2 mJy per beam. The extended sources in the W and SE are shown enlarged in Fig.~\ref{j1149b}.     
}
\label{j1149a}
\end{figure}

\begin{figure}[ht]
\centering
\minipage{0.3\textwidth}
\includegraphics[width=\linewidth]{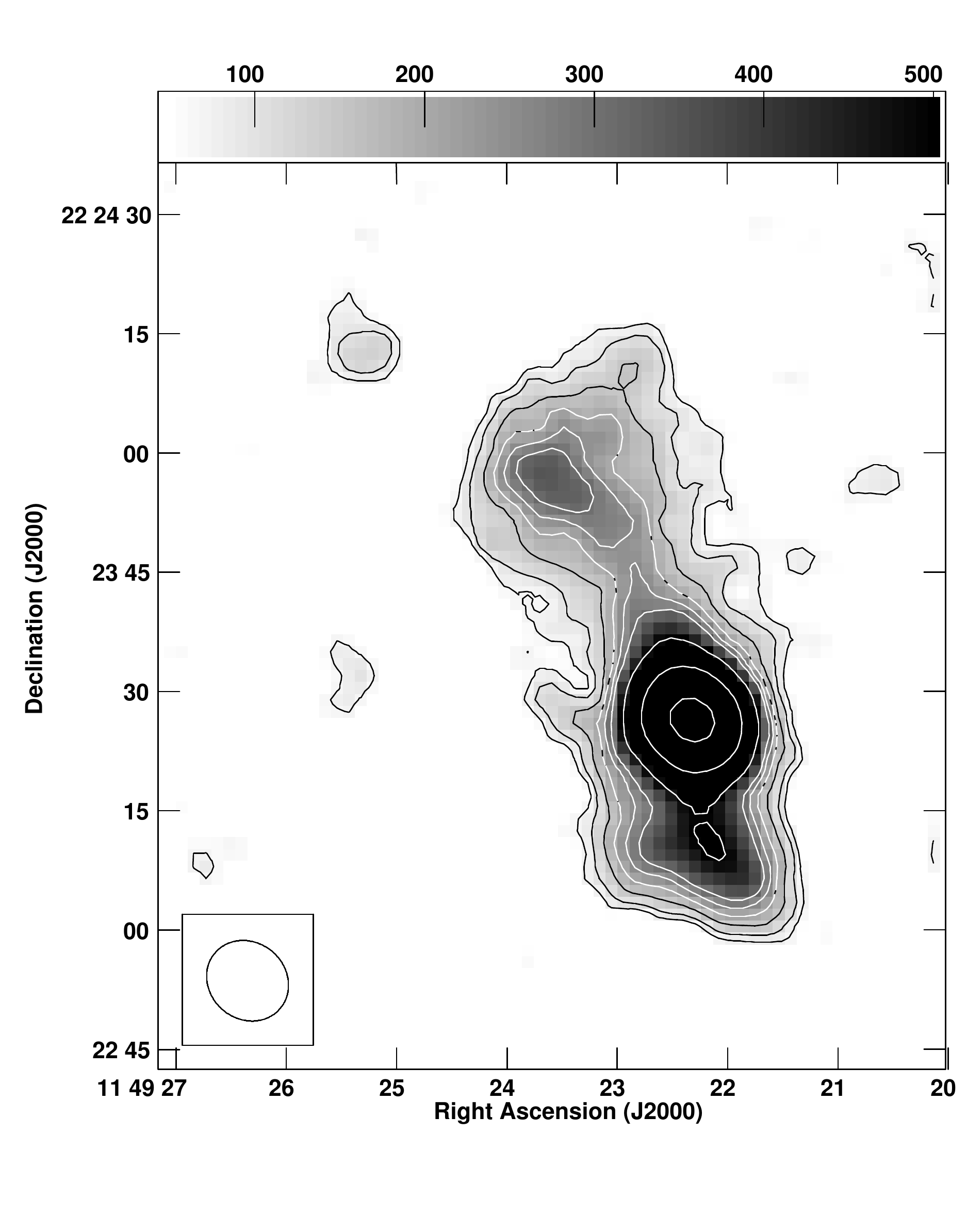}
\endminipage\hfill
\minipage{0.3\textwidth}
\includegraphics[width=\linewidth]{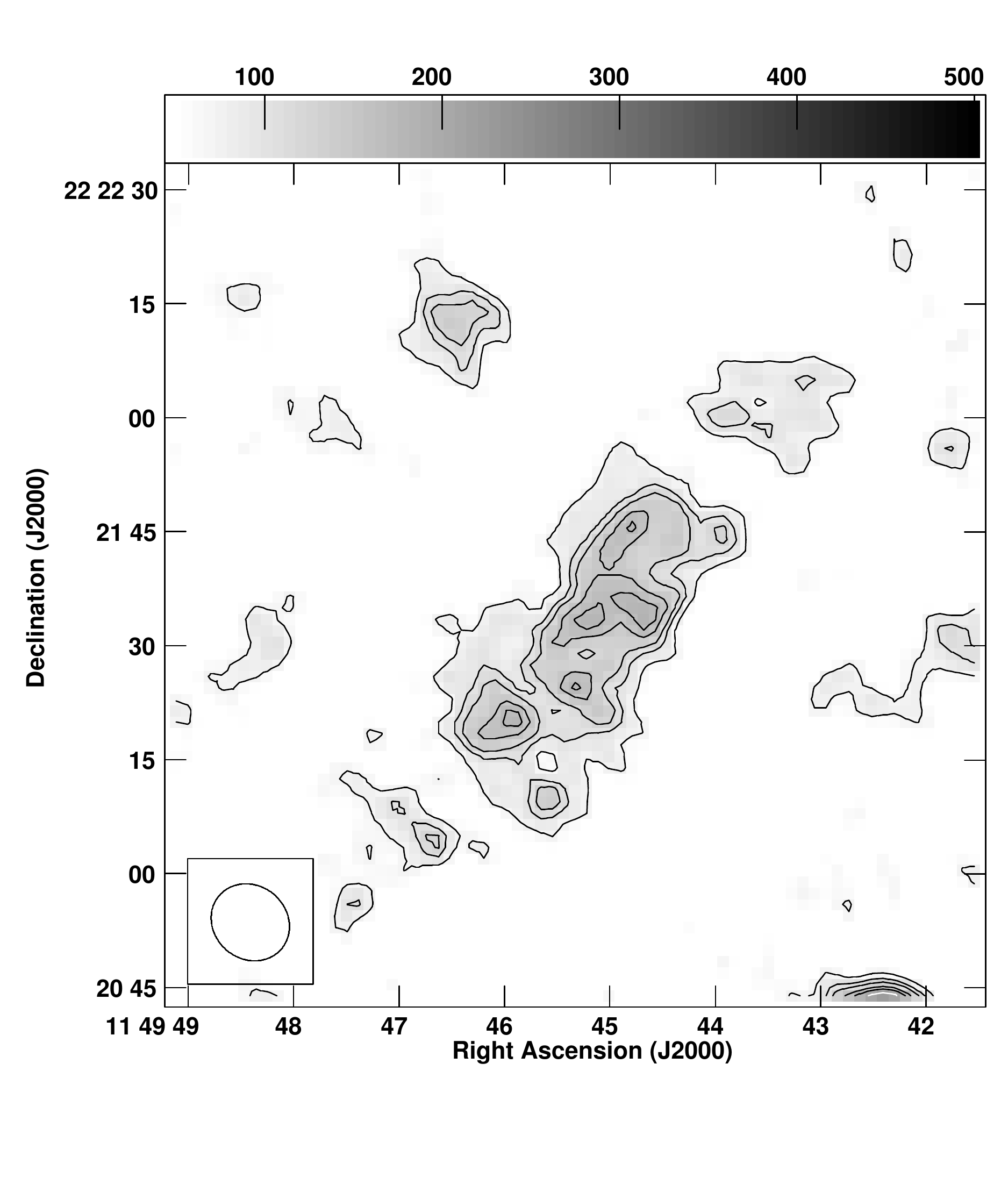}
\endminipage\hfill
\caption{\footnotesize \emph{Top panel}: radio image (1.53 GHz) of the W source in MACS J1149.5$+$2223 that we classify as a radio galaxy.  \emph{Bottom panel}: radio image of the SE source that we classify as a filament. Contours, resolution, and noise level as in Fig.~\ref{j1149a}.}
\label{j1149b}
\end{figure} 

\begin{figure}[ht]
\centering
\includegraphics[scale=0.5, angle = 0]{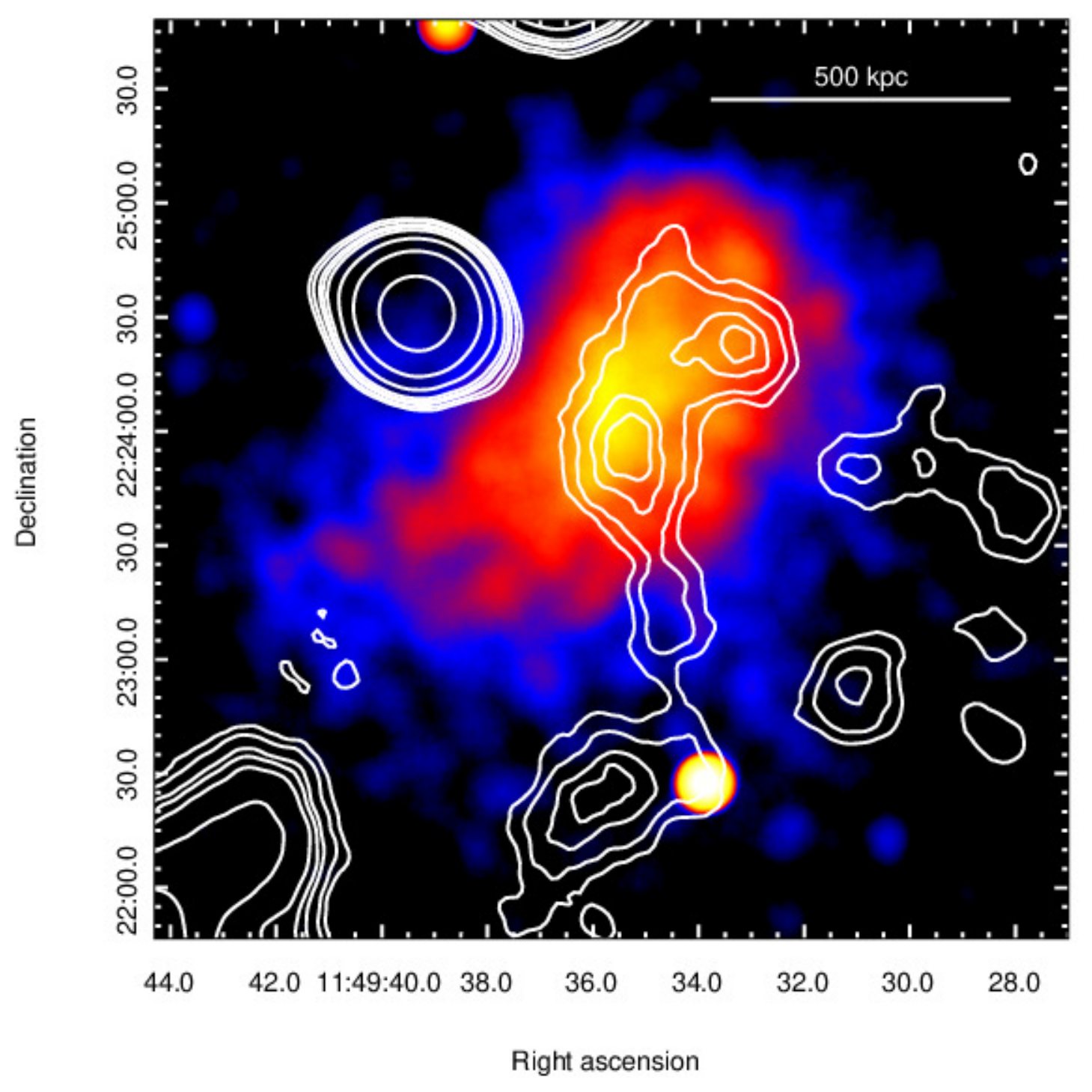}
\caption {Radio contours in the central region of MACS J1149.5$+$2223 overlaid on the X-ray image obtained by Chandra. The radio HPBW is 26.0\arcsec$\times$ 25.1\arcsec (PA=33$^\circ$), and the noise level is 0.048 mJy per beam. Contours indicate 0.1, 0.12, 0.15, 0.17, 0.2, 0.3, 0.5, 1, 3, 5, 10, 15, and 20 mJy per beam. 
}
\label{j1149c}
\end{figure}


{\bf MACS J1149.5$+$2223} (morphology class 4) -- One of the most complex cluster lenses known \citep{Smith2009}, this system was studied in detail by \cite{Bonafede2012}, using GMRT, NVSS, and FIRST data in the radio band and Chandra observations in X-ray regime. Two extended radio sources were found at 3\arcmin and 3.7\arcmin, W and SE of the cluster center, respectively, and classified as peripheral radio relics. Additional diffuse emission at the cluster centre was classified as a radio halo as it follows the cluster X-ray brightness.

We used archival VLA data in the L band in the B, C, and D configurations to create deep and high-resolution images of this cluster. The high-resolution image (Fig.~\ref{j1149a}) shows no diffuse
emission at the cluster centre.  Fig.~\ref{j1149b} (top and bottom) shows enlargements of the two diffuse sources (W and SE) visible in Fig.~\ref{j1149a} and discussed by \cite{Bonafede2012}.

In the images obtained with the B array at the highest resolution (not shown here), the W source shows central emission and a two-sided structure typical of a radio galaxy. The position of the central peak of flux $(2.23\pm 0.01)$ mJy is RA = 11\h 49\m 22.308\s and Dec. = +22\deg 23\arcmin 26.19\arcsec, coincident with a galaxy at RA = 11\h 49\m 22.312\s and Dec. = +22\deg 23\arcmin 26.55\arcsec and $z = 0.174$ in the NASA/IPAC Extragalactic Database (NED). We identify this foreground galaxy as the optical counterpart of the extended radio source. Its total flux density is 6.77 mJy in the high-resolution image and 9.02 mJy at low resolution (Fig.~\ref{j1149b}, top). Its total radio power is 7.5 $\times$ 10$^{23}$ W Hz$^{-1}$ at the quoted redshift, and the spectral index between 323 MHz and 1.56 GHz is $\sim$0.4.

The SE component is well resolved in our high-resolution images and shows no evidence of nuclear or jet structure (Fig.~\ref{j1149b}, bottom). We note that its morphology is very peculiar for a relic source. Relics are produced through shocks induced by cluster mergers and are expected to be perpendicular to the merger axis, while here the diffuse emission is elongated toward the cluster center. Moreover the diffuse emission does not show the transverse structure present in many relics due to the shock in the ICM. We suggest that this source could be a filamentary structure along the merger axis as discussed, e.g., by \cite{Giovannini2013} for A3411 and A3412. Its total flux density is $(1.68\pm 0.03)$ mJy at high resolution and $(3.76\pm 0.03)$ mJy in the low-resolution images with a spectral index of 1.17 between 323 MHz and 1.5 GHz. The total radio power is 4.42 $\times$ 10$^{24}$ W Hz$^{-1}$ and the size $\sim$570 kpc.

At low resolution (see~Fig. \ref{j1149c}) diffuse emission is observed in the cluster centre. Its shape is that of a halo source following the X-ray emission but with a double peak. From this region, a possible filament extends in the direction of the SE filament discussed before. Since the X-ray emission is extended in the same direction, all of these features could be related to the same large-scale structure.

The total flux density in the central region of the cluster is $(0.9\pm 0.1)$ mJy (corresponding to a radio power of 1.02 $\times$10$^{24}$ W Hz$^{-1}$) across a size of $\sim$400 kpc. Because of the large difference between the source morphology in our image at 1.5 GHz and the one published at 323 MHz \citep{Bonafede2012} we do not estimate its spectral index. 

We conclude from our analysis that MACS J1149.5$+$2223 hosts a faint radio halo with a peculiar morphology. In the SE region a radio filament has been detected (Fig.~\ref{j1149b}, bottom) that may be connected to a complex structure extending to the cluster center, marginally visible in Fig.~\ref{j1149c}. Deeper images are necessary for a more detailed study. No relic source has been identified in our images.

{\bf MACS J1206.2$-$0847} (no figure presented; morphology class 2) -- A joint lensing/X-ray analysis by \citet{Ebeling2009} identified this cluster as a likely post-collision merger proceeding along the line of sight. A detailed follow-up study by \cite{Young2015} confirms the general dynamic equilibrium of this system but also raises the possibility of a more complex morphology as evidenced by offset centroids between the Bolocam data, the diffuse X-ray emission, and the BCG. The authors state that deeper multi-band radio data are necessary to understand the nature of diffuse emission W of the BCG seen in GMRT data at 610 MHz. Since the resolution of the D-configuration JVLA data obtained by us is insufficient to investigate this structure, we also analyzed archival public data in the B/C configuration of this cluster at 1.5 GHz. Our final image with a HPBW of 29.9\arcsec $\times$ 26.5\arcsec (PA 14$^\circ$) shows a point-like source coincident with the BCG (RA 12\h 06\m 12.15\s, DEC --08\deg 48\arcmin 03.1\arcsec) with a flux density of $(108.8\pm 0.2)$ mJy and a nearby point-like faint source (12\h 06\m 10.73\s, --8\deg 47\arcmin 16.9\arcsec;  $(6.6\pm 0.1)$ mJy). Since, at a noise level of 0.1 mJy per beam, we did not detect the diffuse emission W of the BCG seen by \cite{Young2015} we take its nature to be uncertain and do not consider it in this work.
 
{\bf MACS J1319.9$+$7003} (no figure presented; A1722; morphology class 2) -- Observations of this cluster with the GMRT at 610 MHz by \cite{Kale2015} did not detect any diffuse emission but found two point-like sources in the peripheral region of the cluster, with no obvious optical counterpart. Our image confirms both the presence of these sources and the lack of extended emission. The noise level in our image is 0.1 mJy per beam for a HPBW of 22\arcsec $\times$ 18\arcsec at PA 0$^\circ$.

\begin{figure}[ht]
\centering
\minipage{0.4\textwidth}
\includegraphics[width=\linewidth]{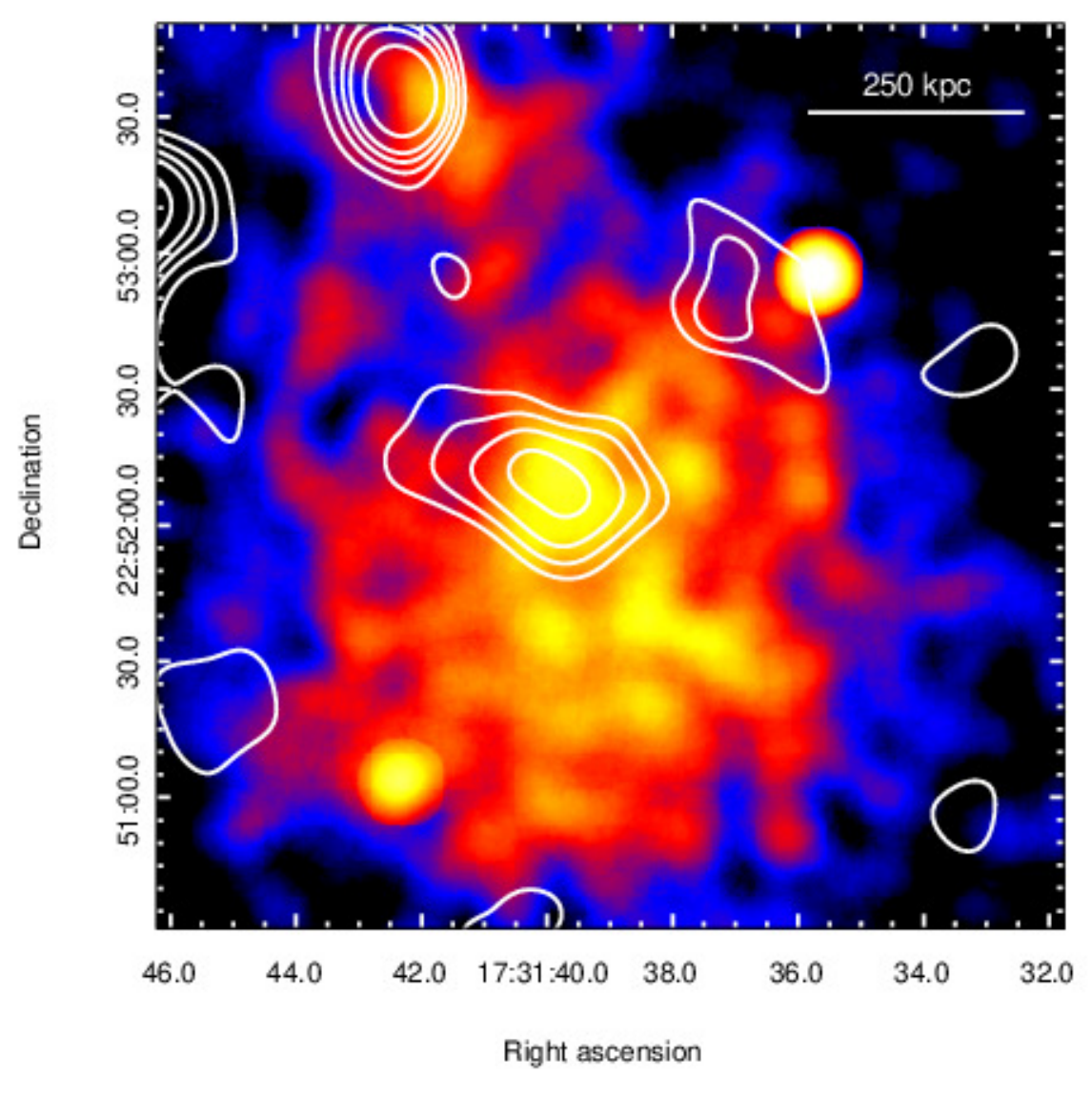}
\endminipage\hfill
\caption{\footnotesize Radio contours for the central region of MACS J1731.6$+$2252 after subtraction of discrete sources, superposed on the Chandra X-ray image. The HPBW is 25\arcsec, and the noise level is 0.3 mJy per beam.  Contours are placed at 0.5, 1, 1.5, 2, 3, and 5 mJy per beam.}
\label{j1731}
\end{figure} 

{\bf MACS J1347.5$-$1144} (no figure presented; morphology class 1) -- This system is the most X-ray luminous cluster discovered in the ROSAT All-Sky Survey \citep{Schindler1995}. It was studied in detail by \cite{Gitti2007}, who report the detection of a mini-halo around the central cool core. They also note an elongation of the diffuse radio emission in the direction of a hot X-ray-bright subclump, suggesting that energy for the electron re-acceleration might be provided by this sub-merger event.

{\bf MACS J1423.8$+$2404} (no figure presented; morphology class 1) -- Our C-array images of this distant ($z = 0.5431$) relaxed cluster show a point-like source at the cluster centre, identified as the BCG by \cite{Hlavacek2013}. Another point-like source is present in a peripheral region of the X-ray emission detected by Chandra. The HPBW of our image is 24.9\arcsec $\times$ 16.7\arcsec at P.A.\ 80$^\circ$; the noise level is 0.02 mJy per beam, and the source flux density is $(5.47\pm 0.02)$ mJy.

{\bf MACS J1427.6$-$2521} (no figure presented; morphology class 1) -- In agreement with the NVSS data, our D-array image shows an unresolved faint source of flux density $(4.1\pm 0.1)$ mJy at the center of this relaxed cluster (14\h 27\m 39.4\s, --25\deg 21\arcmin 02\arcsec).

{\bf MACS J1532.8$+$3021} (no figure presented; RXJ1532.9$+$3021; morphology class 1) -- Our C-configuration data are in agreement with results published by \cite{Giacintucci2014}. A small mini-halo at the cluster centre measures $\sim$100 kpc in radius. The flux densities for the central galaxy and the diffuse radio emission (mini-halo) are $(20.1\pm 0.1)$ mJy and $(4.4\pm 0.3)$ mJy, respectively.
 
{\bf MACS J1720.2$+$3536} (no figure presented; morphology class 1) -- Our image from C-configuration data shows an unresolved source at the cluster center (RA = 17\h 20\m 16.75\s, DEC = 35\deg 36\arcmin 26.3\arcsec) with a flux density of $(17.2\pm 0.3)$ mJy, in agreement with the findings of \cite{Hlavacek2013}.

{\bf MACS J1731.6$+$2252} (morphology class 4) --  A previous radio study of this active merger by \cite{Bonafede2012} failed to detect diffuse emission in GMRT data. We did not observe this cluster but analyzed archival data taken in the L-band with the C array and confirmed the presence of two relatively strong discrete sources at the cluster centre: one at 17\h 31\m 39.82\s, +22\deg 51\arcmin 56.8\arcsec with a flux density of $(13.9\pm 0.3)$ mJy and the other about 50\arcsec to the West at 17\h 31\m 35.84\s, +22\deg 52\arcmin 04.1\arcsec with a total flux density of $(104.0\pm 0.3)$ mJy.

We created a high-resolution image without short baselines and subtracted clean components of the two central discrete sources to generate a new image with a HPBW of 25\arcsec and a noise level of 0.3 mJy per beam. We identify diffuse emission seen in this image in the cluster center (Fig. \ref{j1731}) as a small halo radio source with a total flux density of $(3.23 \pm 0.4)$ mJy and an angular size of $\sim$45\arcsec. Note that no residual is visible at the position of the strong Western source, confirming the accuracy of the discrete-source subtraction.

{\bf MACS J1931.8$-$2634} (no figure presented; morphology class 1) -- Our low-resolution radio data for this cluster agree with the results of \cite{Giacintucci2014} who found a diffuse mini-halo source at the position of the BCG. We estimate a flux density of $(50\pm 4)$ mJy for the diffuse component of this source.

\begin{figure}[ht]
\centering
\includegraphics[scale=0.65, angle = 0]{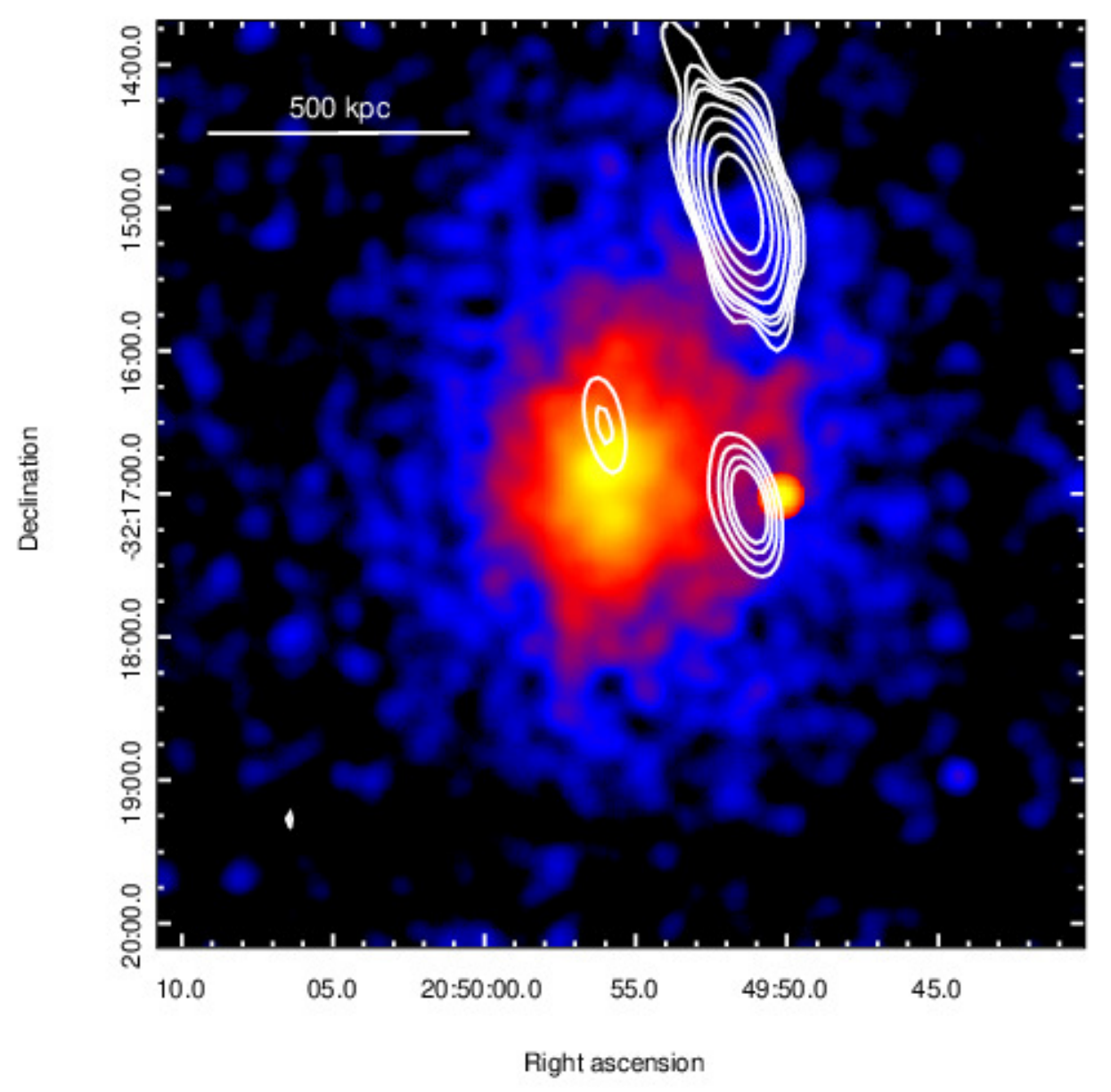}
\caption {Radio contours in the central region of MACS J2049.9$-$3217 superposed on the Chandra X-ray image. See text for discussion. The radio HPBW is 42\arcsec $\times$ 16\arcsec at PA 15$^\circ$, and the noise level is 0.07 mJy per beam. Contours indicate (0.8, 1.5, 3, 6, 9, ...)$\times$0.22 mJy per beam.
}
\label{j2049}
\end{figure}


\begin{figure}[ht]
\centering
\includegraphics[scale=0.60]{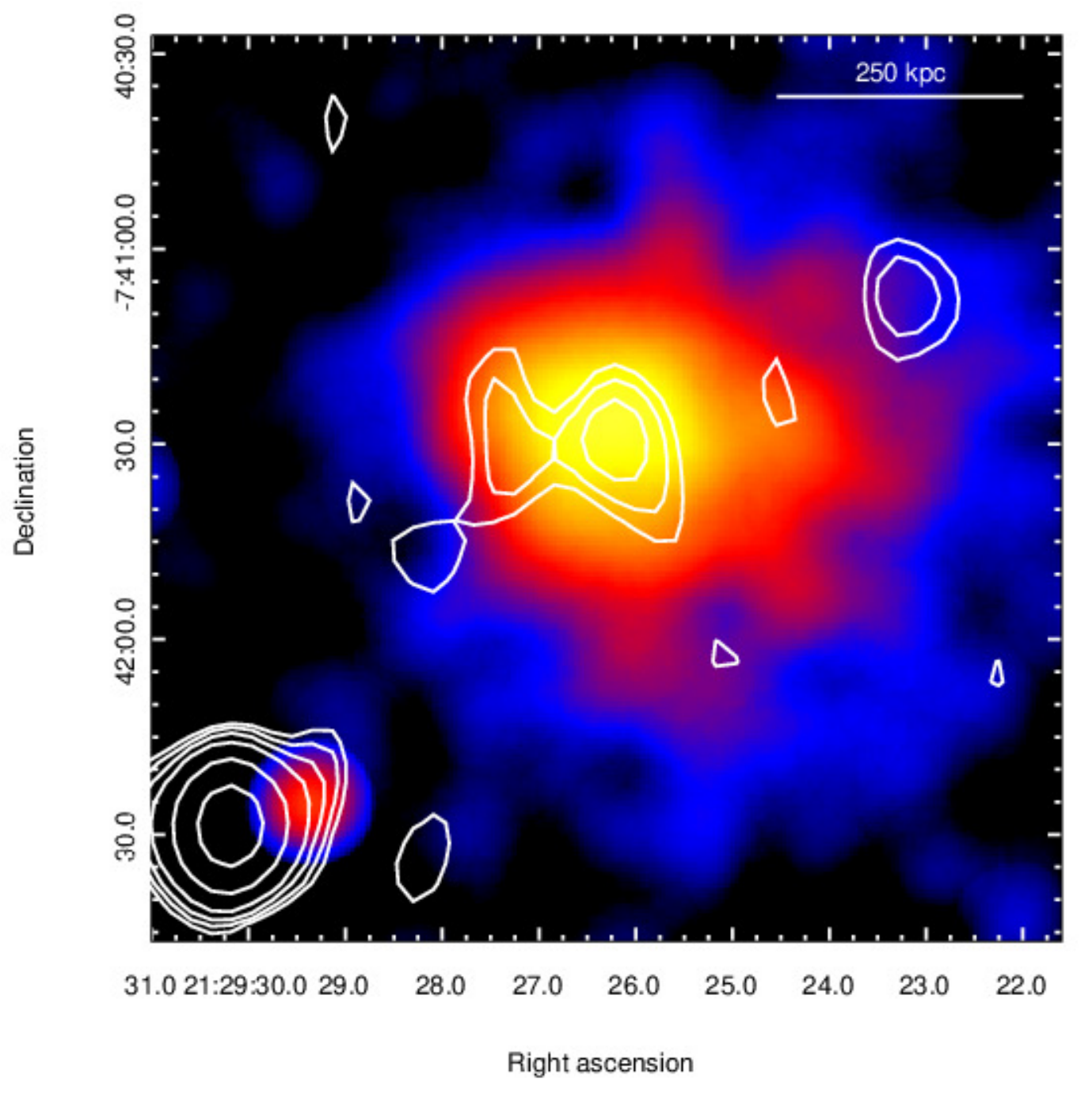}
\caption {Radio contours from JVLA observations of the cluster MACS J2129.4$-$0741, overlaid on the Chandra X-ray image. The HPBW is 16.6\arcsec $\times$ 13.1\arcsec at PA 4$^\circ$, and the noise level is 0.03 mJy per beam. The shown contours indicate (3, 3.5, 4, 6, 9, 12)$\times$ 0.03 mJy per beam. 
}
\label{j2129}
\end{figure}

{\bf MACS J2049.9$-$3217} (morphology class 3) -- No previous radio observations exist for this cluster. In our images, we detect a faint unresolved source near the cluster center at 20\h 49\m 56.07\s, $-$32\deg 16\arcmin 31.4\arcsec with a flux density of $(0.45\pm 0.05)$ mJy. Two other sources are located in more peripheral cluster regions: one at 20\h 49\m 51.19\s, $-$32\deg 17\arcmin 04.9\arcsec with a flux density of $(1.57\pm 0.05)$ mJy, and another, stronger source toward the North with a flux density of $(32.91\pm 0.07)$ mJy (see Fig.~\ref{j2049}). Based on their positions, these could be a halo and two relic sources. However, since both appear unresolved in our images and thus could be discrete sources, we identify the source at the cluster center as an unresolved cluster source and the two peripheral sources as unrelated to the cluster.

\begin{figure}[ht]
\centering
\minipage{0.4\textwidth}
\includegraphics[width=\linewidth]{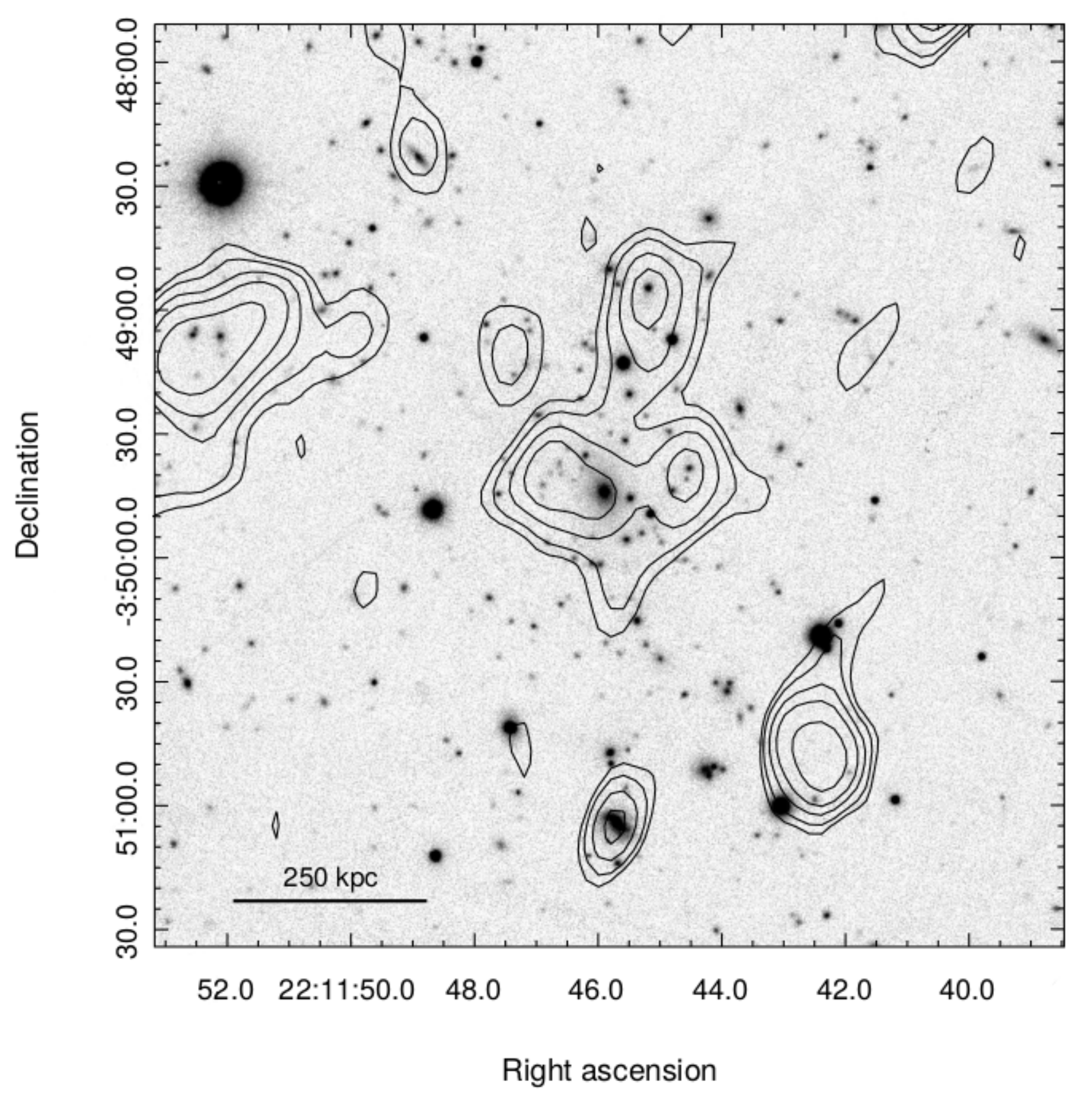}
\endminipage\hfill
\minipage{0.4\textwidth}
\includegraphics[width=\linewidth]{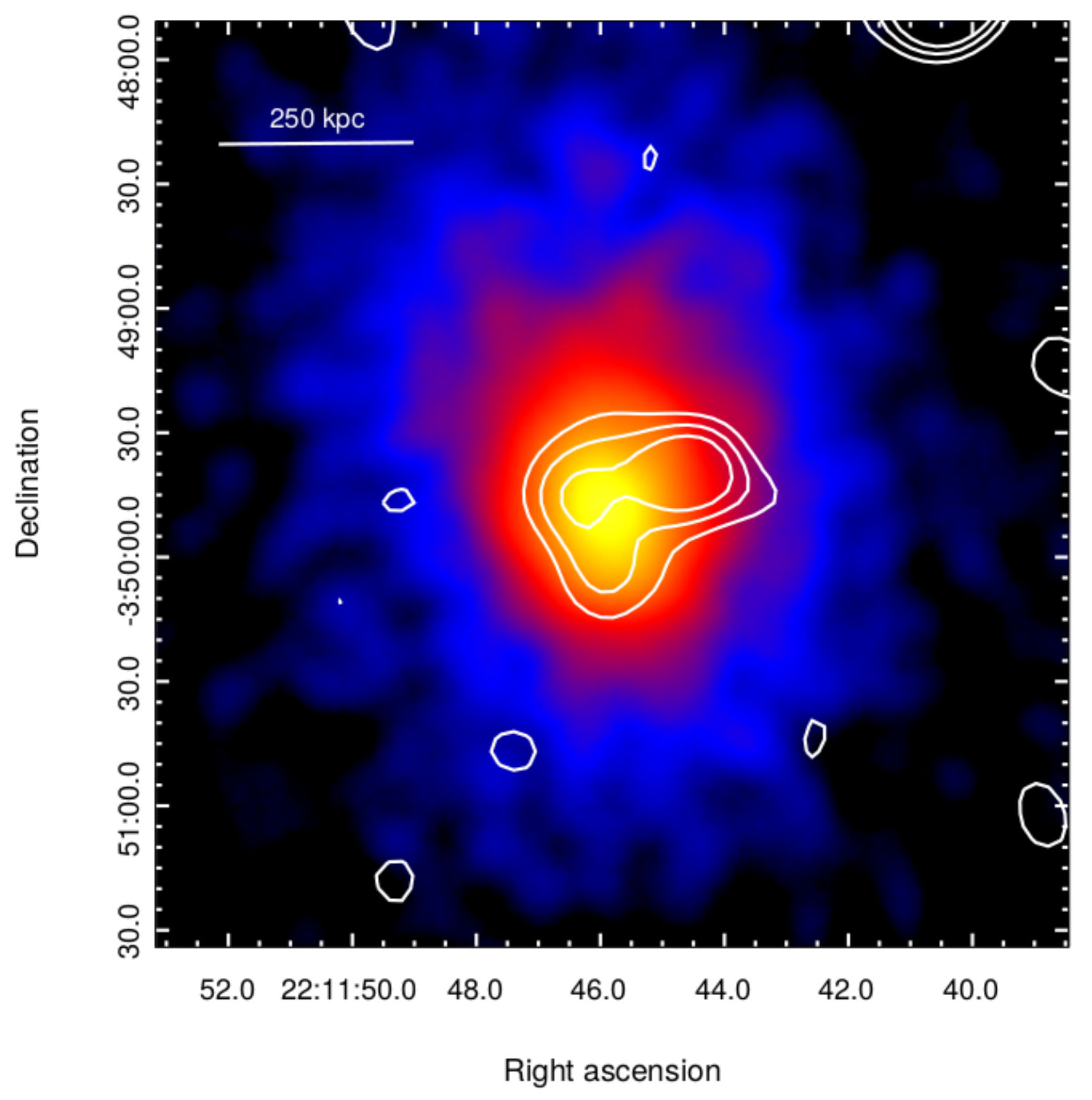}
\endminipage\hfill
\caption{\footnotesize \emph{Top panel}: total intensity of MACS J2211.7$-$0349 at 1.53 GHz superposed on the optical image. Discrete sources are labeled as A, B, C. The image has an FWHM of 20\arcsec $\times$ 13\arcsec at PA $-14^\circ$. The contour levels are (1.5, 2, 3, 4.5)$\times\sigma$ with rms noise $\sigma$ of 0.04 mJy per beam. \emph{Bottom panel}: radio image of the central region of MACS J2211.7$-$0349 after subtraction of discrete sources. The HPBW is 25\arcsec, and the noise level is 0.03 mJy per beam. Contours mark $-$0.1, 0.1, 0.15, and 0.2 mJy per beam. The radio contours are shown overlaid on the Chandra X-ray image.}
\label{j2211}
\end{figure} 

{\bf MACS J2129.4$-$0741} (morphology class 3) -- The C-array images of this dynamically disturbed cluster at $z = 0.5889$ show a resolved faint structure; our D-array data are severely corrupted by the Sun.

From the image presented in Fig.~\ref{j2129}, we measured a total flux of the diffuse source of 0.33 mJy corresponding to a radio power of 4.70 $\times$ 10$^{23}$ W Hz$^{-1}$. We tentatively classify this source as a faint, small radio halo.

{\bf MACS J2140.2$-$2339} (no figure presented; MS2137.3$-$2353; morphology class 1) -- Because of strong RFI we were not able to produce a useful image from our D-configuration data (this target was not observed in the C configuration). A pointlike source is detected in the NVSS data and hence classified as unresolved by us. However, when comparing the NVSS data with a high-resolution image by \cite{Yu2018}, we note a flux excess in the NVSS (3.8 mJy vs.\ 1.39 mJy) suggesting the presence of a possibly extended radio structure near the BCG of this relaxed cluster.  

\begin{figure}[ht]
\minipage{0.4\textwidth}
\includegraphics[width=\linewidth]{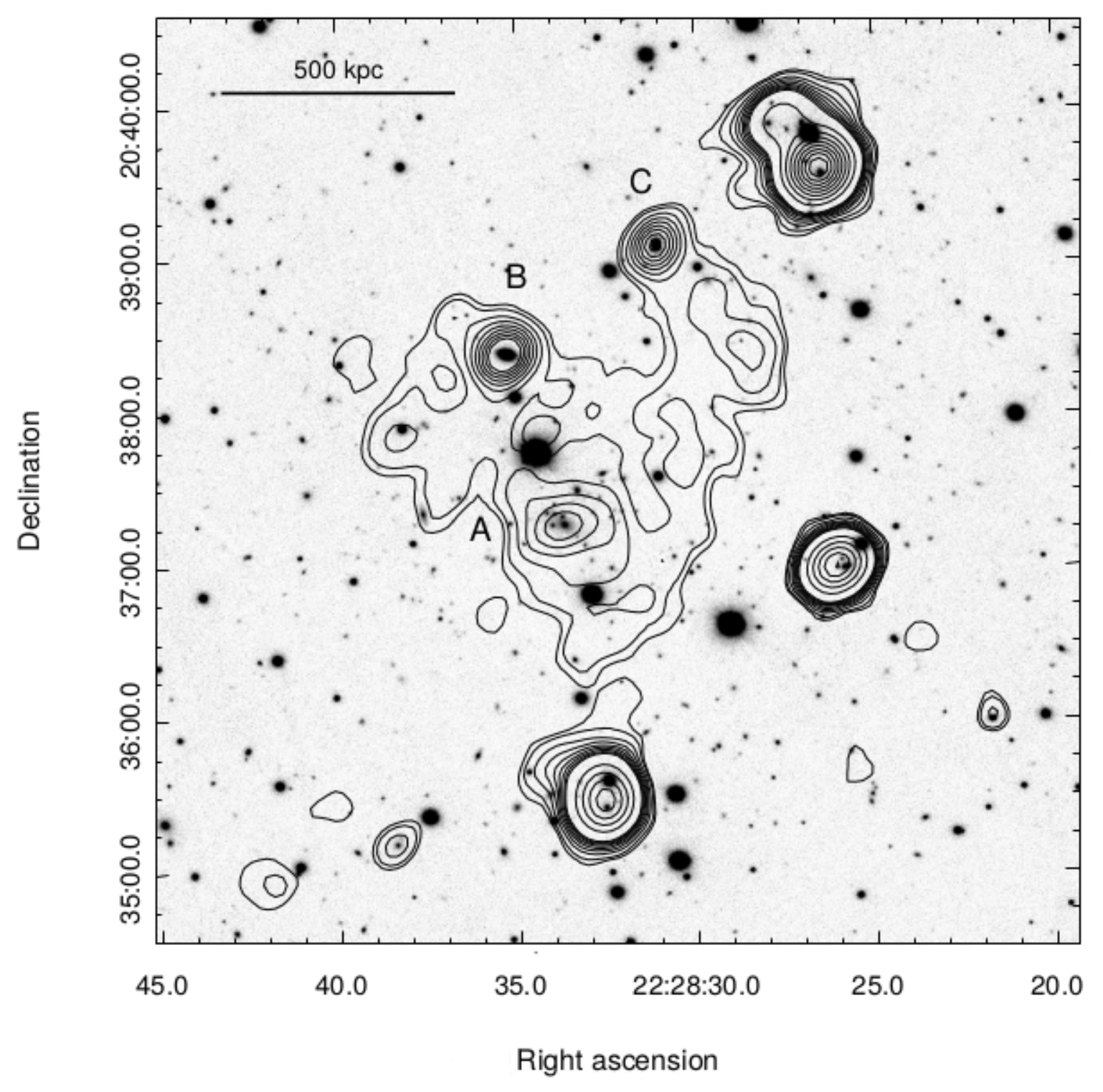}
\endminipage\hfill
\minipage{0.4\textwidth}
\includegraphics[width=\linewidth]{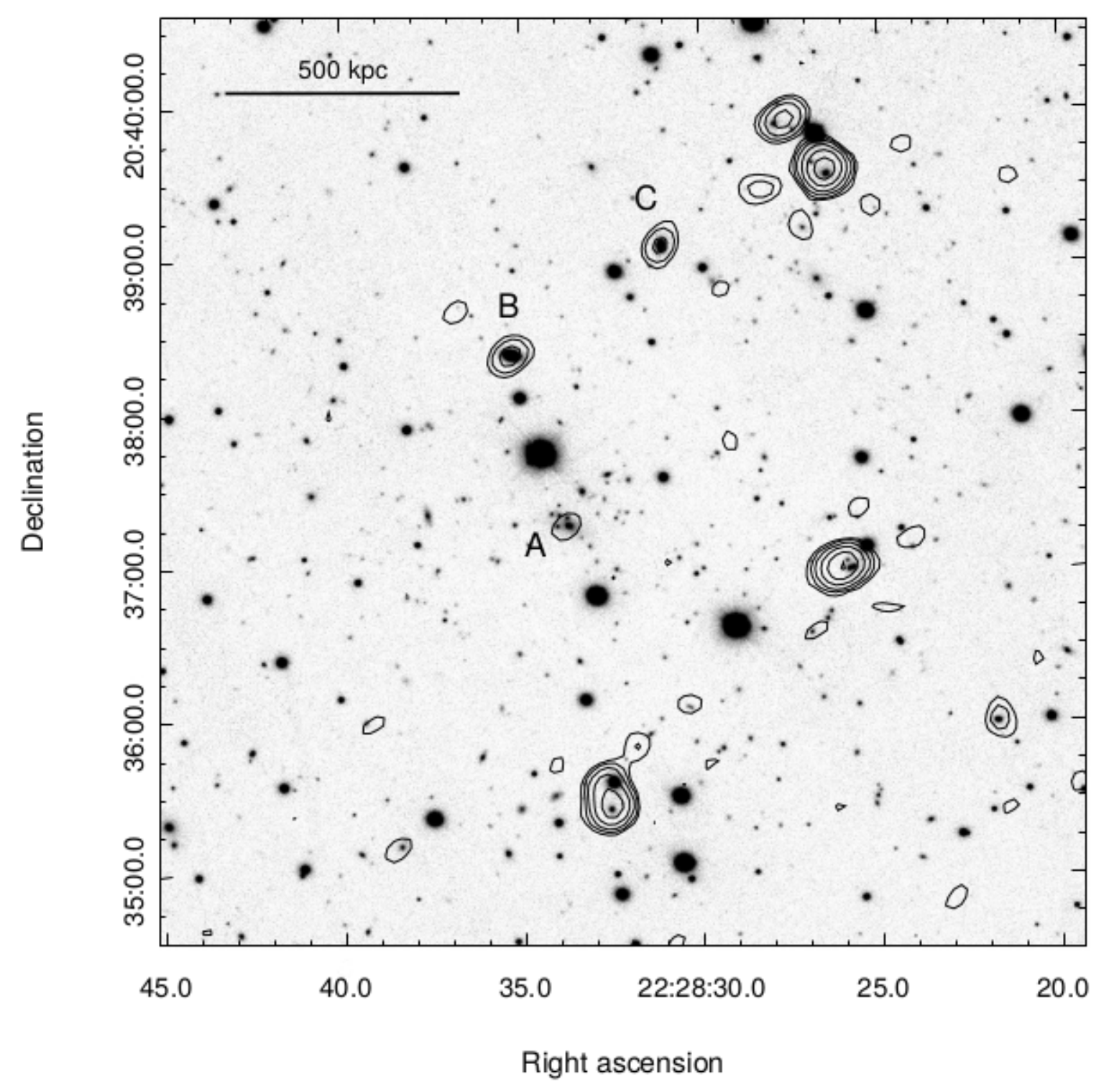}
\endminipage\hfill
\caption{\footnotesize \emph{Top panel}: radio contours of MACS J2228.5$+$2036 at 1.53 GHz overlaid on the optical image. Point sources embedded in the halo are labeled as A, B, C. The image has an FWHM of 19\arcsec $\times$ 16\arcsec at PA $-43^\circ$. Contour levels are (3, 3.5, 4.5, 6, 7, 9, 12, 24, 48, 96)$\times\sigma$ with $\sigma=$0.08 mJy per beam. \emph{Bottom panel}: same as top panel but with HPBW = 13.6\arcsec $\times$ 6.4\arcsec at PA= 58$^\circ$ and a noise level of 0.06 mJy per beam.} 
\label{j2228}
\end{figure} 

{\bf MACS J2211.7$-$0349} (morphology class 2) -- We did not find any radio information on this cluster in the literature. In our images small-size diffuse emission is seen at the cluster center. After subtraction of discrete sources we classify this diffuse emission as a possible halo source (see Fig. \ref{j2211}). The deconvolved source size is $\sim$ 40\arcsec ($\sim$220 kpc), with a flux density of $(0.58\pm 0.05)$ mJy. The radio image has an angular resolution of 25\arcsec with a noise level of 0.03 mJy per beam.

{\bf MACS J2214.9$-$1359} (no figure presented; morphology class 2) -- Our observations of this cluster detected no radio emission down to a detection limit of 0.05 mJy per beam at a HPBW of 30\arcsec.

\begin{figure}[ht]
\includegraphics[scale=0.40, angle = 0]{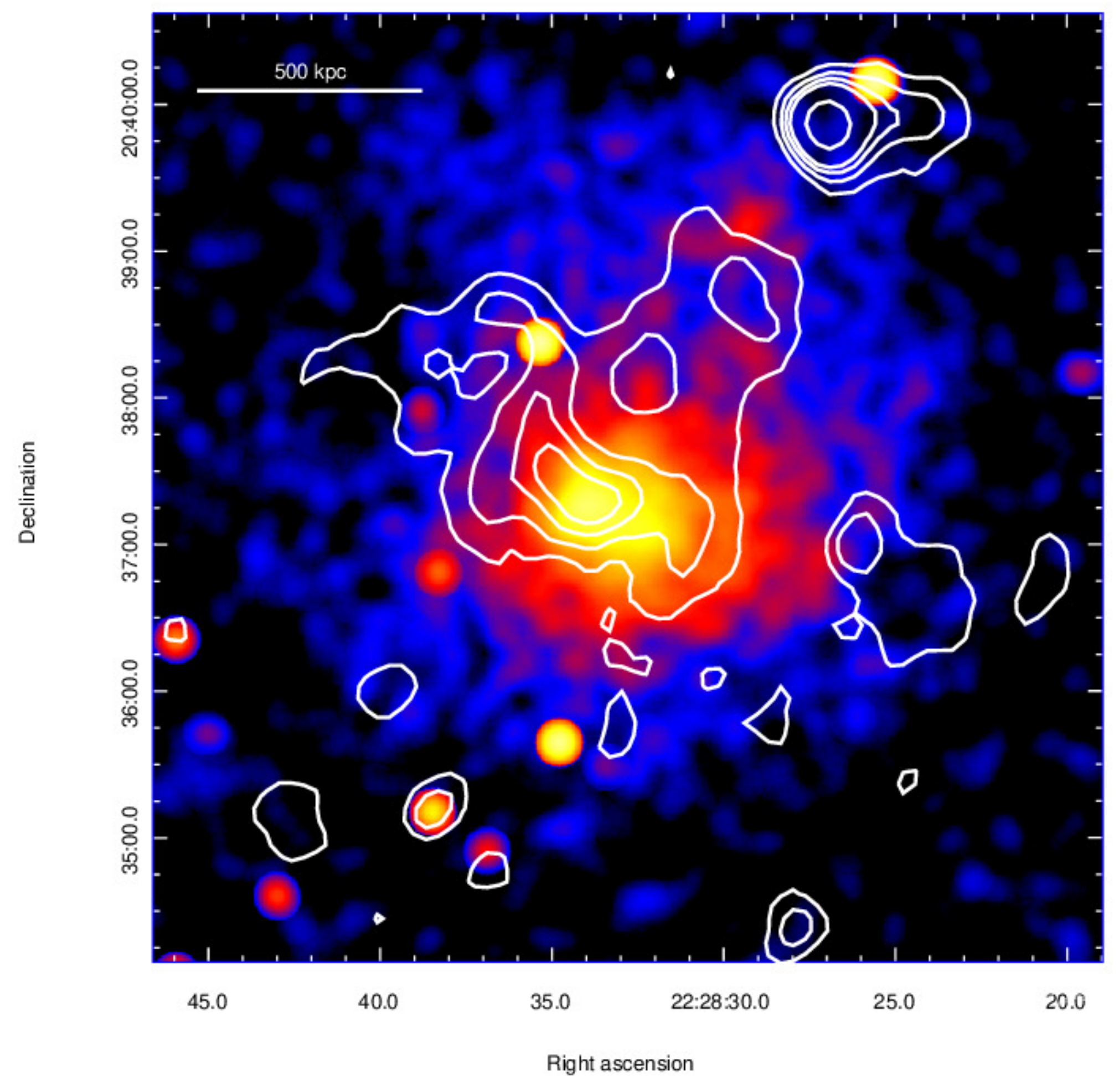}
\caption {Radio contours in the central region of MACS J2228.5$+$2036 after subtraction of all discrete sources. The HPBW is 24.8\arcsec $\times$ 20.6\arcsec at PA $-18^\circ$, and the noise level is 0.03 mJy per beam. Contours are placed at 3, 6, 9, 12, 24, and 48 mJy per beam. The radio contours are shown superposed on the Chandra X-ray image.
}
\label{j2228b}
\end{figure}

{\bf MACS J2228.5$+$2036} (morphology class 4) -- Radio observations of this disturbed cluster with the GMRT at 610 MHz by \cite{Venturi2008} detected no diffuse emission. \cite{Parekh2017} report upper limits at 610 MHz from the rms in the cluster's central region.

The radio contours at 1.5 GHz obtained by us with the JVLA in the C+D configuration are shown in the top panel of Fig.~\ref{j2228} overlaid on the PanSTARRS optical image. We detect diffuse  radio emission of low surface brightness, which we classify as a radio halo, and several discrete radio sources, three of them embedded in the diffuse emission, labeled A, B, and C (bottom panel of Fig.~\ref{j2228}).
By subtracting all discrete sources from the (u,v) data, we obtained the image shown in Fig. \ref{j2228b}. The radio halo is irregular in structure, but centered on the cluster X-ray emission. Its  angular extent is about 200\arcsec, corresponding to a linear size of 1.09 Mpc (among the largest in our sample), and its flux density is $(15.0\pm 0.1)$ mJy. Comparison of this total flux density with the upper limit estimated by \cite{Parekh2017} at 610 GHz (3.13 mJy for a halo of a size $\sim$1 Mpc) yields a spectral index for the halo of less than 1.9. Deeper GMRT observations are necessary to secure a detection at 610 MHz.
  
{\bf MACS J2229.7$-$2755} (no figure presented; morphology class 1) -- Only D-configuration data are available for this relaxed cluster. We detect an unresolved source with a flux density of $(4.3\pm 0.2)$ mJy at the cluster center at 22\h 29\m 45.16\s, --27\deg 55\arcmin 30.6\arcsec.

{\bf MACS J2243.3.3$-$0935} (no figure presented; morphology class 3) -- GMRT observations at 610 MHz reported by \cite{Cantwell2016} detected a radio halo at the cluster center, as well as a potential radio relic candidate. We did not observe this cluster and list the results of \cite{Cantwell2016} in Table 2, after scaling to 1.5 GHz using a spectral index of 1.1.

\begin{figure}[ht]
\centering
\includegraphics[scale=0.50, angle = 0]{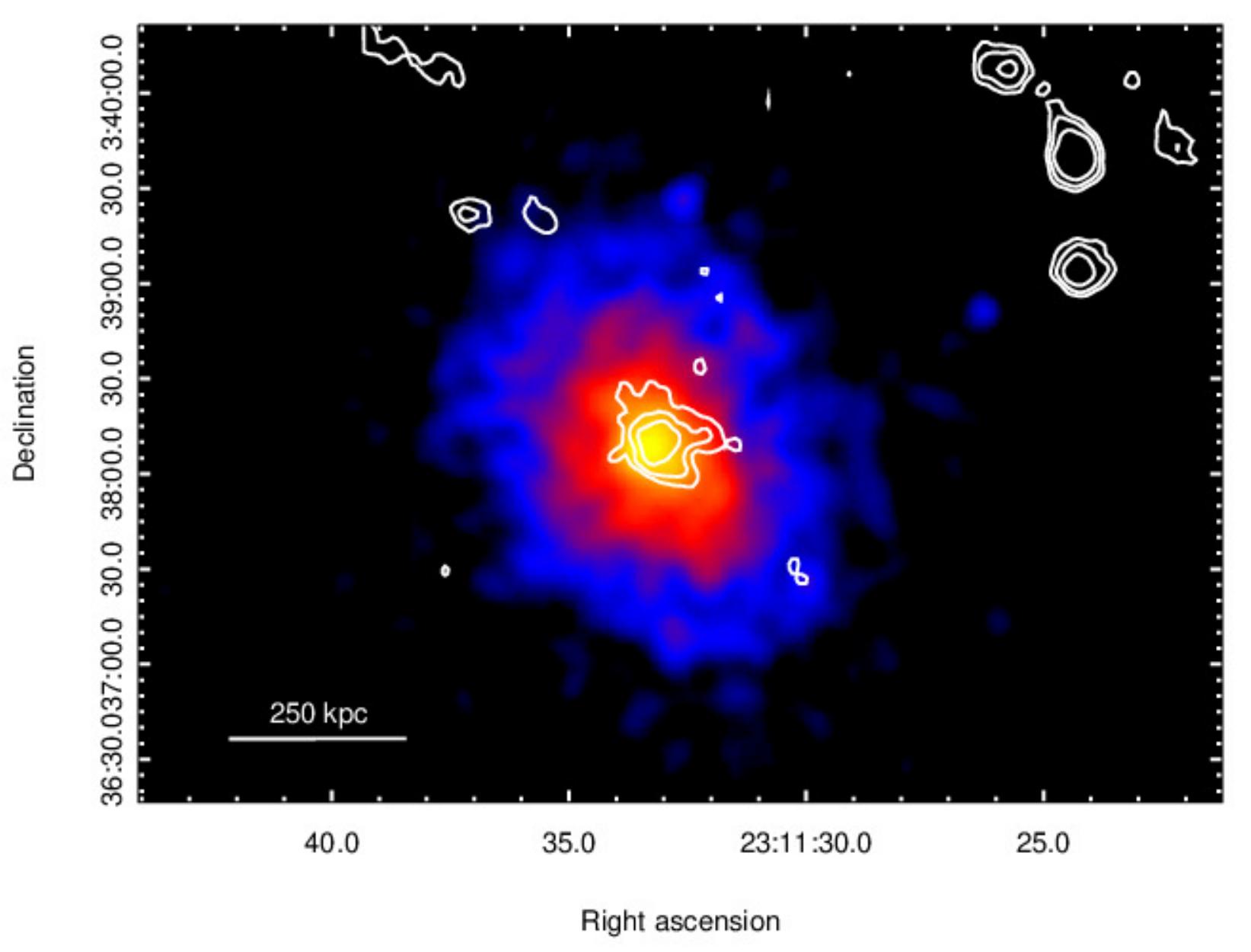}
\caption{Radio contours of the central region of MACS J2311.5$+$0338 (A2552) overlaid on the Chandra X-ray image. The HPBW is 15\arcsec. The noise level is 0.017 mJy per beam. Contours are shown at 0.05, 0.07, and 0.1 mJy per beam.} 
\label{j2311}
\end{figure}

{\bf MACS J2245.0$+$2637} (no figure presented; morphology class 1) -- \cite{Giacintucci2017} do not report any detection of diffuse radio emission at the center of this cluster. Our images show unresolved emission at the cluster center, possibly associated with the brightest cluster galaxy at position 22\h 45\m 04.64\s, +26\deg 38\arcmin 05.0\arcsec and featuring a flux density of $(4.5\pm 0.1)$ mJy, in agreement with \cite{Giacintucci2017}. 

 
{\bf MACS J2311.5$+$0338} (A2552; morphology class 3) -- A possible radio halo was discovered in this cluster by \cite{Kale2015}, using GMRT observations at 610 MHz. We used our observations combined with archival JVLA data in the L band in the B and C configuration. Using B-configuration data we found four discrete sources in agreement with \cite{Kale2015}. After subtraction of these sources, we detect in the combined image diffuse emission in a small region at the cluster center (Fig.~\ref{j2311}). The total flux density (0.32 mJy) and size (150 kpc) suggest a radio halo that is smaller and fainter than the tentative halo source detected by \cite{Kale2015}.

\section{Discussion}

Most clusters show evidence of diffuse emission in the radio band. Various origins and types of emission have been established in the literature: 1) diffuse radio emission in the central region of clusters is produced by turbulence in the ICM due to cluster mergers, and is referred to as a halo; 2) diffuse radio sources at the cluster periphery are created by shock waves in the ICM triggered by cluster mergers, and are called relics; 3) emission from the BCG at the center of cool-core clusters is sometimes accompanied by the presence of a diffuse mini-halo. All these different kinds of emission demonstrate that relativistic particles are present in the ICM, together with large-scale magnetic fields. We note that the classification of extended sources as halos or mini-halos is not just related to the source size, but also to the physical processes involved in their formation.

\begin{figure}[ht]
\centering
\includegraphics[scale=0.6, bb=60 550 470 800, clip]{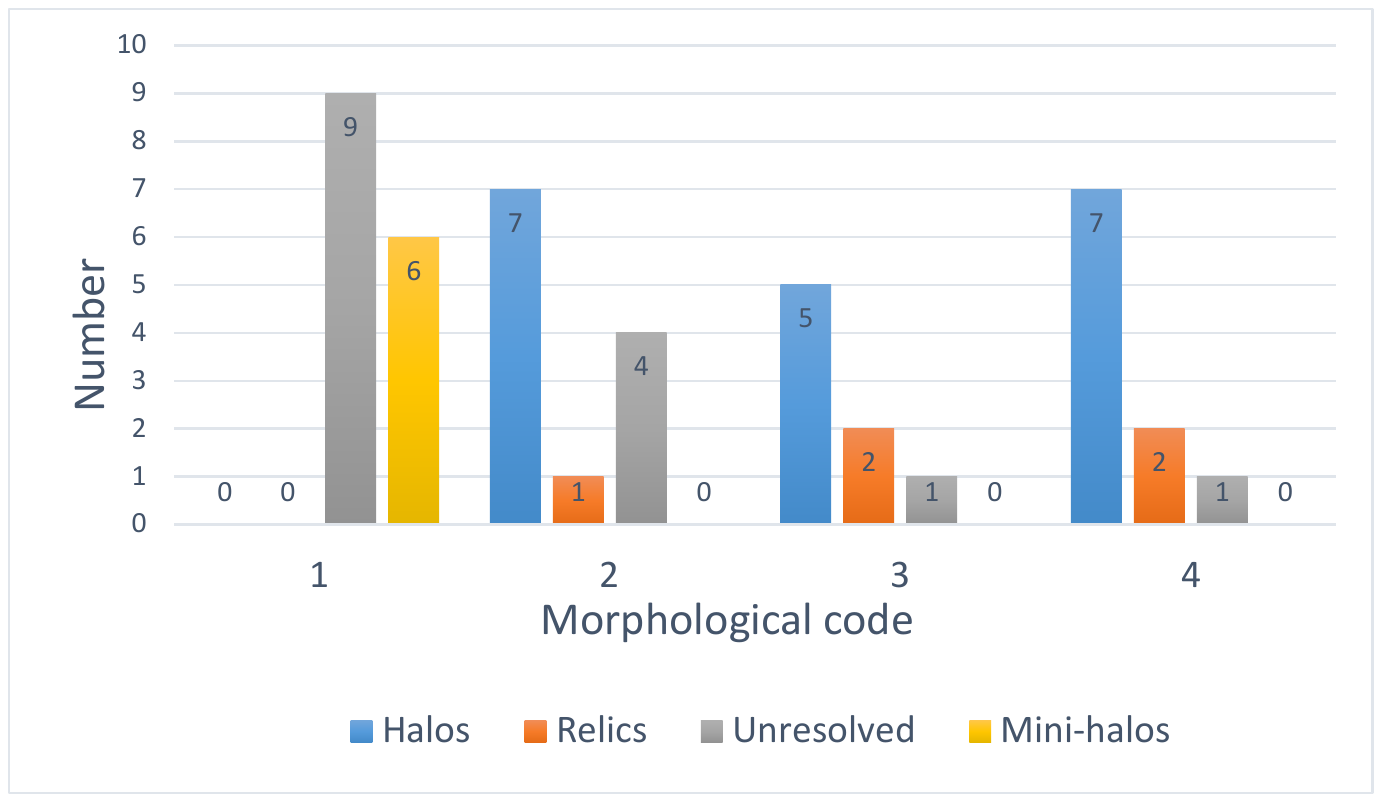}
\caption {Distribution of radio structures detected in our cluster sample as a function of morphological class. Note that clusters hosting two different types of diffuse structures appear twice, while radio-quiet clusters do not appear at all.
}
\label{detections}
\end{figure}

We find no significant difference between the detection statistics for the two cluster samples discussed here. Radio halos and relics are detected in clusters of morphology code 2, 3 and 4, whereas fully relaxed clusters (code 1) show unresolved sources or mini-halos, as summarized in Fig.~\ref{detections}. A detailed discussion is presented in the next subsections.

\subsection{Radio Halos}

A radio-halo source (including three halo candidates) is found in 19 clusters of our sample of X-ray luminous clusters, corresponding to 41\%. This value confirms previous claims that the occurrence of radio halos is high for clusters of high X-ray luminosity \citep{Giovannini2002,Feretti2012}.

This percentage increases if we consider only clusters featuring a disturbed X-ray morphology, i.e., actively merging systems. We detected diffuse halo emission in 12 clusters out of the 17 clusters with a morphological code of 3 or 4. This high percentage (71\%) underlines that halo sources are a common characteristic of merging clusters (in addition, one merging cluster shows no halo but a double relic structure). We detected radio-halo sources also in seven clusters with code 2 that show evidence of ICM disturbance in Chandra X-ray images. 

\begin{figure}
\centering
\includegraphics[scale=0.45, angle = 0]{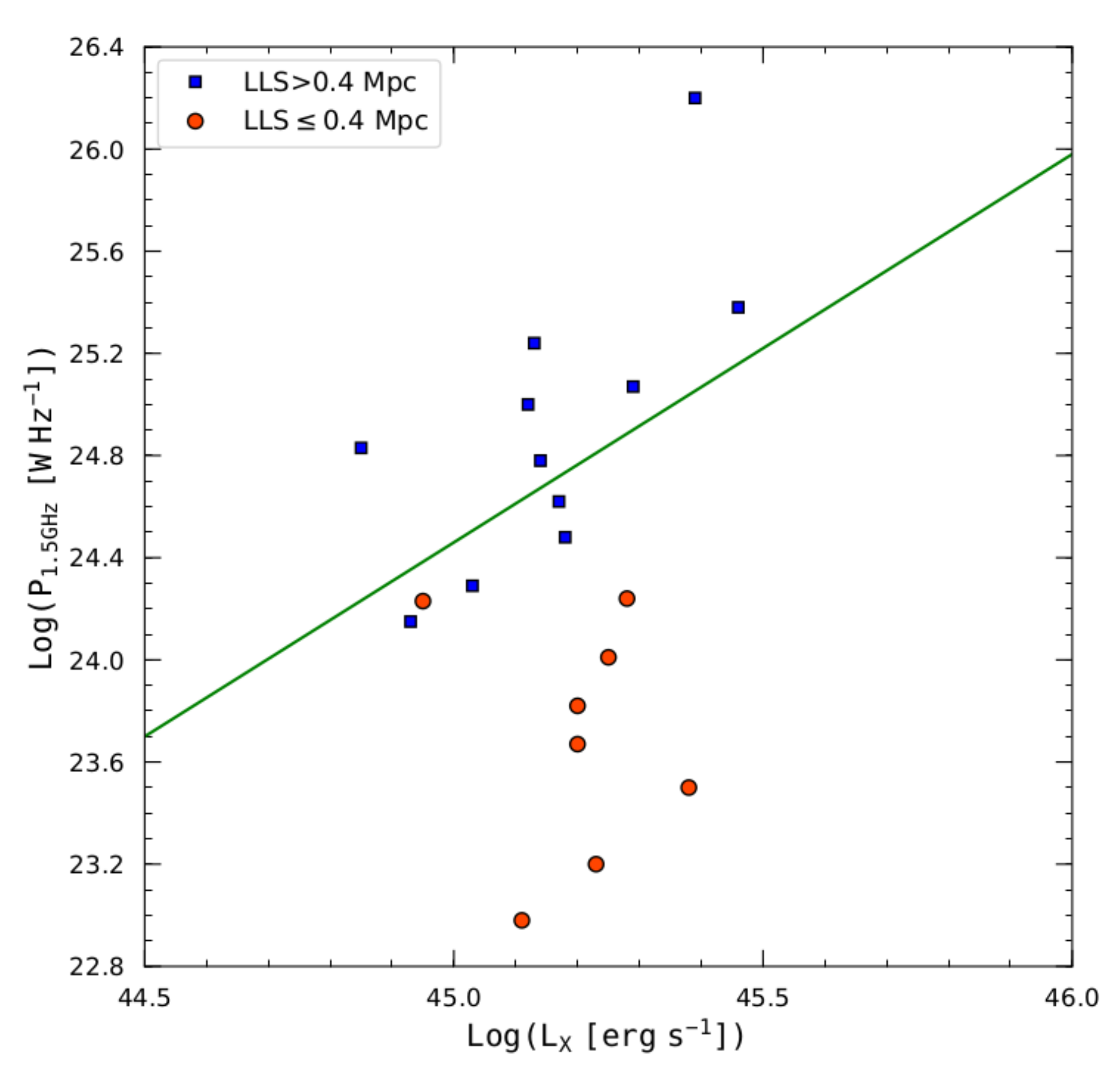}
\caption {Radio-power distribution of radio halos at 1.5 GHz versus cluster X-ray luminosity (0.1--2.4 keV). Blue squares represent radio halos in our sample with a size in excess of 0.4 Mpc, red circles mark radio halos with a size below 0.4 Mpc. The line is the best fit to data for a collection of radio halos in \cite{Feretti2012}, all in clusters at $z<0.3$. The halo with the highest radio power is hosted by MACS J0717+3745.
}
\label{corr}
\end{figure}

In Fig.~\ref{corr} we present the radio-power distribution of radio halos at 1.5 GHz in our sample versus the cluster X-ray luminosity in the 0.1--2.4 keV band. Red circles represent radio halos with a maximal linear size of less than 0.4 Mpc, while blue squares mark more extended radio halos. For comparison, we show the best-fit line obtained for halo clusters at $z<0.3$ from \cite{Feretti2012}. The most extreme outlier is the well known complex cluster MACS J0717$+$3745 (isolated blue square at the highest radio power $\sim$1.6 $\times$ 10$^{26}$ W Hz$^{-1}$), presumably because of its peculiar triple-merger activity \citep[see comments and Fig.~8][]{Feretti2012} and the presence of not only a radio halo, but also of a relic, a bridge, and a radio arc \citep{Bonafede2018}.
  
A comparison of the distribution in the P$_{1.5}$ -- L$_X$ plane between nearby ($z<0.3$) and more distant clusters ($z>0.3$) shows that large radio halos ($>$0.4 Mpc) follow the same approximate relation at high redshift as at low redshift. By contrast, small radio halos ($\le$0.4 Mpc) show a different distribution (red circles).

\begin{figure}
\centering
\includegraphics[scale=0.5, bb=20 30 490 300, clip]{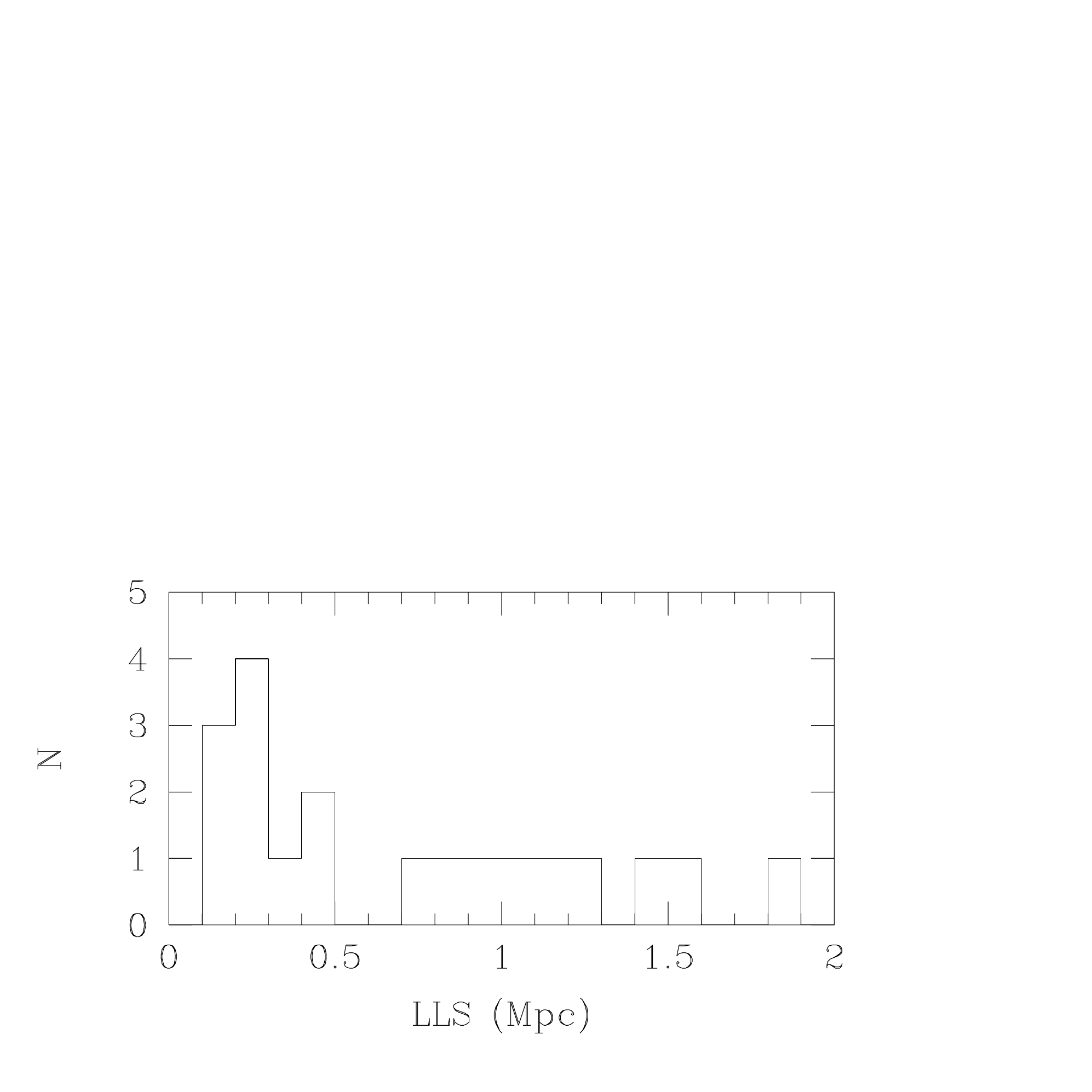}
\caption {Distribution of the maximum linear size of radio halos in the clusters from our sample.}
\label{hist1}
\end{figure}

We detect, for the first time, an excess of small radio halos in very X-ray luminous clusters. Specifically, the number of powerful radio halos larger than 1 Mpc in our sample is $\sim$15\%  (Fig.~\ref{hist1}), whereas \cite{Feretti2012} find $\sim$44\% for their sample of nearby ($z<0.3$) clusters. These small halos show a different behaviour, in that their radio power is lower than 
expected from their high X-ray luminosity. Among the eight clusters with a small radio halo, only MACS J1731.6$+$2252 is in agreement with the radio power -- X-ray luminosity relation. We note that, although low-power halos are also present in low-redshift clusters \citep{Feretti2012}, the ones combining low radio power and small size also feature low X-ray luminosity. By contrast, seven of the eight low-power radio halos found here are located in very X-ray luminous clusters ($L_{\rm X}>10^{45}$ erg s$^{-1}$).

\begin{figure}
\centering
\includegraphics[scale=0.4, angle = 0]{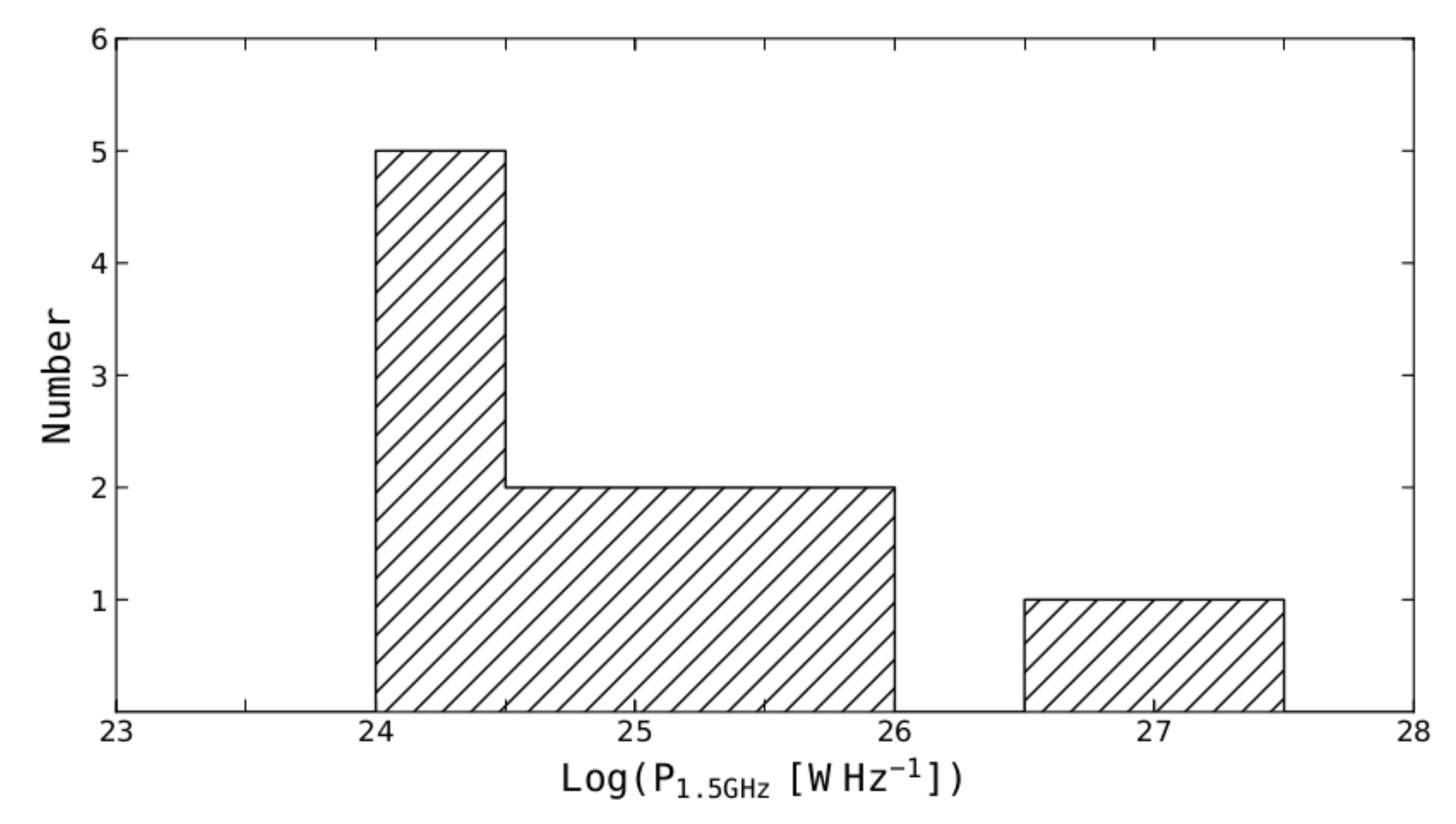}
\caption {Radio-power distribution at 1.5 GHz of unresolved sources at the center of relaxed clusters (morphology class 1 and 2)}
\label{hist2}
\end{figure}

The radio powers of the seven outliers in the P$_{1.5}$ -- L$_X$ relation fall below the best-fit line shown in Fig.~\ref{corr} by factors between 5 and 40. Although it is possible that our observations detect only the brightest region of these halos and miss the extended outer regions (see also the discussion on the cosmological surface-brightness dimming in Sect.~3), the undetected regions are of very low surface brightness and unlikely to contribute more than 50\% of the total radio power, far too little to explain the observed large shortfall. Even allowing for dispersion in the P$_{1.5}$ -- L$_X$, the seven outliers remain very faint and inconsistent with the correlation.

The existence of low-power halos in very X-ray luminous mergers  has never been predicted, and no such systems are seen in previously published data. Deeper and multi-frequency observations are necessary to confirm our findings and to fully understand the non-thermal properties of these sources.
Our results suggest that the halo radio power could be due both to the presence of an ongoing major merger and to the evolutionary history of the cluster over longer timescales. This hypothesis is supported by the work of \cite{Mann2012}, who found strong evidence for merger evolution with redshift, suggesting that the increase of the merger fraction starts approximately at $z = 0.4$. Similarly, it was shown that clusters strongly evolve from $z\sim 0.5$ to now, on average doubling their masses during this time \citep{Boylan2009,Haines2018}.  The merger history is central to understanding radio halos, since the energy injected during each merger contributes to the creation of relativistic particles, as shown in studies of the evolution of the ICM that consider all mergers during the cosmological history of a cluster \citep[merger trees;][]{Cassano2005}. 

We thus speculate that most of the nearby clusters are ``evolved'' systems in the sense that a large number of previous mergers produced a large pool of relativistic electrons. The acceleration of this population by the intense turbulence in the ICM during a major merger gives rise to giant radio halos. At higher redshift, we expect to find ``less evolved'' clusters, in which a smaller number of previous mergers created far fewer relativistic electrons, likely concentrated at the cluster center where turbulence is more intense and efficient. In these clusters, even a major merger would only produce a faint (and small) halo.

Our results need to be confirmed in two main regards: i) the sample of low-power halos contains three objects classified as `candidates'  (MACS J2129.4$-$0741, MACS J2211.7$-$0349, MACS J2311.5$+$0338, see Sect.~3); ii) more sensitive radio data are required to properly image the faint halos. We also acknowledge that some contamination could be present from extended discrete sources not recognized as such in the existing images. We note, however, that in this case the radio power of the halo would be even lower when corrected for contamination, creating an even larger discrepancy with the best-fit relation shown in Fig.~\ref{corr}. 

To summarize, the presence of a radio halo in a cluster, whether it is giant and powerful or small and faint, could be related not only to an ongoing major merger, but also to the past merger history of the cluster. The X-ray luminosity in a merger is a very good indicator of the occurrence of a major merger between high mass systems. We can consider the halo radio power (in relation to the cluster X-ray luminosity) as an indicator of these systems' evolutionary history. If our hypothesis is correct, the small, faint radio halos found in our sample will grow with time, as clusters continue to evolve, and may eventually give rise to giant radio halos. 
A study on larger complete samples at high redshift would be important to test this interpretation and improve our understanding of non-thermal cluster properties and their correlation with cluster evolution. 

Using clusters from the work presented here and from \cite{Feretti2012}, we did not find any significant correlation between the halo radio power and redshift, or between the halo size and redshift. We confirm the correlation between halo radio power and size, presented by \cite{Feretti2012}. The present sample adds information on the low-power and small halos in bright and distant clusters, in agreement with the size distribution shown in Fig.~\ref{hist1} and radio powers in Fig.~\ref{corr}.

\subsection{Clusters with a peripheral relic source}

Only five out of 46 clusters in our sample contain a diffuse relic source. MACS J0014.3$-$3022 (A2744) and MACS J1131.8$-$1955 (A1300) show strong merger activity, a powerful central halo and diffuse radio relics at the cluster periphery. In particular MACS J0014.3$-$3022 (A2744) shows four peripheral relic sources, suggesting a large number of expanding shock waves after a major merger event involving massive clusters. A relic source is also seen in the peripheral region of MACS J0520.7$-$1328, a relaxed cluster with an evident peripheral minor merger that is unable to destroy the cooling cluster core but creates shocks in the peripheral region. The cluster MACS J0025.4$-$1222 shows a double relic structure detected in GMRT observations, but not by us, and GMRT observations also show both a radio halo and a potential relic candidate in the cluster MACS J2243.3$-$0935 (which was not observed by us).

The rarity of relic sources in our sample, which contains many X-ray luminous clusters undergoing merging processes, is unexpected, given the known higher percentage of relics in nearby clusters \citep[e.g.][]{Feretti2012}. Like for radio halos, we speculate that the cluster merger history is crucial, i.e., that the number of previous merger events is not large enough in most high-redshift clusters to create a pool of relativistic electrons that extends out to the peripheral regions. As a result, the number of relativistic electrons is too low in the outskirts of many distant clusters to create relics from the interaction with cluster shocks. 

\subsection{Unresolved radio emission and mini-halos}

Thirteen relaxed clusters (nine with morphological code 1 and four with code 2) show central radio emission only from the BCG. Moreover, in six additional clusters (code 1) the active BCG is surrounded by a radio mini-halo. This high fraction (19/46 = 41\%) of strong radio emission from the BCG indicates that at this relatively high redshift (the most distant cluster is MACS J0744.8+3927 at $z = 0.6976$) feedback already occurs between the cluster cool core and BCG activity. As expected in this scenario, the BCG is rather powerful; indeed most BCG show radio powers in the range of FRII radio galaxies (see Fig.~\ref{hist2}).

The small number of mini-halo sources detected in our observations is not unexpected, because of the relatively low resolution and sensitivity. Since the angular resolution of the observations is typically of the order of 20\arcsec (which corresponds to a linear extent of 90 to 140 kpc at the redshift of our targets), it is possible that some sources classified by us as unresolved may in reality include diffuse emission around the true pointlike source. Moreover, longer observations and better (u,v) coverage are necessary to avoid dynamic-range problems in detecting low-brightness structures near a powerful source. 

Two disturbed clusters (morphological code 3 and 4) were classified by us as showing only unresolved radio emission. The first one, MACS J2049.9$-$3217, is a complex system with other, unrelated radio sources in the field (see Section 3.1). The second one, MACS J0404.6$+$1106, shows evidence of a very recent merger: the BCGs of the two merging clusters are clearly visible, and one of them is still active in the radio band, suggesting that at least the core of the cluster with the active BCG was relaxed before the merger and is not yet affected by the merger.

\subsection{Radio-quiet clusters}

We label clusters as ``radio quiet'' if neither halo, relic, mini-halo, nor BCG radio emission is detected. In our total sample of 46 clusters, only four do not show any evidence of diffuse non-thermal emission or radio emission from the central dominant galaxy. Two of them (MACS J1319.9$+$7003 and MACS J2214.9$-$1359) are classified as code 2, i.e, close to relaxed. Since BCGs in relaxed clusters often show restarted activity \citep[e.g.,][]{Liuzzo2010}, the dominant galaxies in these clusters may currently be in their quiescent phase. The two other clusters are MACS J0035.4$-$2015 and  MACS J0911.2$+$1746, both disturbed systems (code 3 and 4, respectively). The lack of radio emission from these two clusters could suggest either a very recent merger which has not yet generated  diffuse emission, or the presence of diffuse emission of surface brightness that falls below our sensitivity limit.

\section{Conclusions}

In this paper, we have analyzed the radio emission of 46 high-redshift ($z>0.3$) clusters of galaxies, to compare their properties with those of low-redshift systems and to investigate evolutionary features from radio and X-ray information. The objects under study are very luminous X-ray clusters belonging to two statistically complete samples from the Massive Cluster Survey (MACS): the first sample consists of  34 clusters with nominal X-ray fluxes S$_{\rm 0.1-2.4 keV}\ge 2 \times 10^{-12}$ erg s$^{-1}$ cm$^{-2}$ in the ROSAT Bright Source Catalogue and redshifts of $0.3 < z <0.5$ \citep{Ebeling2010}. The second sample includes the 12 most distant MACS clusters at $z >0.5$, presented by \cite{Ebeling2007}.  The clusters in the total sample are characterized by a  wide range of morphologies, parameterized by a morphological code ranging from 1 to 4, with no obvious bias in favor of either relaxed or merging systems. 

The radio information was obtained from new JVLA observations of the sample at 1.5 GHz, from JVLA archive data for a few sources, or from the literature for a few well known clusters. Deeper and higher-resolution data will be crucial to confirm the faintest radio sources. X-ray data, needed for the comparison between extended radio features and the X-ray surface-brightness distribution, were extracted from the Chandra Data Archive. 
\par\noindent
Our results can be summarized as follows: 

1) Most clusters show either diffuse radio emission or radio emission from the brightest galaxy at the cluster center.  We present images for the most interesting objects, and a short description of each cluster. We newly discovered ten unresolved or slightly resolved sources at the cluster center and nine radio halos. Moreover, for seven clusters our new data significantly improve the knowledge of the non-thermal emission. 

2) All relaxed clusters of morphology class 1 show central radio emission associate with the BCG. In six of them, the BCG is surrounded by a radio mini-halo. Unresolved emission is also detected in four clusters of morphology class 2. This result indicates that, at the relatively high redshift of our sample, feedback already occurs between cool cluster cores and active BCGs. The radio power of the BCG is high, mostly in the range of FRII radio galaxies. 

3) Clusters of morphology code 2, 3, and 4 host radio halos and relics, as well as a few unresolved sources. 

4) A radio-halo source (including three halo candidates), is found in 19 clusters corresponding to a percentage of 41\%, in agreement with the very high X-ray luminosity of all clusters under study. The percentage increases to 72\% if only clusters with an evidence of disturbed X-ray structure are considered, thus confirming that halo sources are a  common characteristic of merging clusters. 

5) While powerful radio halos with size of at least 0.4 Mpc follow the same radio-power vs X-ray-luminosity relation at low and high redshift, we detect for the first time  seven low-power radio halos, all with size $\le$ 0.4 Mpc, that show a different behaviour, in the sense that their radio power is far lower than expected from their host clusters' high X-ray luminosity. 

6) A diffuse relic source is found in only five out of the 46 clusters, three of which also host a radio halo. It is possible that the number of relativistic electrons in peripheral regions is too low to allow the formation of relic sources at these higher redshifts.

7) The low number of detected relic sources and the detection of seven low-power radio halos suggest redshift evolution in the properties of diffuse sources. In particular, since the respective clusters are all very X-ray luminous and characterized by an ongoing major merger, we suggest that the radio power (and size) of radio halos (and relics) could be related not only to the present merger state of the cluster, but also to the number of previous merger events in the history of the system. The study of a larger sample of high-$z$ clusters, supported by numerical simulations,  is necessary to test this possibility.

\section*{Acknowledgements}
We are grateful to an anonymous referee for constructive comments that helped to improve this paper. The National Radio Astronomy Observatory is operated by Ass. Univ., Inc., under cooperative agreement with the National Science Foundation. This research has made use of the NASA/IPAC Extragalactic Database (NED) which is operated by the Jet Propulsion Laboratory, California Institute of Technology, under contract with the National Aeronautics and Space Administration. AB acknowledges support from the ERC-Stg n 714245 DRANOEL and from the MIUR grant FARE SMS.

\label{lastpage}

\end{document}